\documentclass[letterpaper,11pt]{article}

\usepackage{psfig, amsmath, epsfig, amssymb,array}
\usepackage{sidecap}
\usepackage{graphicx}
\usepackage{bm}
\def\etmiss{E\!\!\!\!\slash_{T}}

\def\pslash{\not{\hbox{\kern-4pt $p$}}}
\def\qslash{\not{\hbox{\kern-4pt $q$}}}
\def\lv{\not{\hbox{\kern-4pt $L$}}}
\def\lsim{\mathrel{\raise.3ex\hbox{$<$\kern-.75em\lower1ex\hbox{$\sim$}}}}
\def\gsim{\mathrel{\raise.3ex\hbox{$>$\kern-.75em\lower1ex\hbox{$\sim$}}}}
\def\ifmath#1{\relax\ifmmode #1\else $#1$\fi}

\newcommand{\nc}{\newcommand}
\nc{\postscript}[2]{\setlength{\epsfxsize}{#2\hsize}\centerline{\epsfbox{#1}}}

\nc{\beq}{\begin{equation}}   \nc{\eeq}{\end{equation}}
\nc{\bea}{\begin{eqnarray}}   \nc{\eea}{\end{eqnarray}}
\nc{\baa}{\begin{array}}      \nc{\eaa}{\end{array}}
\nc{\bit}{\begin{itemize}}    \nc{\eit}{\end{itemize}}
\nc{\ben}{\begin{enumerate}}  \nc{\een}{\end{enumerate}}
\nc{\bce}{\begin{center}}     \nc{\ece}{\end{center}}

\nc{\non}{\nonumber}

\begin{document} 

\baselineskip=17pt


\thispagestyle{empty}
\vspace{20pt}
\font\cmss=cmss10 \font\cmsss=cmss10 at 7pt

\begin{flushright}
\today \\
UMD-PP-10-003, UCB-PTH-09/27\\
\end{flushright}

\hfill

\begin{center}
{\Large \textbf
{Distinguishing Dark Matter Stabilization Symmetries Using \\
Multiple Kinematic Edges and Cusps}}
\end{center}

\vspace{15pt}

\begin{center}
{\large Kaustubh Agashe$\, ^{a}$, Doojin Kim$\, ^{a}$, Manuel 
Toharia$\, ^{a}$ and
Devin G. E. Walker $\, ^{b,c,d}$} \\
\vspace{15pt}
$^{a}$\textit{Maryland Center for Fundamental Physics,
     Department of Physics,
     University of Maryland,
     College Park, MD 20742, U.S.A.}
\\
$^{b}$\textit{Department of Physics, University of California, Berkeley, CA 94720, U.S.A}
\\
$^{c}$\textit{Theoretical Physics Group, Lawrence Berkeley National Laboratory, Berkeley, CA 94720, U.S.A.}
\\
and
\\
$^{d}$\textit{Center for the Fundamental Laws of Nature, Jefferson
  Physical Laboratory, Harvard University, Cambridge, MA 02138,
  U.S.A.} 
\end{center}

\vspace{5pt}

\begin{center}
\textbf{Abstract}
\end{center}
\vspace{5pt} {\small \noindent 
We emphasize that the stabilizing symmetry for dark matter (DM) particles does not have to be the 
commonly used parity ($Z_2$) symmetry.  We therefore examine the potential of the colliders to 
distinguish models with parity stabilized DM from models in which the DM is stabilized by other 
symmetries.  We often take the latter to be a $Z_3$ symmetry for  illustration.  We focus on signatures 
where a {\em single} particle, charged under the DM stabilization symmetry decays into the DM and 
Standard Model (SM) particles.  Such a $Z_3$-charged ``mother'' particle can decay into {\em one or two} 
DM particles along with the {\em same} SM particles.  This can be contrasted with the decay of
a $Z_2$-charged mother particle, where only one DM particle appears.
Thus, if the intermediate particles in these decay chains are {\em
off}-shell, then the reconstructed invariant mass of the SM
particles exhibits two kinematic edges for the $Z_3$ case but only one
for the $Z_2$ case.  For the case of {\em on}-shell intermediate
particles, distinguishing the two symmetries requires more than the
kinematic edges.  In this case, we note that certain decay chain
``topologies'' of the mother particle which are present for the $Z_3$ case (but
absent for the $Z_2$ case) generate a ``cusp'' in the invariant mass
distribution of the SM particles.  We demonstrate that this cusp is
generally invariant of the various spin configurations.  We further
apply these techniques within the context of explicit models.} 
\vfill\eject
\noindent


\section{Introduction}

There is compelling evidence for the existence of dark matter (DM) in the universe \cite{Bertone:2004pz}.  These observations can be explained by the postulating of new stable particles.  A consensus picture of the nature of such a particle is provided by a host of astrophysical, cosmological and direct detection experiments:  A viable DM candidate must be electrically neutral and colorless, non-relativistic at redshifts of $z \sim 3000$ and generate the measured relic abundance of  $h^2\, \Omega_{\mathrm{DM}} = 0.1131 \pm 0.0034$~\cite{Hinshaw:2008kr}.  
Additionally a Weakly Interacting Massive Particle (WIMP) is a very well-motivated paradigm \cite{Bertone:2004pz}.  Consider DM particles as relics which were once in thermal equilibrium with the rest of the universe.  It is well-known that the measured relic abundance is correlated with the dark matter annihilation cross section~\cite{Kolb:1990vq} by 
\begin{equation}
h^{2}\,\Omega_\mathrm{DM} \,\simeq \frac{0.1 \,\,\mathrm{pb} \cdot c}{\langle \sigma v \rangle}.
\end{equation}
The annihilation cross section of a pair of dark matter particles into a 
two particle final state goes as
\begin{equation}
\langle \sigma v \rangle \sim \frac{g^4}{ 8\pi } \frac{1}{ M^2},
\end{equation}
where $g$ denotes the couplings and $M$ the masses of the particles in the dark sector.
This cross-section is naturally of the right value for $g \sim {\cal O}(1)$ and $M \sim 100$ GeV.  Moreover, many extensions of the SM at the weak scale, most of which are invoked primarily as solutions to other problems of the SM (most notably the Planck-weak hierarchy problem), contain such stable WIMPs.  Because of this possibility, it may be possible to detect DM directly via scattering off nuclei or indirectly via detection of its (SM) annihilation products \cite{Bertone:2004pz}.

Such a scenario also makes the idea of dark matter amenable for testing at the high-energy colliders.  It is possible to produce only DM particles directly at colliders, but then we do not have any visible signal since the DM particles will simply escape these detectors without interacting.  Instead we investigate events where the dark matter is produced (indirectly) along with visible SM particles from the decays of particles charged under both dark matter stabilization 
(but heavier than the DM) and the unbroken SM symmetries.  The existence of such ``mother'' particles is a feature of almost all models of physics beyond the SM that contain stable WIMPs. 

To date, a tremendous amount of effort has been made to reconstruct such events at the upcoming Large Hadron Collider (LHC) in order to determine the masses of the DM, the mother particles and possibly
intermediate particles in the decay chains.  For example, see references~\cite{Lester:1999tx,
Cheng:2008mg,Matchev:2009iw,Han:2009ss}. Most of this work has been for parity ($Z_2$) stabilized dark matter.  This is because the most popular models, e.g. supersymmetric (SUSY),
little Higgs and extra-dimensional
scenarios~\cite{Jungman:1995df, Lee:2008pc, Cheng:2003ju, Servant:2002aq, Agashe:2007jb},
all ensure the dark matter candidates remain stable by employing a $Z_2$ stabilization symmetry. Importantly these models have served as a guide of expected signatures of dark matter at the LHC \cite{ATDR, CTDR}.

In this paper we emphasize that any discrete or continuous global
symmetry can be used to stabilize dark matter.\footnote{Gauge
symmetries alone cannot be used to stabilize dark matter.  See the
discussion in reference \cite{Walker:2009en}.}  Furthermore,
because all fundamental particles in nature are defined by how they
transform under various symmetries, most of the popular ($Z_2$) models
actually consider only one type of DM candidate!  It is therefore 
critical to determine \textit{experimentally}, i.e.,~without
theoretical bias, the nature of the symmetry that stabilizes dark
matter.  We embark on a program of study to distinguish
models in which the DM is stabilized by a $Z_2$ discrete symmetry from
models in which the DM is stabilized with other symmetries.  A
beginning effort was made in reference~\cite{Walker:2009ei} which
focused on signatures with long-lived mother particles, i.e., which
decay to DM and the SM particles {\em outside} of the detector.  In this
paper we study the complementary possibility of mother particles
which decay to DM and the SM particles {\em inside} of the detector.  

Our main idea is that the final states and the ``topology'' of the
decay of a mother particle are (in part) determined by the DM stabilization symmetry.
Thus reconstructing the visible parts of these decay chains
will allow us to differentiate 
a model of DM stabilized with a non-$Z_2$ symmetry from one
where DM is stabilized with a $Z_2$ symmetry.  
In this paper we begin to explore such signatures.  
Our conclusions seem generic for most stabilization symmetries that
are not parity symmetries; however, for definiteness, we focus on the
case of a $Z_3$ symmetry.  When illustrating the signatures we will
generically refer to any model stabilized with $Z_2$ and $Z_3$
stabilization symmetry simply as $Z_2$ and $Z_3$ models, respectively.  

More specifically to see differences between $Z_2$ and $Z_3$ models,
we focus on the \textit{kinematic edges} and
\textit{shapes} of invariant mass distributions  
of the SM particles resulting from the decay of a \textit{single}
mother particle charged under the SM and the DM 
stabilization symmetries.  We note the possibility of one or two DM particles
in each decay chain being allowed by the $Z_3$ symmetry 
(along
with SM particles which can, in general, be different in the two decay chains).
%
%
%
Whereas, in $Z_2$ models, decays of a mother particle in given SM
final state cannot have two DM particles in the decay chain and hence typically has only one DM particle.  Thus, 
\begin{itemize}
\item If all the intermediate particles in the two decay chains are \textit{off-shell}
and the SM particles in the two decay chains are the {\em same},
then we show that there
are two $Z_3$ kinematic edges 
in the invariant mass distribution of this SM final state
at approximately 
$M_\mathrm{mother} - m_\mathrm{DM}$ and $M_\mathrm{mother} -
  2\,m_\mathrm{DM}$.  Models with $Z_2$ stabilized dark matter have
  only one endpoint approximately given by $M_\mathrm{mother} -
  m_\mathrm{DM}$.  
\end{itemize}

In the case of on-shell intermediate particles, the decay of such a mother in a $Z_3$ model can similarly result in
double edges due to the presence of one or two DM in the final state. 
However, in this case the endpoint also depends on the masses of
intermediate particles. Thus it is possible to obtain multiple edges
even from decay of a single mother particle in a $Z_2$ model due to different intermediate particles to the same final
state. Hence, multiple edges are not a robust way to distinguish between
$Z_3$ and  $Z_2$ symmetries in the case of on-shell intermediate
particles. 
For the case of on-shell intermediate particles, we thus use shapes of
invariant mass distributions instead of edges. In particular, 
\begin{itemize}
\item We find a unique decay chain topology with two SM particles
  separated by a DM particle (along with another DM at the end of the decay
  chain) which is generally present for $Z_3$  models but absent for the $Z_2$ case.  Based on pure
  kinematics/phase space, this topology leads to a ``cusp" (i.e.,
  derivative discontinuity) in the invariant mass distribution of the
  SM particles. 
\end{itemize}
Of course for a generic model, it is possible to have a ``hybrid"
scenario where elements from the \textit{on-shell} and
\textit{off-shell} scenarios are present.  

An outline of the paper is as follows:  In the next section, we begin
with the case of off-shell intermediate particles in a decay chain.
There we show how differing kinematic edges can be used to
distinguish $Z_2$ from $Z_3$ models.  In section~\ref{sec:onshell} we move on to the case of intermediate
particles being on-shell.  There we show the existence of a ``cusp"  in $Z_3$ models for
certain topologies; further we show that this cuspy feature survives even
when taking spin correlations into consideration.  We then discuss a
couple of explicit models -- one based on warped extra dimensional 
framework~\cite{Agashe:2004ci, Agashe:2004bm}
and another using DM stabilization symmetry from spontaneous breaking~\cite{Walker:2009en};
see also reference~\cite{Ma:2007gq} for another example of a $Z_3$-model.  Here DM is not stabilized by $Z_2$ symmetries. In the second model, we show our signal is invariant under
basic detector and background cuts.  We next conclude and briefly
enumerate how $Z_2$ models  can fake signals from $Z_3$ models.  We also mention
future work to better reconstruct and distinguish $Z_3$ from
$Z_2$ models, e.g., using the {\em two} such decay chains present in each full
event. 
%


\section{Off-shell intermediate particles}
\label{sec:offshell}

In this paper, we mostly study the decay  of a {\em single} heavy particle, which is charged
under the dark matter stabilization and SM symmetries, into SM
and dark matter candidate(s) {\em inside} the detector. Henceforth, we denote such heavy particles by
``mother'' particles.  In this section we assume that all intermediate
particles (if any) in this decay chain 
are \textit{off}-shell. This off-shell scenario has been frequently
studied by the ATLAS and CMS collaborations~\cite{ATDR, CTDR} for SUSY 
theories (which is an example of a $Z_2$ model).

We consider constructing the invariant mass distribution of the
(visible) decay products.  
Unlike for the $Z_2$ case, for $Z_3$ models a mother particle $A$ can 
decay into one or two DM particles along with 
(in general different)
SM particles. 
We mostly assume, just for simplicity, that there
exist two 
%
%
visible particles ($a$, $b$ or $c$, $d$) in the final state as
shown below (note however that the same argument is relevant to the general cases where
more than two visible particles are emitted)\footnote{See Figure \ref{z3kmod} for appearance of 
these two types of decays, including the required interactions, in the context of an explicit model.}:
\newpage
\begin{figure}[h!]
\vspace{-.35cm}
  \centering
  \includegraphics[height=3.1cm,clip]{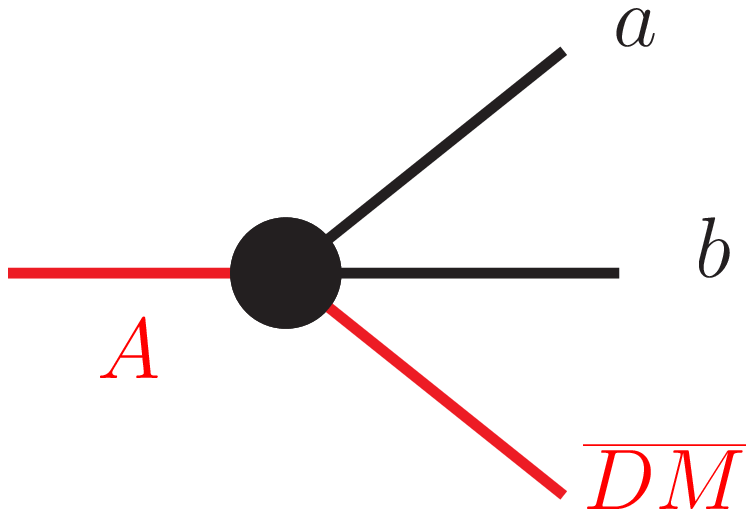} 
 \hspace{2cm} \includegraphics[height=3.3cm,clip]{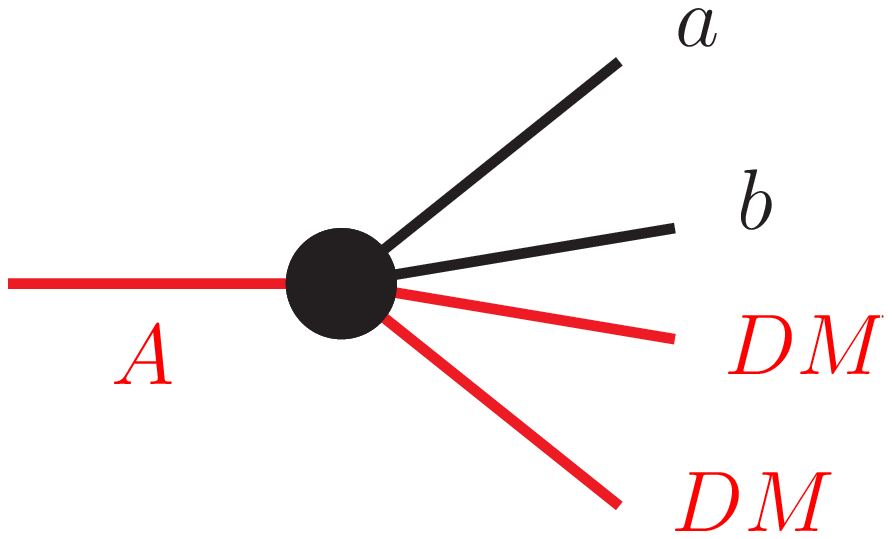}
\vspace{-2cm}
\bea \label{type1}
\eea
\vspace{.1cm}\
\end{figure}

\noindent
Here (and henceforth) the ``blob'' denotes intermediate particles in the decay which are off-shell.
Also, upper-case letters/red/solid lines denote particles charged
under the DM symmetry ($Z_3$ or $Z_2$) and lower-case letters/black/solid lines denote
SM (or ``visible'', as opposed to DM) particles, including, for example, a $W$
boson.  Such an unstable SM particle decays further into SM fermions, at least some of which are
observed by the particle detector.  

In order to avoid any possible confusion, we would like to explain the
above diagrams.
Under the $Z_3$ symmetry, a particle/field $\phi$ transforms
as 
\begin{eqnarray}
\phi & \rightarrow & \phi \; \exp \left(  \frac{ 2 \pi i q }{3} \right)
\end{eqnarray}
where $q = 0$ (i.e., $Z_3$-neutral) or $q = 1, 2$
(non-trivial $Z_3$-charge). Suppose the lightest of the $Z_3$-charged particles (labeled
$\phi_0$) has charge
$q = 1$ (similar argument goes through for charge $q = 2$ for $\phi_0$). 
Clearly, its anti-particle ($\bar{ \phi }_0$)
has (a different) charge $q = -1$ (which is equivalent to $q = 2$) and has same mass as $\phi_0$.
Then, solely based on $Z_3$ symmetry considerations,
all other (heavier) $Z_3$-charged particles can decay into
this lightest $Z_3$-charged particle (in addition to $Z_3$-neutral particles, including SM ones).
To be explicit, a heavier $Z_3$-charged particle with charge $q = 1$ can decay into either
(single) $\phi_0$ or {\em two} $\bar{ \phi }_0$'s
(and $Z_3$-neutral particles). Taking the CP conjugate of the preceding statement,
we see that a heavier $Z_3$-charged 
particle with the other type of charge, namely $q = 2$, is allowed to decay into {\em two} $\phi_0$'s
or single $\bar { \phi }_0$.
Of course, $\phi_0$ cannot decay
and thus is the (single) 
DM candidate in this theory.
%
%
We denote this DM particle and its anti-particle
by DM and $\overline{\mathrm{DM}}$, respectively, in above diagram 
(and henceforth),
although we do not make this distinction 
in the text since DM
and anti-DM particles are still degenerate.\footnote{Of course,
which of the two particles is denoted anti-DM is a matter of convention.
Also, as a corollary, the 
DM particle should be Dirac fermion or complex scalar in a $Z_3$ model.} 

For simplicity, we assume that the SM (or visible) parts of the event can
be completely reconstructed.\footnote{We explore the effects of basic
  background and detector cuts at the LHC for a simple model in
  section~\ref{subsec:extract}.  There we show the effects discussed
  in this section remain after cuts for the background.} 
Considering the invariant masses $m_{ab}$ and $m_{ cd }$, which are
formed by the two SM particles $a$, $b$ and $c$, $d$ in each decay chain, 
one can easily derive the minimum and the maximum kinematic
endpoints of the distributions of $m_{ab}$ and $m_{ cd }$ which are given by 
\cite{Byckling:1973bk}:   
\begin{eqnarray}
m_{ab}^{\textnormal{min}} &=& m_a +m_b, \\
m_{ab}^{\textnormal{max}} &=&
M_{\textnormal{mother}}-m_{\textnormal{DM}}\ \ \ \ \ \  \ \  \Big(\textnormal{Left
  process of Eq.~(\ref{type1})}\Big),\label{biggeredge} \\  
m_{ cd }^{\textnormal{min}} &=& m_c +m_d, \\
m_{cd}^{\textnormal{max}} &=&
M_{\textnormal{mother}}-2\,m_{\textnormal{DM}} \ \ \ \ \ \ \Big(
\textnormal{Right process of
  Eq.~(\ref{type1})} \Big)\label{smalleredge}.  
\end{eqnarray}
Physically, the lower limit corresponds to the case when the two
visible particles $a$, $b$ (and similarly $c$, $d$) are at rest in their center-of-mass
frame so that they move with the same velocity in any Lorentz
frame. The upper limit corresponds to the case in which the DM
particle(s) are at rest in the overall center-of-mass frame of the
final state. Both maxima are independent of the masses of the virtual
intermediate particles.   
The 
point is that the upper endpoints in the two distibutions are different.

\subsection{Double edge}

An expecially striking/interesing case is when the SM particles in the two 
decay chains are identical:
\begin{figure}[h!]
\vspace{-.35cm}
  \centering
  \includegraphics[height=3.1cm,clip]{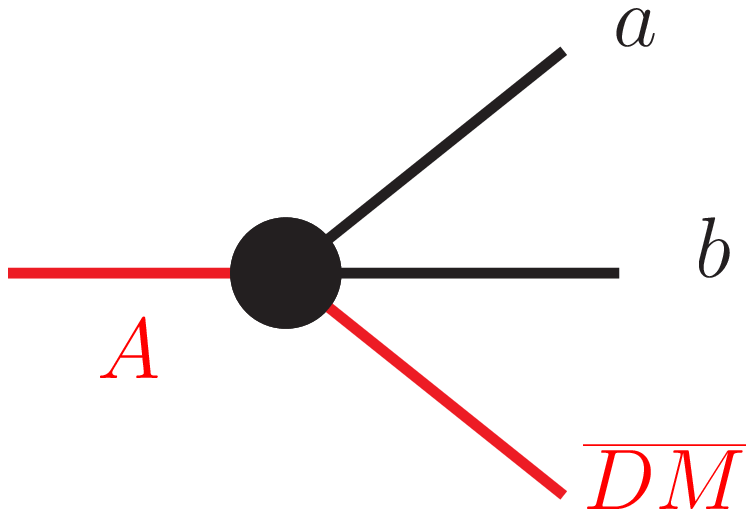} 
 \hspace{2cm} \includegraphics[height=3.3cm,clip]{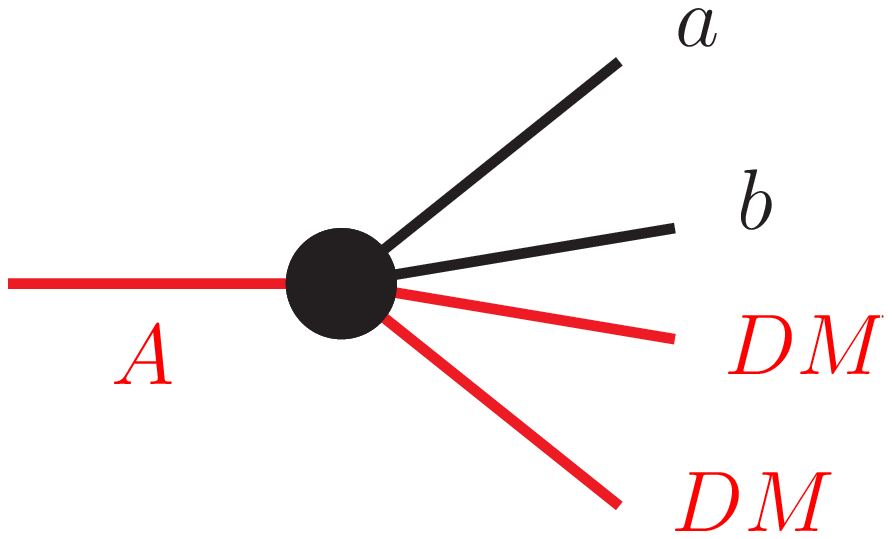}
\vspace{-2cm}
\bea \label{type1}
\eea
\vspace{.1cm}\
\end{figure}
%

\noindent As we show below, it is possible to obtain a {\em double} edge
in the distribution of this SM final state.
We begin with presenting a basic idea of this phenomenon, 
before going on to more details.

\subsubsection{Basic Idea}

Taking into account the fact that the visible particles of both decays are the same and
assuming that both subprocesses are allowed, the experimental
distribution $(1/\Gamma)\ d\Gamma/dm_{ab}$ will contain events of both
processes. In such a combined distribution, clearly, the endpoint of
Eq.~(\ref{smalleredge}) -- denoted
now by $m_{ab}^{\prime\,\,\textnormal{max}}$ -- will become an {\it edge} in the middle of the
distribution, which along with the overall kinematic endpoint given by
Eq.~(\ref{biggeredge}), will give rise to a {\it double edge} signal.   
Assuming the two edges
are visible,
it is interesting that we can determine {\em both} the DM and mother particle masses
by simply inverting Eqs. (\ref{biggeredge}) and (\ref{smalleredge}):
\begin{eqnarray}
m_{\textnormal{DM}}&=&m_{ab}^{\textnormal{max}}-m_{ab}^{\prime\,\,\textnormal{max}}, \label{MDM} \\
M_{\textnormal{mother}}&=&2m_{ab}^{\textnormal{max}}-m_{ab}^{\prime\,\,\textnormal{max}}.
\label{Mmother}
\end{eqnarray}
In particular, the distance between the two edges is
identified as the DM mass.

In contrast to the cases just considered, in $Z_2$ scenarios only one
or three  
DM particles (i.e., not two) are allowed in a single decay chain due
to $Z_2$-charge conservation (unless the process is triggered with an
uncharged mother particle \cite{Han:2009ss}).  Independently of phase-space considerations, 
we note that in $Z_2$ models the decay chain with three DM
particles should be highly suppressed with respect to the one DM
case.
The reason for such an expectation is that 
a decay with three DM in the final state requires a vertex with {\em four} (in general different) 
$Z_2$-charged particles which is typically
  absent, at least at the renormalizable level in most models.\footnote{Compare
  this situation to the $Z_3$ case, where appearance of two DM in a decay chain comes
  from vertex with {\em three} $Z_3$-charged particles which is more
  likely to be present, especially at the renormalizable level.}
Therefore with only one possible decay process (in terms of
the number of DM in the final state) we can only observe a
single kinematic endpoint in the invariant mass distributions in a $Z_2$ model.

\subsubsection{Details}

Of course the visibility of such a signal depends on the shapes of the
distributions of each subprocess as well as their relative decay
branching fractions.   
\begin{figure}[t]
	\centering
		\includegraphics[width=8.0truecm,height=7.0truecm,clip=true]{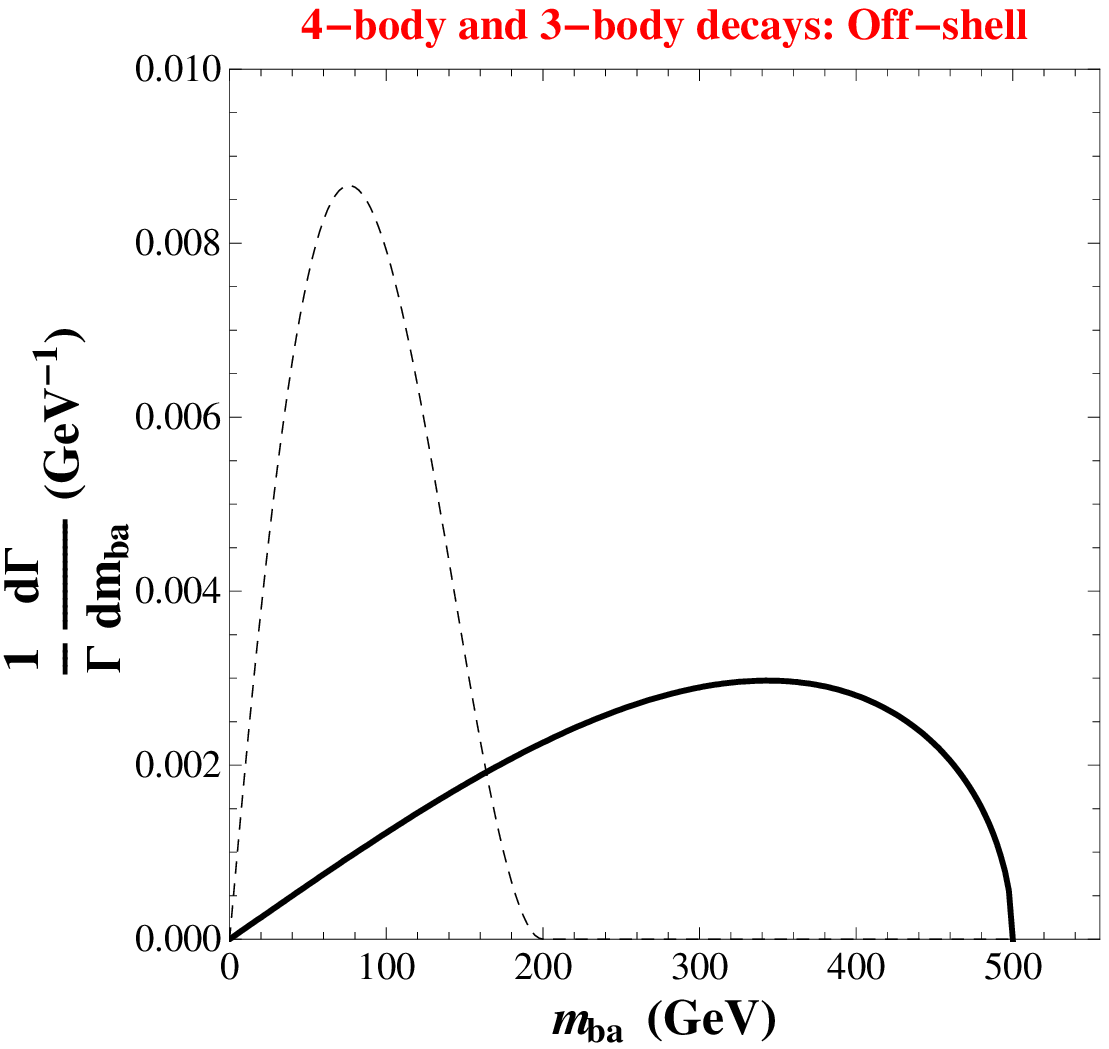}
                \hspace{0.2cm}
                \includegraphics[width=8.0truecm,height=7.0truecm,clip=true]{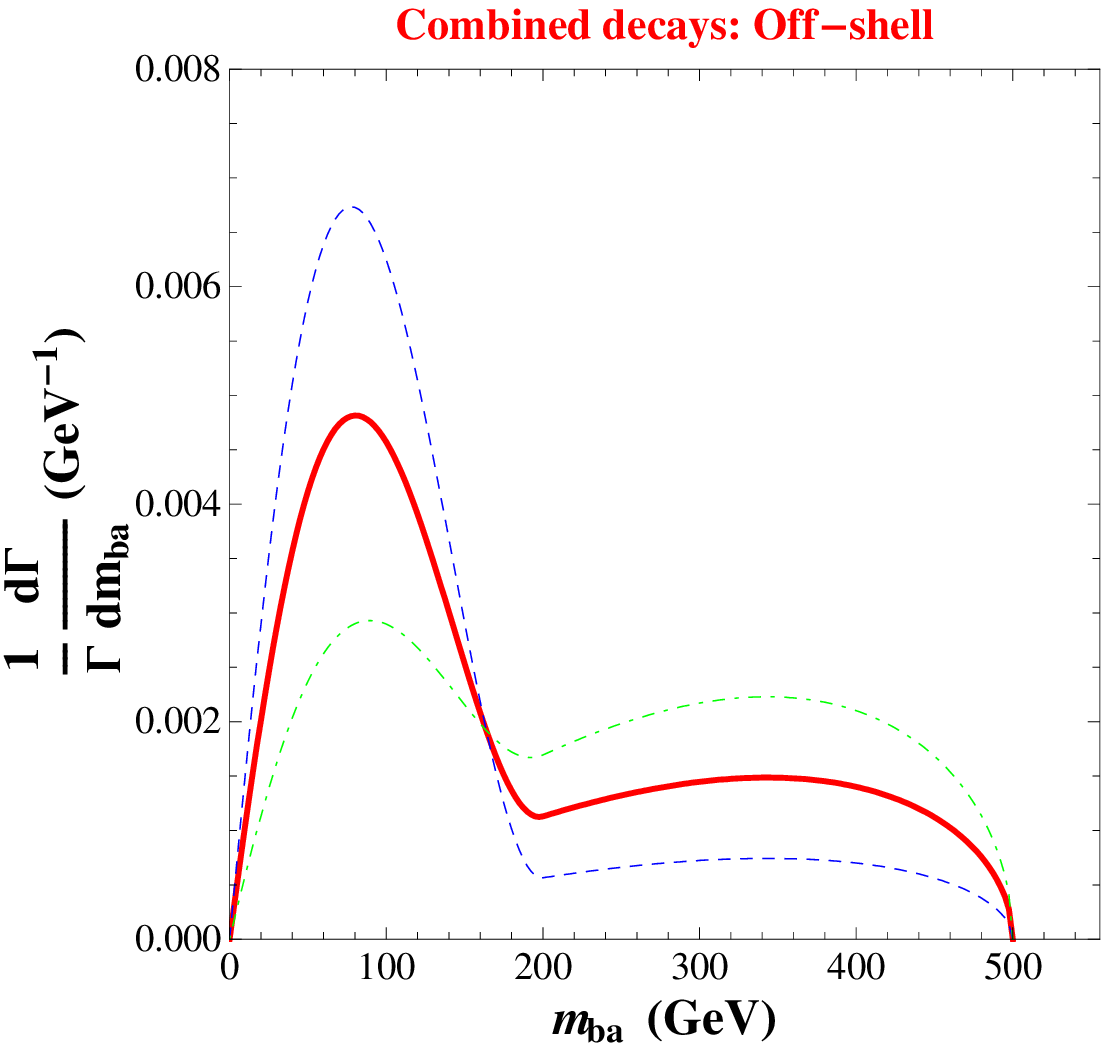}
	\caption{Invariant mass distribution
          $(1/\Gamma)\ d\Gamma/dm_{ab}$ for the processes of
          Eq.~(\ref{type1}). The masses of the mother particle $A$ and
          of the DM particles are $m_A=800$ GeV and $m_{\rm DM}=300$
          GeV and the SM particles $a$ and $b$ are assumed to be
          massless. The solid and dashed curves on the left panel
          represent the distributions for the 3-body decay and the 4-body decay,
          respectively. On the right panel, blue/dashed (highest peaked), red/solid, and green/dot-dashed (lowest  
          peaked) curves show the combined distributions with branching ratios
          of 3-body to 4-body given by 1:3, 1:1, and 3:1,
          respectively.  }  
	\label{fig:OffshellDist}
\end{figure}
The solid curve and the dashed plot in the left panel of Figure \ref{fig:OffshellDist}
illustrate the generic shape of the distributions for the two processes of
Eq.~(\ref{type1}) based only on pure kinematics,
i.e., no effects of matrix element and spin-correlations.
(Such effects might be important and we will return to this issue in the context of
specific models to show that multiple edges can still ``survive''
after taking these effects into consideration.)
Because of the phase-space structure of the processes one realizes that the distribution in
the case of 3-body decays is more ``bent'' towards the right (i.e.,
larger values of invariant mass) whereas for the 4-body decays the peak of the
distribution 
leans more towards the left (i.e., smaller values of invariant mass). Because of this feature,
the combination of the two distributions can give rise to two visible
{\it edges} (as long as the relative branchings of the
two decays are of comparable size). This is shown in the right panel
of Figure \ref{fig:OffshellDist} in which we show the combined
invariant mass distribution of the two visible SM particles, for three
different relative branching fractions of the two subprocesses.
Based on the
location of the {\it edges} in right panel of Figure
\ref{fig:OffshellDist} and Eqs. (\ref{MDM}) and (\ref{Mmother}), 
the mass of DM particle must be
about $300$ GeV and the mass of the mother particle must be about
$800$ GeV, which are of course the masses used in the example.  

\begin{figure}[t]
	\centering
		\includegraphics[width=9.0truecm,height=7.0truecm,clip=true]{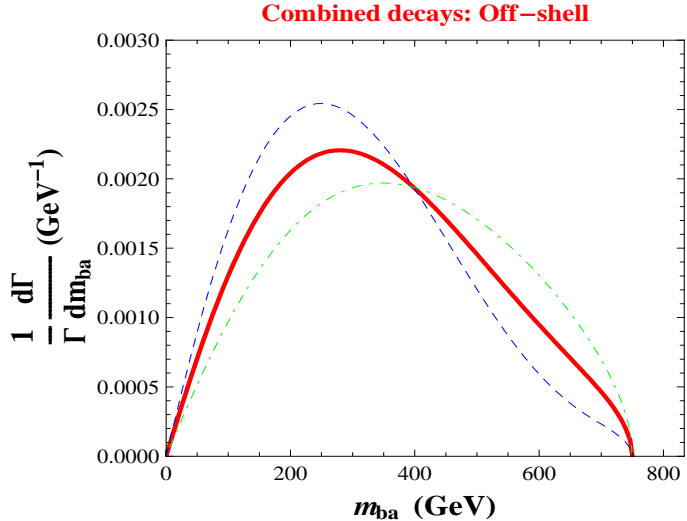}
               	\caption{Same as the right panel of Figure
                  \ref{fig:OffshellDist} but using a smaller DM mass, $m_{\rm DM}=50$
                  GeV. The {\it edge} in the middle of the
                  distribution is no longer apparent.}
	\label{fig:OffshellDist2}
\end{figure}

Whether or not the double-edge signal is
clear (and hence we can determine the DM and mother masses) also depends on the DM mass which must be relatively sizable compared
to the mass of the mother particle. For example, if we take a DM mass
of $50$ GeV instead of $300$ GeV that we assumed above, with the mother mass fixed at $800$ GeV,
we observe from Figure \ref{fig:OffshellDist2} that the plotted
distribution does not provide a good measurement of
$M_{\textnormal{mother}}$ and $m_{\textnormal{DM}}$. 

Let us return to the
issue of the relative branching fraction for each
subprocess. 
The decay into two DM particles should be generically phase-space suppressed
relative to the decay into just one DM particle, So, based on pure phase-space suppression,
the branching ratio of the decay into two DM might be much smaller than 
the decay into one DM (unlike what is chosen in the figures above). Hence, it might be difficult to observe a double-edge signal. 
However, in 
specific models this suppression could be compensated by larger
effective couplings so that the two decays have comparable 
branching ratio, and therefore, the double-edge is visible as in
Figure~\ref{fig:OffshellDist}.   

In fact, another possibility is that the two decay chains for the $Z_3$
case, i.e., with one and two DM particles, do {\em not} have
{\em identical} SM final states, but there is some overlap
between the two SM final states. For example, 
\begin{figure}[h!]
  \centering
  \includegraphics[height=3.cm,clip]{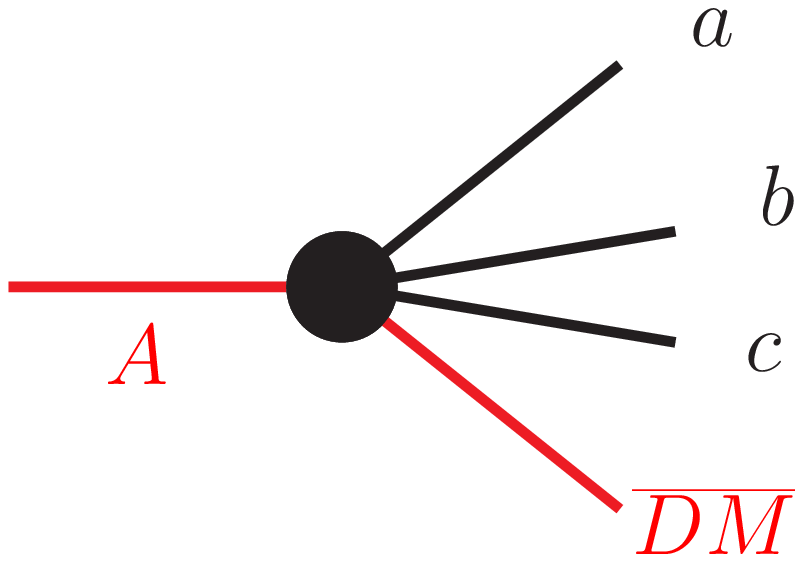} 
 \hspace{2cm} \includegraphics[height=3.cm,clip]{OffShDecZ3.eps}
\vspace{-2cm}
\bea
\eea
\vspace{.2cm}\
\label{4and4body}
\end{figure}
\newline
\newline
If we assume that particle $c$ is (at least approximately) massless, then the
maximum kinematic endpoint of $m_{ a b }$ in the first of the
above-given two reactions is still $M_{mother}-m_{DM}-m_c\approx M_{
  mother } - m_{ DM }$. In this situation {\em both} the reactions have $4$-body
final states and hence could be easily have comparable rates, at least based on
phase-space (\textit{c.f.} Earlier we had $3$-body vs $4$-body 
by requiring the same two-body SM final state for the two
reactions found in~(\ref{type1})). On the other hand, although the two rates are
now comparable, it might actually
be
harder to observe a double edge because the
shape of the two individual distributions are both peaked towards the
left (i.e., smaller values of invariant mass) and even if they have different end-points, the
combined distribution might not show as clearly a double edge as the earlier case where the two shapes are apparently distinct.

\subsection{Different Edges in Pair Production}
\label{different}

Finally, 
what if there are no common SM particles between the final states
of the decay chains with one and two DM particles
so that we do not obtain a double edge?
In this case, 
one can consider another analysis, by making the further assumption
that the same mother particle $A$ is pair-produced in each event, and that the decay
products of each $A$ are now distinct and very light or massless, \textit{i.e.},
\begin{figure}[h!]
  \centering
  \includegraphics[height=4.2cm,clip]{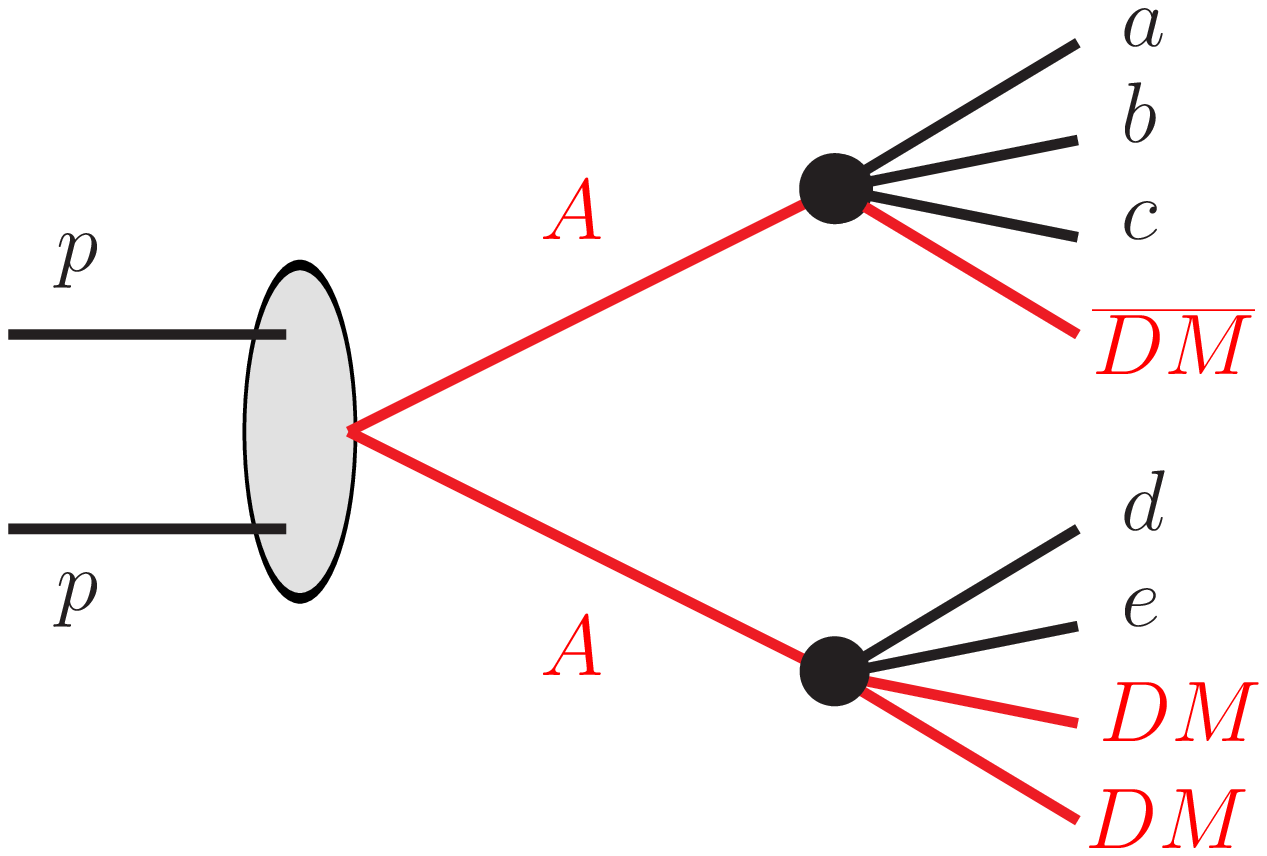} 
\vspace{-3cm}
\bea
\eea
\vspace{1.3cm}\
\end{figure}
\newline
\newline
Here we have chosen three SM particles ($a$, $b$, $c$) in the decay chain 
with one DM
just so that both decay chains involve a $4$-body final state.
In this situation one can restrict to events
with all five SM particles ($a$,...$e$) particles
in the final state\footnote{If we include other events which have $a$, $b$, $c$
or $d$, $e$ on {\em both} sides,
we still get the different edges that we discuss below, 
but as we will mention later, such events will not
allow us to get rid of ``faking'' $Z_2$ models.},
but use both sides of the event, i.e., obtain the
full invariant mass distribution of the visible particles of each
(distinct) side. 
In the interpretation of these results in the context of a
$Z_3$ model, 
the difference between
the endpoints of each separate distribution will give the dark matter mass, and
like before, the mass of the mother particle $A$ can be found using a
combination of the two end-points,
i.e., $m_{ \textnormal{ DM } } = m_{ abc }^{ \textnormal{ max } } - 
m_{ d e }^{\textnormal{max}}$ 
and $M_{\textnormal{mother}}=2m_{abc}^{\textnormal{max}}-m_{de}^{ \textnormal{ max } }$.

\section{On-shell Intermediate Particles}
\label{sec:onshell}

In this section, we consider the case where the mother particle decays
into SM and DM via intermediate particles which are all
\textit{on}-shell. 
Again, like in section~\ref{sec:offshell} all
particles are assumed to decay inside the detector. 
In this case, the
endpoints of invariant mass  
distributions will depend on the masses of these intermediate
states as well as the masses of the mother and the final state particles. 
Both in the $Z_2$ and $Z_3$ cases there will be more
possibilities for the upper endpoints because of the possibilities of ``Multiple
topologies''  and ``Different Intermediate Particles'' (to be explained below) for
the same visible final state. 
Since even for the $Z_2$ case it is possible to obtain multiple edges, finding multiple edges is not anymore a robust
discriminator between $Z_2$ and $Z_3$ unlike the off-shell decay
case.  We then discuss a topology of the decay chain which does allow us to distinguish
between the two models.

\subsection{Additional sources of Multiple Edges}

Here we discuss how it is possible to obtain multiple edges even if we
do {\em not} combine decays of mother particle into one and two DM particles.

\subsubsection{Multiple Topologies}
\label{sec:multitopo}

For $Z_3$ models we can expect multiple endpoints from the decays of
the same mother particle into a
given SM final state by combining the two decay chains with one DM and two DM particles,
respectively, just as in the case of the decays with off-shell intermediate particles.
However, this is not the only way of obtaining multiple
endpoints, i.e., such a combination of decay chains with one and two DM is not essential.  The reason is that there are multiple
possible topologies even with the completely identical final state
if it contains {\em two} DM, due to the various possibilities 
for the locations of two DM particles relative to the other SM particles in a decay chain. For example, for the case of a 4-body 
decay process (i.e., two SM and two DM particles) there will be {\em
  three} different possibilities: 
\begin{figure}[h!]
\vspace{-.2cm}
  \centering
\includegraphics[width=7.8cm,height=6.cm,clip]{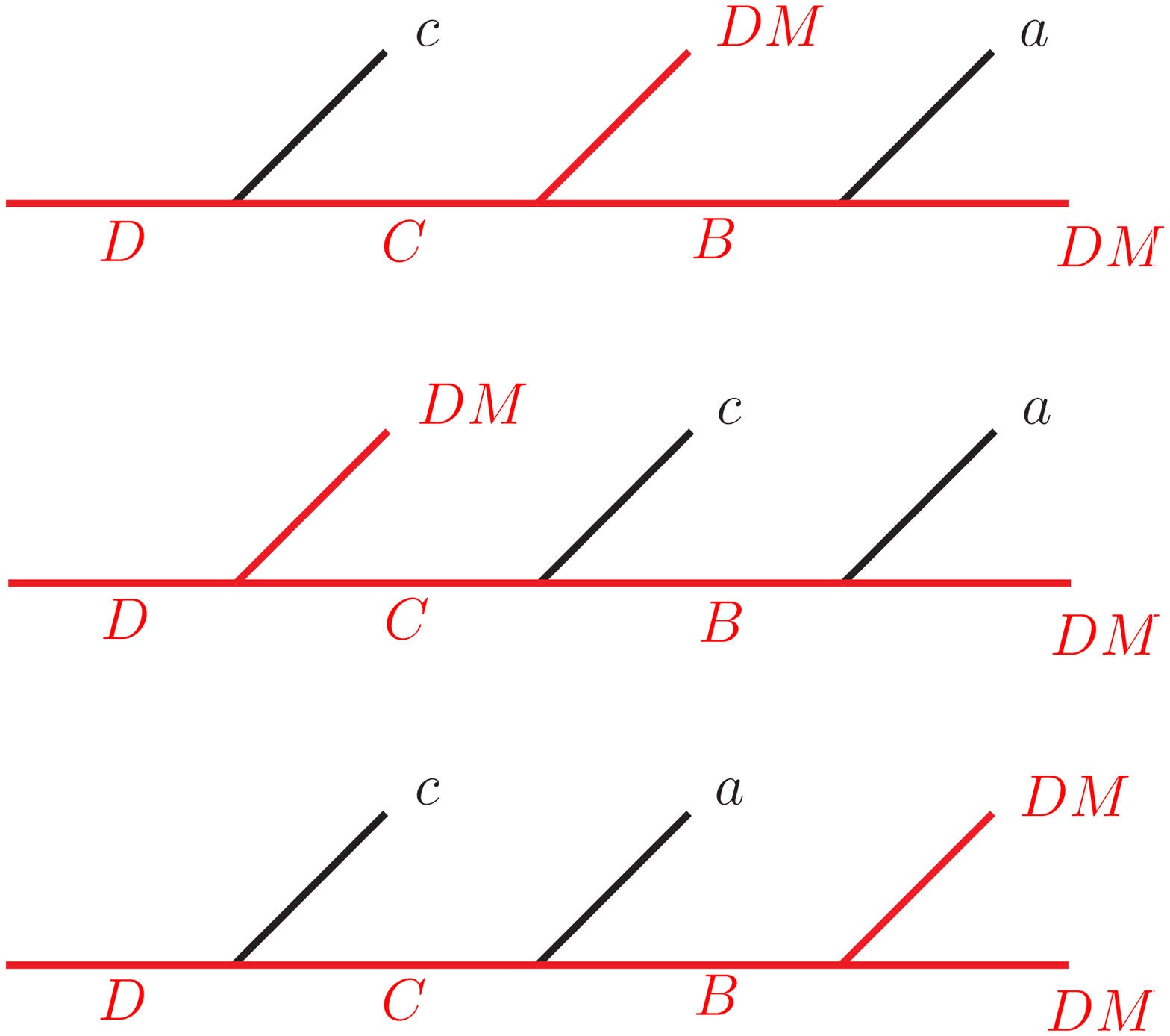}
\vspace{-6.3cm}
\bea\label{topo1}
\eea
\vspace{.7cm}\
\bea\label{topo2}
\eea
\vspace{.7cm}\
\bea\label{topo3}
\eea
\vspace{-.3cm}\
\end{figure}

\noindent Note that (as above) decay cascades involve a ``charged-charged-charged'' (under $Z_3$ symmetry)
vertex (in addition to ``charged-charged-neutral'' vertices) in order to 
contain two DM particles in the final state.  

Assuming that the visible particles are massless, $m_a=m_c=0$, the
upper endpoints for each topology are given by (See Appendix A) for details.): 
\begin{eqnarray}
(m_{ca}^{\textnormal{max}})^2&=&\frac{2(m_D^2-m_C^2)(m_B^2-m_{DM}^2)}{m_B^2+m_C^2-m_{DM}^2-\lambda^{1/2}(m_C^2,m_B^2,m_{DM}^2)}
  \hspace{1cm}  \Big(\textnormal{for
    Eq.~(\ref{topo1})}\Big) \label{eqfortopo1} \ \ \\ 
(m_{ca}^{\textnormal{max}})^2&=&\frac{(m_C^2-m_B^2)(m_B^2-m_{DM}^2)}{m_B^2}
  \hspace{3.5cm} \  \Big(\textnormal{for
    Eq.~(\ref{topo2})}\Big) \label{eqfortopo2} \ \ \\
(m_{ca}^{\textnormal{max}})^2&=&\frac{(m_D^2-m_C^2)(m_C^2-m_B^2)}{m_C^2}
  \hspace{3.8cm} \  \Big( \textnormal{for Eq.~(\ref{topo3})}\Big)\ \ 
\end{eqnarray}
where $\lambda$ is the well-known kinematic triangular function given in the form of
\begin{eqnarray}
\lambda(x,y,z)=x^2+y^2+z^2-2xy-2yz-2zx. \label{eqn:kintri}
\end{eqnarray} 
The main point is that kinematic endpoints are functions of the masses of the
mother, the DM and the intermediate particles, and moreover,
this dependence changes according to different topologies.
Thus, 
even if the intermediate particles involved
in these decays of a given mother particle are the same, one will still
obtain multiple endpoints.\footnote{Of course, the different possible decay topologies can,
in general, have different intermediate states.}  Finally, if we combine decay chains with one and two DM  in the final state (even if the latter has
just one topology), 
the difference between the two endpoints will not lead to a
direct measurement of the DM mass like in the off-shell decay case
because again, the mass of intermediate particles is 
one of the main ingredients to determine the endpoints.

In $Z_2$ models the decay topologies must have a single DM particle and that too at the end
of the decay chain because the vertices in the decay cascade are of the form
``odd-odd-even'' (under the $Z_2$ symmetry).\footnote{Note that
a similar argument applies to decay chains in $Z_3$ models with only {\em one} DM 
in the final state.} Nevertheless, there can still be different topologies 
because of different ordering of the visible states. For example: 
\begin{figure}[h!]
  \centering
\includegraphics[width=7cm,height=1.8cm,clip]{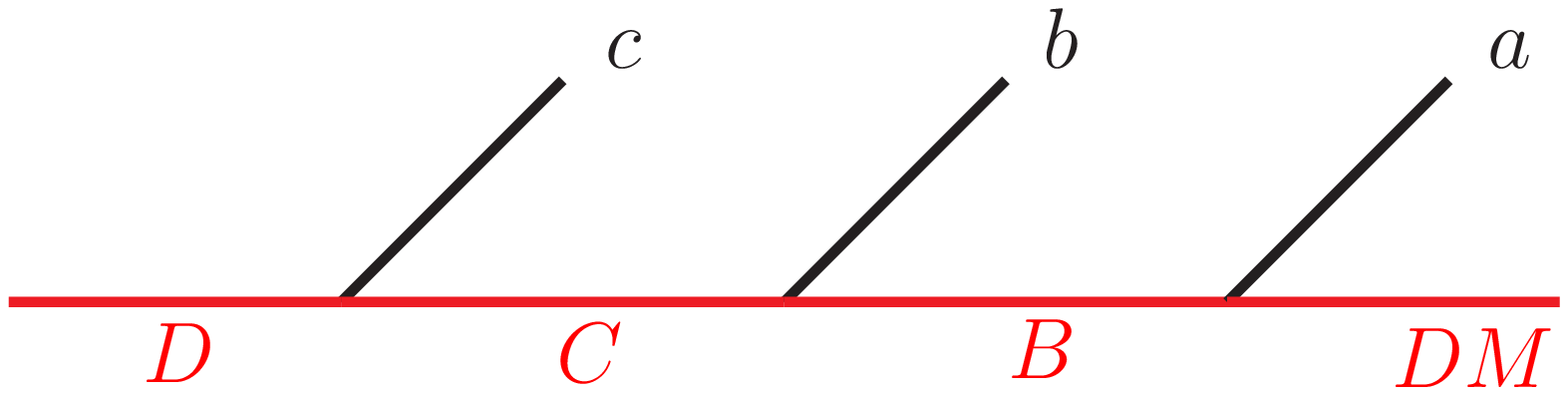}\  \\
\vspace{.2cm}
\includegraphics[width=7cm,height=1.8cm,clip]{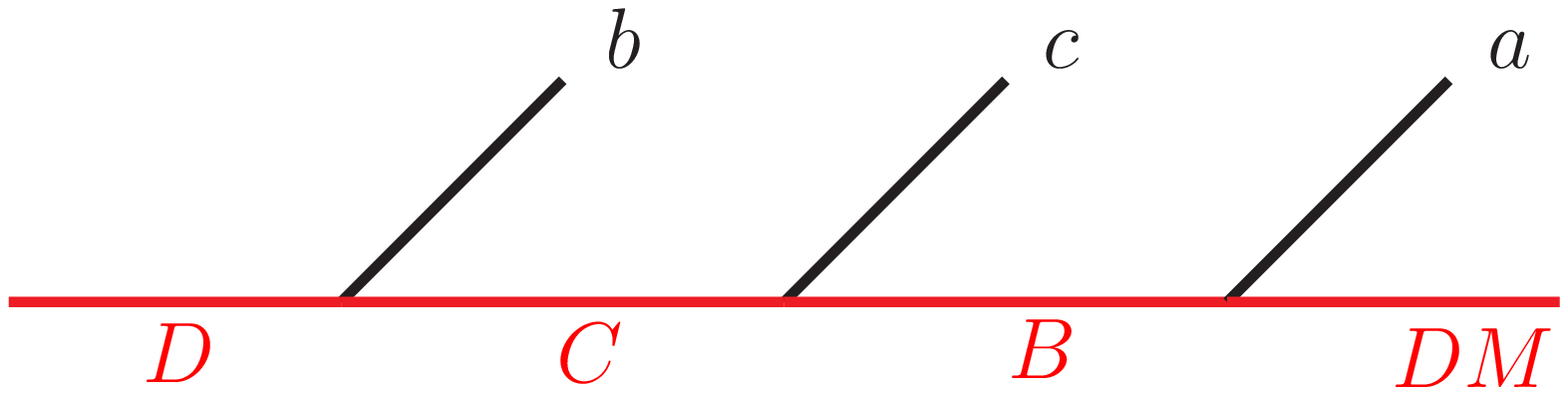}\\ \vspace{.2cm} .\\ \vspace{-.3cm}.\\ \vspace{-.3cm}.\\ 
\vspace{-4.8cm}
\bea
\label{thrZ2}
\eea
\vspace{.4cm}\
\bea \label{threeZ2}
\eea
\vspace{.5cm}
\end{figure}
\newline
Obviously, the endpoints for a given invariant mass distribution, say
$m_{ca}$, will be different for each of these two topologies, and
actually they can be obtained from Eqs.~(\ref{eqfortopo1}) and
(\ref{eqfortopo2}) by just replacing $m_{DM}$ in the denominator of
Eqs.~(\ref{eqfortopo1}) by $m_b$ and leaving Eqs.~(\ref{eqfortopo2})
unchanged (and where $m_a$ 
and $m_c$ are still assumed to vanish).

\subsubsection{Different Intermediate Particles for Same Final State}

In addition, even if the topology and the order of visible particles
are the same, there is the possibility of multiple paths for
the \textit{same} mother particle to decay into the same (SM and DM) final
state by involving different intermediate particles. We will obtain multiple endpoints 
in this case because of the dependence of the endpoints
on the masses of intermediate particles (as mentioned
above). This argument is valid for both the
$Z_2$ and $Z_3$ models (and one or two DM for the latter case): for a final state with
two SM and one DM, we can have 
\begin{figure}[h!]
  \centering
\includegraphics[height=1.8cm,clip]{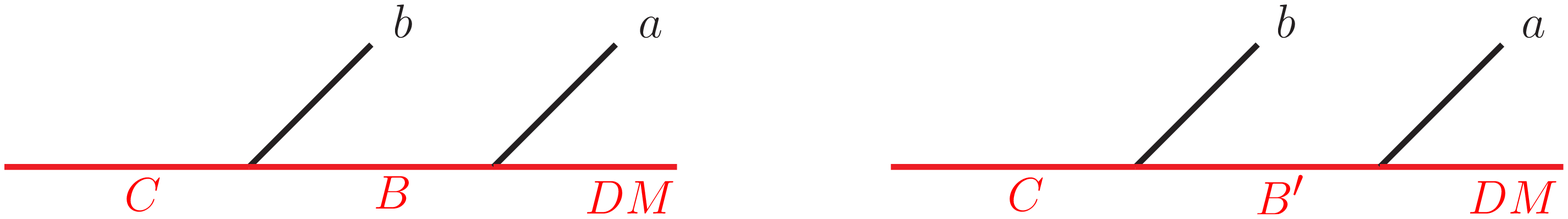}\ 
\vspace{-1.5cm}
\bea\label{Z23body}
\eea
\vspace{.1cm}\
\end{figure}
\newline
For example, in SUSY, the decay chain $\chi^0_2
\rightarrow l^+ l^- \chi^0_1$ can proceed via intermediate right- or left-handed
slepton. Since the masses of intermediate right- and left-handed
sleptons are in general different, multiple endpoints are expected.  

\subsection{Cusp Topology}
\label{sec:newtopo}

So far, we have learned that for on-shell intermediate particle cases
the multiple edge signal is not a good criterion to distinguish $Z_3$
from $Z_2$. Instead, we focus on {\em shapes} of these distributions. Consider 
the topology which can be present in $Z_3$ models (but absent in the $Z_2$ case) with two visible 
SM particles separated by a DM particle\footnote{Note that in general $D$ might come
from the decay of another $Z_3$-charged particle and similarly, at the end
of the decay, $A$ might not be the DM, that is, it could itself decay
further into DM particles and other visible states as long as $Z_3$-charge
conservation is respected. The ``...'' to the left of $D$ and to the right of
$A$ signify this possibility.}, i.e., 
\begin{figure}[h!]
  \centering
\includegraphics[height=2cm,clip]{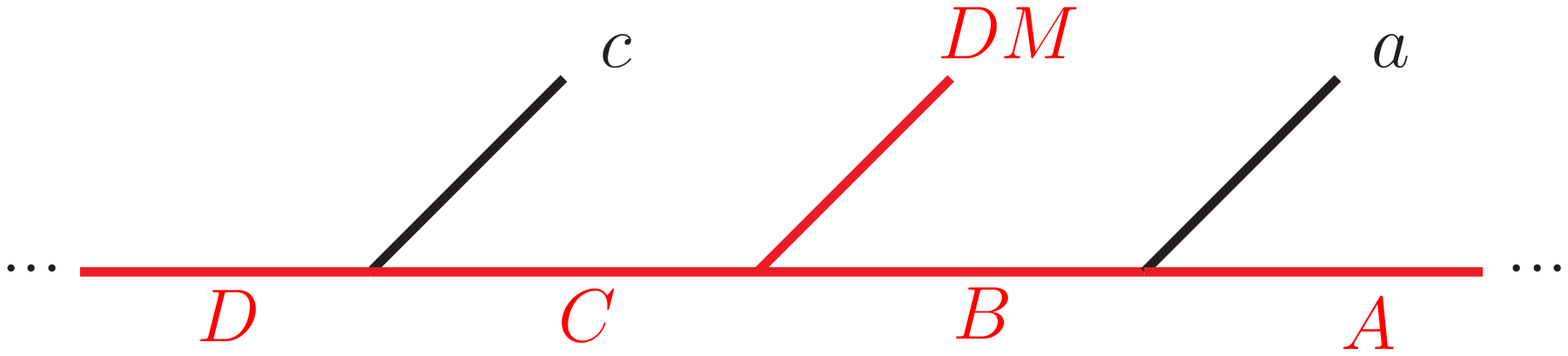}
\vspace{-1.4cm}
\bea
\label{newtopo} 
\eea
\vspace{.1cm}
\end{figure}
\newline
We assume massless SM particles (i.e., $m_a=m_c=0$) and the mass hierarchy $m_D > m_C > m_B > m_A$. Also, we neglect
spin-correlation effects in this section. We sketch the derivation of 
the distribution of the $ac$ invariant mass here and refer the reader to the Appendix A for details. The differential distribution $\displaystyle
\frac{1}{\Gamma} \frac{\partial \Gamma}{\partial m_{ac}^2}$ that we
want to study can be obtained for this ``new'' topology easily by noting that the differential
distribution $\displaystyle \frac{1}{\Gamma} \frac{\partial^2
  \Gamma}{\partial u\partial v}$ must be flat, where the variables are defined as follows
\bea
u=\frac{1-\cos{\theta_{c\,\mathrm{DM}}^{(C)}}}{2} \hspace{1cm} {\rm and}\hspace{1cm} v=\frac{1-\cos{\theta_{ca}^{(B)}}}{2},
\eea
with $\theta_{ca}^{(B)}$ being the angle between $c$ and $a$ in
the rest frame of $B$, and $\theta_{c\,\mathrm{DM}}^{(C)}$ being the angle between $c$ and
DM in the rest frame of $C$ \cite{Miller:2005zp}. 
\begin{figure}[t]
	\centering
		\includegraphics[width=8.0truecm,height=7.0truecm,clip=true]{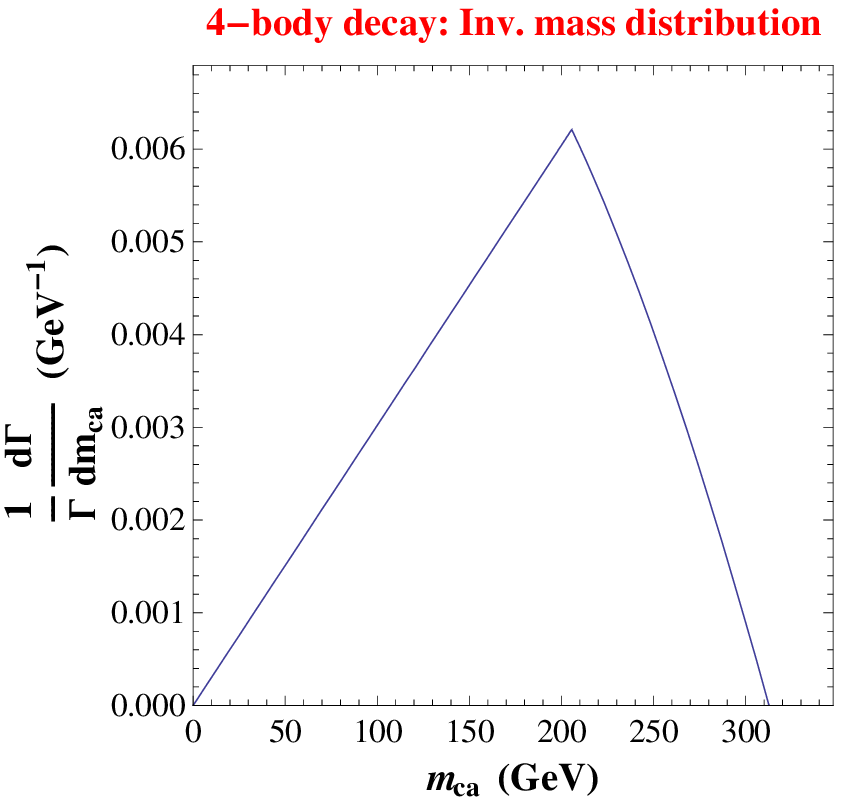}
                \hspace{0.2cm}
                \includegraphics[width=8.0truecm,height=7.0truecm,clip=true]{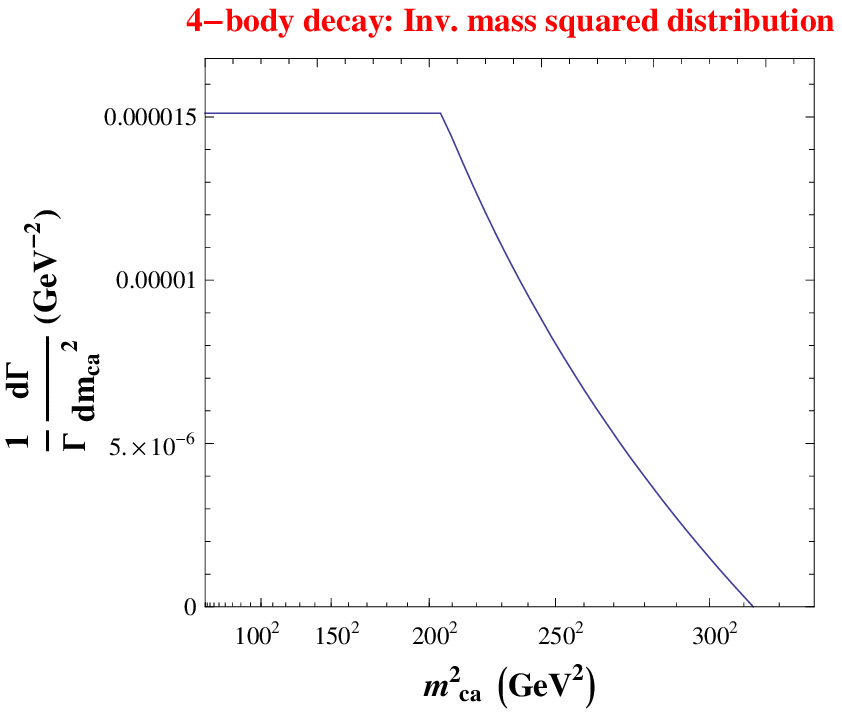}
	\caption{The panel on the left shows the distribution in $m_{ca}$ while the right hand panel shows the distribution in $m^2_{ca}$ from the decay chain of Eq.~(\ref{newtopo}).  The masses of mother particle, two intermediate particles, and DM particles are $800$ GeV, $700$ GeV, $400$ GeV, and $200$ GeV, respectively and the SM particles are assumed massless.  A ``cusp" due to the topology of Eq.~(\ref{newtopo}) is clear in both distributions.}
	\label{fig:4BodyNewTopo} 
\end{figure}
In addition, we have $\ 0\ <\ u,v\ <\ 1$. Thus, we can write
\bea
\frac{1}{\Gamma} \frac{\partial^2  \Gamma}{\partial u\partial v}=\theta(1-u)\theta(u)\theta(1-v)\theta(v)\label{thetadist}
\eea
One further finds that 
\bea
m_{ca}^2=m_{ca}^{max} (1-\alpha u) v, 
\eea
where $m_{ca}^{max}$ is given in Eq.~(\ref{eqfortopo1}) with
$m_{ DM }$ in the numerator replaced by $m_A$, and so we can make a change of variables from the differential
distribution of Eq.~(\ref{thetadist}) and obtain the distribution $\displaystyle \frac{1}{\Gamma}
\frac{\partial^2  \Gamma}{\partial u\partial m_{ca}^2}$, which can
then be integrated over $u$ to finally obtain the distribution with respect to
$m_{ca}$ \footnote{Note that $\ \displaystyle
  \frac{1}{\Gamma}\frac{\partial\Gamma}{\partial m}= 2
  m\ \frac{1}{\Gamma}\frac{\partial\Gamma}{\partial m^2}$.}: 
\begin{eqnarray}
\frac{1}{\Gamma}\frac{\partial \Gamma}{\partial
  m_{ca}}&=& 
\left\{\begin{array}{l}
\displaystyle
\frac{2m_{ca}}{(m_{ca}^{\textnormal{max}})^2\ \alpha}\ \ln
\frac{m_C^2}{m_B^2} \hspace{1.8cm}\;\; \textnormal{for\hspace{1.cm}
  \ $0\ <\ m_{ca}\ <\ \sqrt{1-\alpha}\ 
  m_{ca}^{\textnormal{max}}$}\label{reg1} \\ 
\\ 
\displaystyle
\frac{2m_{ca}}{(m_{ca}^{\textnormal{max}})^2\ \alpha}\ \ln
\frac{(m_{ca}^{\textnormal{max}})^2}{m_{ca}^2}\hspace{.9cm} \;\;\;
\textnormal{for\hspace{.5cm} \
  $\sqrt{1-\alpha}\  m_{ca}^{\textnormal{max}}\ <\ m_{ca}\ <\ m_{ca}^{\textnormal{max}}$}\label{reg2} 
\end{array} \right.
\label{cusp}
\end{eqnarray} 
where $m_{ca}^{max}$ is given in Eq.~(\ref{eqfortopo1}) and 
\begin{eqnarray}
\displaystyle
\ \alpha=\frac{2\lambda^{1/2}(m_C^2,m_B^2,m_{\mathrm{DM}}^2)}{m_B^2+m_C^2-m_{\mathrm{DM}}^2+\lambda^{1/2}(m_C^2,m_B^2,m_{\mathrm{DM}}^2)} .\label{alpha}
\end{eqnarray}
From these results we can easily see that the new topology introduces two
different regions in the $m_{ca}$ distribution with a ``cusp'' at the
boundary connecting both regions, located at $\ \sqrt{1-\alpha}\ 
  m_{ca}^{\textnormal{max}}$. 
Figure \ref{fig:4BodyNewTopo} shows  the same distribution in both
panels, but with respect to
$m_{ca}$  on the left panel and with respect to $m_{ca}^2$  on the right
panel. As we will argue later, the second option seems better suited once spin
correlations are taken into account, but in both plots, one observes that the cuspy feature
is quite clear. 

\subsubsection{Two Visible Particles}

Consider first the simple case of {\em only} two
visible particles in a decay chain. In the $Z_3$
reaction of Eq.~(\ref{newtopo}), $D$ is then the mother particle and $A$ is DM. Clearly,
we would
find a cusp in the invariant mass distribution of the two visible
particles in the $Z_3$ model, but not for the $Z_2$ model since the two visible particles must always be adjacent to each other in the latter case.\footnote{Note that we are considering decay of 
a $Z_3$ or $Z_2$-{\em charged} mother. 
A $Z_2$-uncharged, i.e., {\em even}, mother
{\em is} allowed to decay into two DM and can give a cusp in the invariant mass
of two visible particles from such a decay \cite{Han:2009ss}.}
Thus, the presence/absence of cusp could be used to distinguish $Z_3$ and
from $Z_2$ models.

\subsubsection{Generalization to More than Two SM Particles in Decay Chain}

Of course, in general in both $Z_2$ and $Z_3$ models there will be more than two
visible
particles with possibly some of them being
identical, and this will undoubtedly complicate the analysis. For example, in the 
reaction of Eq. (\ref{newtopo}), $a$,
$c$, or both can be produced at some other place of the same decay
chain in addition to the locations shown there, e.g.,  
\begin{figure}[h!]
\vspace{-.2cm}
  \centering
\includegraphics[height=4.2cm,clip]{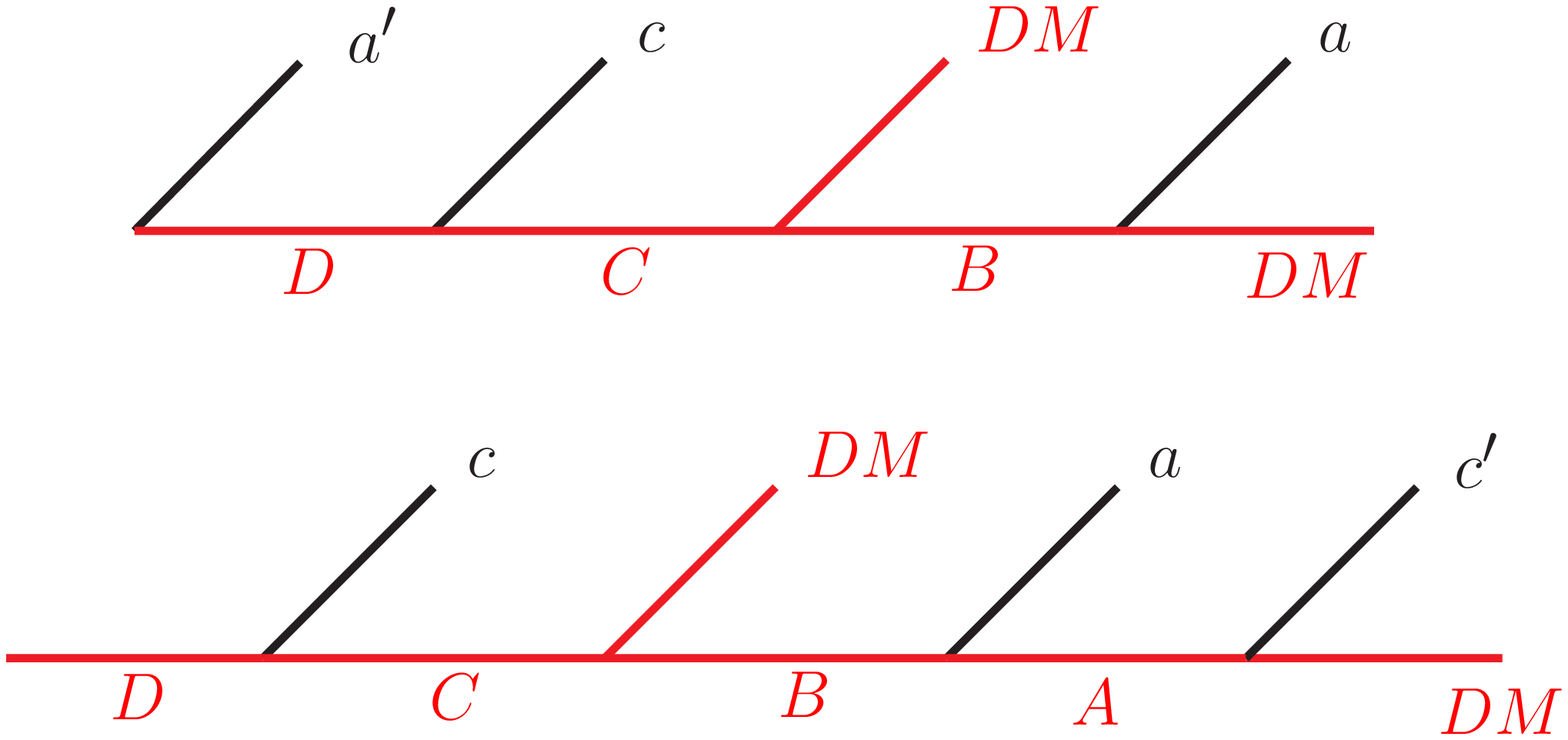}
\vspace{-4.2cm}
\bea
\eea
\vspace{.7cm}
\bea
\eea
\vspace{-.2cm}
\end{figure}
\newline
Here $a^{ \prime }$, which is an identical particle to $a$, is assumed to come
from the immediate left of $D$, and $c^{ \prime }$, which is an identical
particle to $c$, is assumed to come from the immediate right of
$A$. Note that there is no DM between $a^{ \prime }$ and $c$ (unlike between $a$ and $c$) 
in first reaction above (similarly between $a$ and
$c^{ \prime }$ vs between $a$ and $c$ in second reaction above), and that $a$ and $a^{ \prime }$, and 
$c$ and $c^{ \prime }$ are identical. Therefore, in both these examples, it is clear that we will obtain a more complicated distribution in
$m_{ac}$ than the one studied previously. 

Nevertheless, the method described previously to disentangle
the $Z_2$ from the $Z_3$ cases (when having two visible particles), can still
be generalized to the situation of many visible particles in a decay
chain. For example, let us consider the case of {\em three} 
visible SM particles in the final state for both $Z_3$ and $Z_2$ models. 
We will obtain a cusp even
in the $Z_2$ case when considering the invariant mass of two
\textit{not} ``next-door neighbor'' visible particles such as in
$m_{ac}$ for the decay process in Eq. (\ref{thrZ2}).  The reason is that,
even though the precise topology of Eq. (\ref{newtopo}) is absent in a $Z_2$ model,
a similar one is generated by the presence of a SM particle (i.e., $b$) in-between two other 
SM particles (i.e., $a$ and $c$) as in Eq. (\ref{thrZ2}).
Thus the analysis performed earlier for Eq. (\ref{newtopo})
applies in this case, but with the 
DM mass set to zero (assuming SM particle $b$ is massless).

However, this type of degeneracy between $Z_2$ and $Z_3$ can be 
resolved by considering all of the three possible {\it two-(visible) particle} invariant mass
distributions. In the $Z_3$ case with two $DM$ particles in the final
state, two of these three invariant mass distributions will have cuspy features
whereas only one such invariant mass distribution will have a cusp in the $Z_2$ case. 
The reason is again that in the $Z_3$ case, since one more particle is
added to the decay products compared to the $Z_2$ case (i.e., we have
two invisible and three visible particles),  
there will be final state particles (visible or not) in-between the two visible particles
for two of the three pairings. 
This feature remains true for more visible particles, i.e., in general
we will obtain more cusps in the invariant mass
distributions in a $Z_3$ model than in a $Z_2$ model.      
 
\subsection{Spin Correlations}
Once spin correlations are involved, the derivative discontinuity
(cusp) might appear unclear. Nevertheless, it may
still be possible to distinguish a $Z_3$ model from a $Z_2$ model by employing the
fitting method which we will show in the rest of this section. The 
basic idea is that the distribution $d\Gamma/dm_{ca}^2$ of three-body
decays in $Z_2$ (i.e., one DM particle and two visible particles) can (almost)
always be fitted into a quadratic function in
$m_{ca}^2$, whereas the distribution of the new topology of $Z_3$
cannot not be fitted into a single quadratic function, that is, two different functions are
required for fitting each of the two sub-regions of the distribution. Let us see
how this works for a $Z_2$ model (i.e., one DM and two visible particles) and a $Z_3$ model 
(i.e., two DM and two 
visible particles) in turn.  

\subsubsection{$Z_2$ case: 1 DM + 2 Visible}

\begin{figure}[t]
  \centering
  \includegraphics[height=8cm,clip]{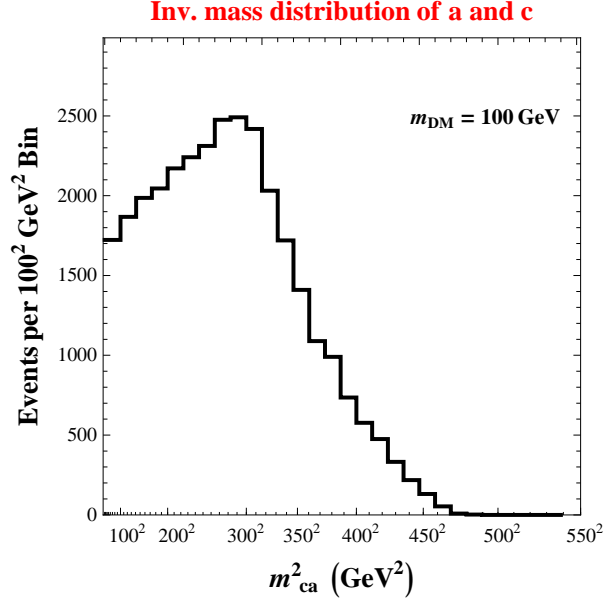} 
\caption{Invariant mass distribution of particles $a$ and $c$, from the
  decay chain shown in Eq.~(\ref{newtopo}), including
  spin correlations, and such that the intermediate particle $C$ has spin
  1 and the intermediate particle $B$ has spin 1/2, and the couplings
  are chiral. The ``cusp" in this distribution appears more defined than in Figure~\ref{fig:4BodyNewTopo} where spin correlations were not considered.}
\label{spincorrgood}
\end{figure}

We can again make use of the same angular variables considered earlier
for the case of  this 3-body decay cascade, shown for example in
Eq.~(\ref{Z23body}). According to the references \cite{Miller:2005zp} and
\cite{Kraml:2005kb}, the normalized distribution including spin-correlations is given by  
\begin{eqnarray}
\frac{1}{\Gamma}\frac{\partial \Gamma}{\partial t}=\theta (t) \theta (1-t)f(t)
\end{eqnarray}
where again we have defined the variable $t$ as
\begin{eqnarray}
t\equiv \frac{1-\cos \theta _{ba}^{(B)}}{2}.
\end{eqnarray}
Here $f(t)$ is a function of $t$ and $\theta _{b a}^{(B)}$
is the angle between particles $a$ and $b$ of Eq.~(\ref{Z23body}) in
the rest frame of particle $B$. 
One then notes that $\displaystyle
m_{ba}^2=(m_{b a}^{\textnormal{max}})^2\ t$ which basically means that
the distribution with respect to  the invariant mass $m_{b a}^2$ (which is of our interest) 
is essentially the same as the
one with respect to $t$ above.
This means that the distribution in $m_{b a}^2$ will have the functional form $f$.
According to the reference \cite{Wang:2006hk}, such spin correlation 
functions are just polynomials of $\cos \theta_{b a}$ (i.e., $(1-2t)$). Moreover,
if we restrict our consideration to particles of spin-1 at most, the
maximum order in $t$ of the polynomial is two, which means that the most general form of
$f$ will be  
\begin{eqnarray}
f(t)=c_1+c_2t+c_3t^2.
\end{eqnarray}
In turn, the invariant mass distribution we are interested in must
therefore take the form (in the region between the endpoints enforced by the $\theta$-functions)
\begin{eqnarray}
\frac{1}{\Gamma}\frac{d\Gamma}{dm_{b a}^2}=c^{ \prime }_1+c^{ \prime }_2 m_{b a}^2+c^{ \prime }_3 m_{b a}^4.\label{z2fit}
\end{eqnarray}
With the experimental data we can construct the invariant mass
distribution, and we will be able to determine the three
constants $c^{ \prime }_1$, $c^{ \prime }_2$, and $c^{ \prime }_3$ by fitting into a parabola in the $m_{b a}^2$ variable. 
In other words, for any 3-body decay chain, with or without
spin-correlation, it is always possible to fit the 
invariant mass distribution $\displaystyle
\ \frac{1}{\Gamma}\frac{d\Gamma}{dm_{b a}^2}\ $ into a curve
quadratic in $m_{b a}^2$. \\

\subsubsection{$Z_3$ case: 2 DM + 2 Visible}
\label{subsubsec:Z3}

We now consider the new topology of Eq.~(\ref{newtopo})
including the possibility of spin correlations. As in Section
3.1, we use the same angular variables $u$ and $v$. However, the
normalized distribution with spin correlations become a little more complicated than before 
\begin{eqnarray}
\frac{1}{\Gamma}\frac{\partial^2 \Gamma}{\partial u\partial v}=\theta
(v) \theta (1-u) g(u) \theta(v) \theta (1-v) h(v) 
\end{eqnarray}
where again
\begin{eqnarray}
u\equiv \frac{1-\cos \theta _{c\,\mathrm{DM}}^{(C)}}{2},\ v\equiv \frac{1-\cos \theta _{ca}^{(B)}}{2}.
\end{eqnarray}
Like in the previous section, $g(u)$ and $h(v)$ are spin-correlation
functions (cf. $g=h=1$ without spin correlation discussed earlier) and again the invariant mass squared is given by 
\begin{eqnarray}
m_{ca}^2=(m_{ca}^{\textnormal{max}})^2 (1-\alpha u)\ v.
\end{eqnarray}
where $\alpha$ is the same kinematical constant defined in Eq.~(\ref{alpha}).
As in the analysis without spin correlations, the two types of $\theta$-functions
will split the entire region into two sub-regions, with a cusp at
the separation point, whose location is independent of the
spin correlation effects (since it depends on purely kinematical
constants $\alpha$ and $m_{ca}^{max}$). But unlike the scalar case (i.e., with no
spin correlations), we have now two functions $g(u)$ and $h(v)$ which
can change the shape of the distribution and in principle affect the derivative discontinuity (the cusp).  

In detail,
by the chain rule the previous normalized distribution can be modified
and partially integrated to obtain   
\begin{eqnarray}
\frac{1}{\Gamma}\frac{d\Gamma}{dm_{ca}^2}=\int_0^{u_{\textnormal{max}}}\frac{du}{(m_{ca}^{\textnormal{max}})^2
  (1-\alpha u)}\
g(u)\ h\left(\frac{m_{ca}^2}{(m_{ca}^{\textnormal{max}})^2 (1-\alpha
  u)}\right) \label{eqn:integrate}
\end{eqnarray}
where
\begin{eqnarray}
u_{\textnormal{max}}=\textnormal{Max}\left[1,\frac{1}{\alpha}\left(1-\frac{m_{ca}^2}{(m_{ca}^{\textnormal{max}})^2}\right)\right]. 
\end{eqnarray}
The two possible choices in the definition of $u_{\textnormal{max}}$ above arise when
integrating $\frac{1}{\Gamma}\frac{\partial^2\Gamma}{\partial m_{ca}^2
  \partial u}$ with respect to $u$ due, in turn, to the integration limits
enforced by the $\theta$ functions. This leads to two different regions 
for the differential distribution such that in the first sub-region, we have $0<m_{ca}<\sqrt{1-\alpha}\ m_{ca}^{\textnormal{max}}$ and $u_{\textnormal{max}}=1$, while for the second region, 
we have
$\sqrt{1-\alpha}\  m_{ca}^{\textnormal{max}}<m_{ca}<m_{ca}^{\textnormal{max}}$
and $u_{\textnormal{max}}=\frac{1}{\alpha}\left(1-\frac{m_{ca}^2}{(m_{ca}^{\textnormal{max}})^2}\right)$
\cite{Miller:2005zp}. 
So far, most of the steps are similar to the
case of no spin correlations except for the presence of the factors of spin correlation
functions, $g$ and $h$. 

It would seem that we need to know the {\em precise} form of $g$ and
$h$ in order to proceed further, i.e., in order to perform the 
integration in Eq.~(\ref{eqn:integrate}). However, for
the purpose of determining whether or not there is a cusp, we will show that it
is sufficient to know the fact that those
spin-correlation functions must be second order polynomials in
their argument as mentioned in the analysis of the $Z_2$ case. 
Using this fact we can write down the above
integrand as 
\begin{eqnarray}
\hspace{-1cm}\frac{1}{1-\alpha u}g(u)h\left(\frac{t}{1-\alpha u}\right) &=&
\frac{b_1}{(1-\alpha u)^3}t^2+\frac{1}{(1-\alpha u)^2} (b_2t+b_3t^2)+\frac{1}{1-\alpha u}(b_4+b_5t+b_6t^2)\non \\
&&\  +\ \ (b_7+b_8t)+b_9(1-\alpha u)\label{cusptopo}, 
\end{eqnarray}
where we have introduced the same variable $t \equiv
m_{ca}^2/(m_{ca}^{\textnormal{max}})^2$ used for the 3-body
decays and where the kinematical constants $b_i$ will depend on the specific nature
of the couplings and particles in the decay chain (\textit{i.e.}, they must be
calculated on a case by case basis). 
The terms of the integrand are organized as a power series in 
$\ (1-\alpha u)\ $ -- instead of in $u$ -- because of the simplicity of the former form. 
Integrating then gives 
\begin{eqnarray}
\frac{1}{\Gamma}\frac{d\Gamma}{dt}&=&
\left\{
\begin{array}{l}
b_1^{ \prime} +b_2^{ \prime} t+b_3^{ \prime} t^2 \hspace{4.6cm}\;\; \textnormal{for\hspace{.5cm} \ } 0<t<\sqrt{1-\alpha}  \cr
\cr
b_1^{ \prime \prime } + b_2^{ \prime \prime } t+b_3^{ \prime \prime } t^2+(b_4^{ \prime \prime}+b_5^{ \prime \prime } t+b_6^{ \prime \prime }t^2)\ \log t \hspace{.5cm}\;\;\;
\textnormal{for}\hspace{.5cm} \  \sqrt{1-\alpha}<t<1  
\end{array}\right.\label{z3fit}
\end{eqnarray}
where again, the kinematical constants $b_i^{ \prime } $ and $b_i^{ \prime \prime}$ are specific to each
situation. Thus, even with spin correlations, the functional dependence on
$t \; (\propto m_{ca}^2)$ is quite simple; however, the crucial point is that it is different for each sub-region of
the distribution. In particular this simple dependence in the
distribution of $m_{ca}^2$ (and $not\ $ $m_{ca}$) suggests that it
may be more appropriate to consider the distribution of $m_{ca}^2$ instead of
the distribution of $m_{ca}$. In Figure~\ref{spincorrgood} we show the
$m_{ca}^2$ invariant mass distribution for the decay chain of
Eq.~(\ref{newtopo}), but in the special case where particle $C$ has spin 1
and the intermediate particle $B$ is a fermion, and some of the
couplings are chiral. We used Madgraph~\cite{Alwall:2007st} to generate events taking the particles $a$ and $c$ to be massless and taking $m_{DM}=100$ GeV. One can compare the shape of this distribution
with the one from the right panel of Figure \ref{fig:4BodyNewTopo}
and see that in this case, including the spin correlation makes the
cusp even more apparent.

One of the main differences between the two subregions is that the first one has
no logarithmic dependence in $t$ while the second (in general) does have it. 
Of course, from Eq. (\ref{cusptopo}), we see that this logarithmic term could be
suppressed for the case $b_4=b_5=b_6\sim 0$.
However, even in this special case we would still have to employ different sets of coefficients in the two sub-regions as follows. The functional forms in both the regions are now quadratic in $t$, i.e.,  
\begin{eqnarray}
\hspace{-1cm}&&\frac{1}{\Gamma}\frac{d\Gamma}{dt}=
\left\{
\begin{array}{l}
b_7+\frac{b_9}{2}(2-\alpha)+\left(\frac{b_2}{1-\alpha}+b_8\right)\ t+\left[\frac{b_1(2-\alpha)}{2(1-\alpha)^2}+
  \frac{b_3}{1-\alpha}\right]\ t^2\;\;\;
\textnormal{}\hspace{1cm} \left( 0<t<\sqrt{1-\alpha}\right)   \\
\\ 
\frac{1}{\alpha}\left[b_2+b_7+\frac{b_1+b_9}{2}-(b_2-b_3+b_7-b_8)\ t-(b_3+b_8+\frac{b_1+b_9}{2})\ t^2\right]
\;\;\; \textnormal{}\ \left(\sqrt{1-\alpha}<t<1 \right)
\end{array}\right.
\end{eqnarray}              
Considering just the constant terms, we see that it is possible to obtain
identical functions in the two regions only if $\alpha=1$ and $b_1=b_2=0$. 
However, using Eq.~(\ref{eqn:kintri}) and Eq.~(\ref{alpha}), it can be shown that $\alpha$ is always 
(strictly) less
than 1. 
In other words, it is highly unlikely that the distribution in each
sub-region can be fitted successfully to the same polynomial of order
two in $t$; the cusp will thus survive even in this case. 

\begin{figure}[t]
  \centering
  \includegraphics[height=8cm,clip]{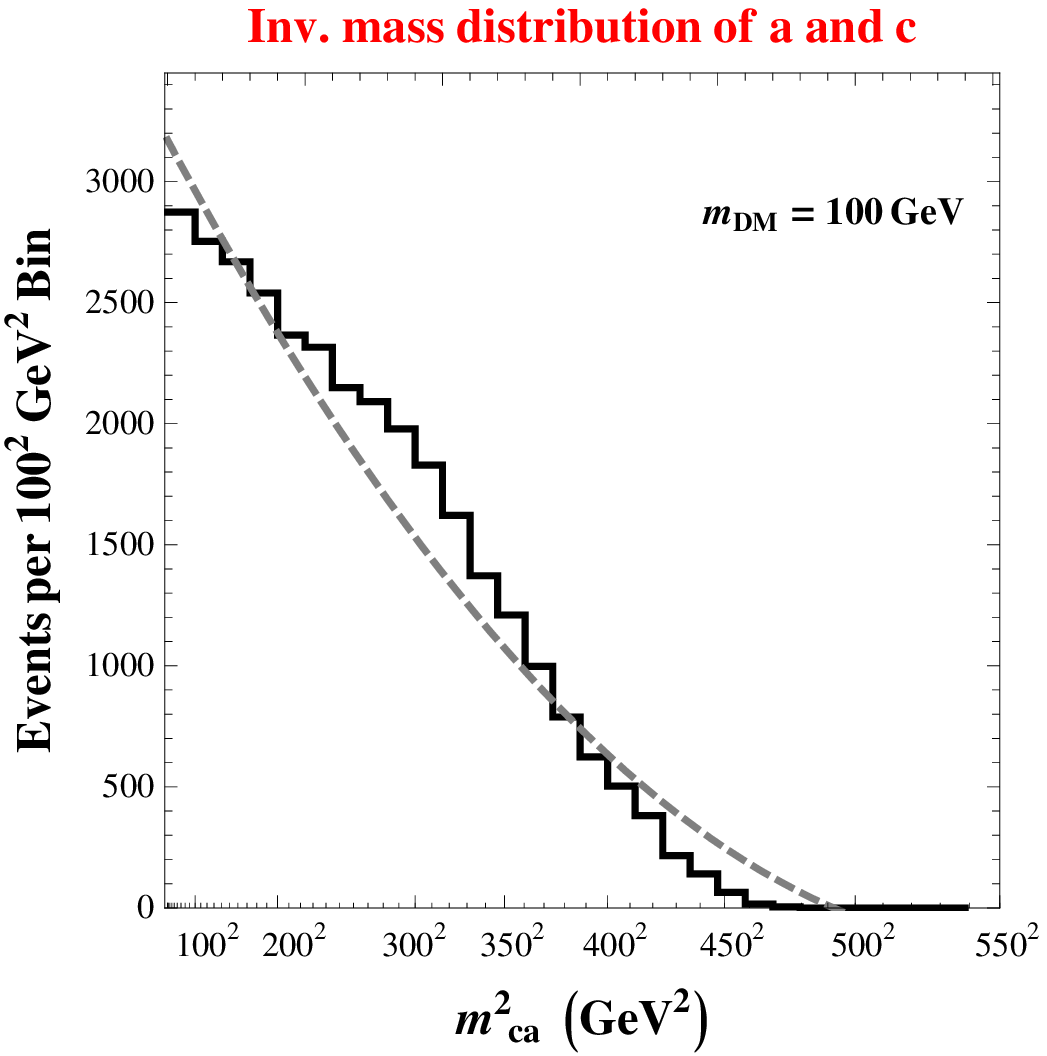} \hspace{.3cm}
  \includegraphics[height=8cm,clip]{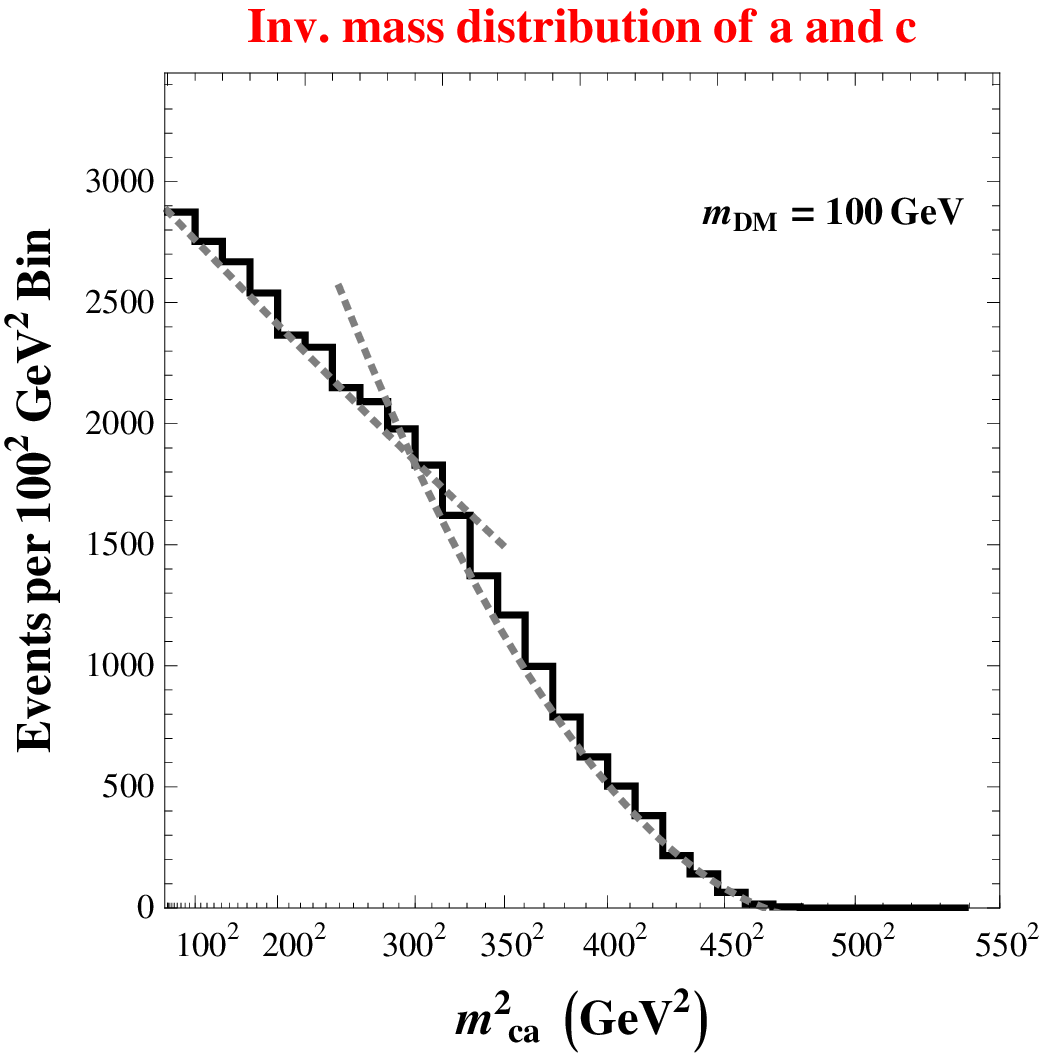} 
\caption{Invariant mass distribution of particles $a$ and $c$, as in
  Figure \ref{spincorrgood}, but with different chiral couplings. The
  cusp position is less apparent in this case but one can see (left panel) that a fit to a
  polynomial of second order as shown in Eq~.(\ref{z2fit}) is not very good (that is, the $Z_2$
  interpretation). On the right panel we  show the same distribution,
  with a different fitting function for the left side of the
  distribution and the right side (see Eq.~(\ref{z3fit})), consistent
  with the existence of a cusp, \textit{i.e.}, the $Z_3$ interpretation. }   
\label{spincorrba}
\end{figure}
In figure \ref{spincorrba} we show the distribution (again obtained 
with Madgraph~\cite{Alwall:2007st}) for the same decay chain as in Figure
\ref{spincorrgood}, but where the chiral structure of some couplings has been
modified from before. 
We see that the cusp feature is now less apparent,
but one also sees that a full fit to a polynomial of order two (left
panel) is not as good as a $multiple$-$region$ fit (right panel), where the
first part of the distribution is fitted to a polynomial of order
two (see first line of Eq.~(\ref{z3fit})), and the right side of the
distribution is fitted to the functional form (with a logarithm) given in the second line
of Eq.~(\ref{z3fit}).

\section{Example Models}

\subsection{Stabilization Symmetries from Spontaneous Symmetry Breaking}
\label{subsec:stabilmodel}

The most popular models of physics beyond the SM focus on solving the weak-Planck hierarchy problem by adding new particles at the {\em weak} scale.  Some of these new particles can be stable as a consequence of a discrete (often a parity) symmetry that is a part of (or imposed on) the theory. Thus these particles can have the correct thermal relic abundance to constitute dark matter, i.e., dark matter is then a ``spin-off'' of solving the hierarchy problem.  It may be possible, however, that the question of the origin of dark matter  is not rooted in {\em first} solving the hierarchy problem.  In this case, it is the thermal relic density which ``fixes" the mass of the dark matter particles to the weak scale.  As well, it is also known that the dark matter and baryon densities today are close in size
\begin{equation}
\Omega_\mathrm{DM} \sim 4.7 \,\Omega_\mathrm{baryon}
\end{equation}
which provides a hint to a possible common origin which may have a solution at the weak scale.

With this background, in reference \cite{Walker:2009en}, a model was introduced in which a $SU(N)$ gauge group is spontaneously broken to the $Z_N$ center.  There the goal was to, in part, determine whether a ``copy" of weak interactions could generate a viable dark matter candidate.  As is well known, the SM electroweak gauge group is spontaneously broken to $U(1)_\mathrm{em}$ which makes the lightest electrically charged particle  (i.e., the electron stable).   By analogy the $SU(N)$ gauge group 
in this model is broken to the 
$Z_N$ center that stabilizes dark matter.  The dark matter candidates which transform, e.g., as a fundamental under the $SU(N)$ are stabilized by the $Z_N$.   A unique, testable feature of these models is the existence of new gauge bosons that are neutral under the SM symmetries but do not couple to SM particles as $Z$ bosons.  These gauge bosons transform as adjoints under the $SU(N)$ and therefore are invariant under the $Z_N$ center.  

The fermionic dark matter candidate mentioned above (i.e., transforming as a fundamental under $SU(N)$) would 
get a mass $m \sim \lambda \,v_\mathrm{new}$.  Here $\lambda$ is a generic $\mathcal{O}(1)$ Yukawa coupling. For this mass to be of order the weak scale as required for the thermal relic abundance~\cite{Walker:2009en}, the vacuum expectation value (vev) that breaks the $SU(N)$ must be of order the SM Higgs vev.  Thus, the $SU(N)$ gauge bosons also have weak scale masses.  As a {\em bonus} (or ``double-duty'') of the new $SU(N)$ vev being similar to the Higgs vev, at finite temperature these gauge bosons have sphaleron solutions and nucleate additional bubbles (in analogy with the weak interactions) that may be relevant for electroweak baryogenesis. 

Let us make a connection to our study of distinguishing $Z_3$ from $Z_2$ using double edges described in section~\ref{sec:offshell} and new topologies described in section~\ref{sec:onshell}.  There, we chose $Z_3$ symmetry mainly for illustration;  in any case,  the key to our signal is the existence of ``charged-charged-charged" couplings under the dark matter stabilization symmetry in the decay chain.  Parity stabilized models have only ``odd-odd-even" couplings.  Any model, not necessarily a $Z_3$, that features such coupling has the potential to generate the signals we discussed earlier.

To this end, we take a ``toy" limit of the model discussed in reference~\cite{Walker:2009en}.  Namely, we consider a scenario where the new gauge bosons are long-lived and register as missing energy ($\etmiss$) in the detectors so that they behave {\em effectively} as dark matter particles, i.e.,
as ``charged'' (even though they were uncharged ``to begin with'').  This assumption will then ``convert''
the ``dark" $SU(N)$ gauge coupling of a $SU(N)$ fundamental fermion into an effective ``charged-charged-charged" coupling which will result in the double kinematic edges as well as the new topologies discussed earlier.  The result will also be to generate a ``hybrid" of the on- and off-shell scenarios presented above. 

Here, we simply want to make an estimate of the robustness of the signal described in sections~\ref{sec:offshell} and~\ref{sec:onshell} in the presence of basic detector and background cuts.  
So, the above toy limit of model in reference~\cite{Walker:2009en} will suffice for such a study of  
exploring the effects of the ``charged-charged-charged'' coupling in a more realistic situation than 
considered in earlier sections.\footnote{Alternatively, extensive model building along the lines of
reference~\cite{Walker:2009en}, which is not the focus of this paper, can provide a ``genuine'', i.e., without assuming
long-lived gauge bosons, ``charged-charged-charged'' coupling.} 
As a first step, following~\cite{Walker:2009en}, we summarize the effective lagrangian for our model.  Later we will discuss a simple production mechanism and discuss cuts consistent with the ATLAS and CMS collaborations.  Results follow afterwards.

For simplicity we chose $N=2$ to make our analysis.  To break the $SU(2)_D \to Z_2$ we require two new additional
Higgses in the ``dark" sector which transform as an adjoint 
\begin{align}
\phi = \begin{pmatrix} \phi_2 \\ \phi_0 \\ \phi_1 \end{pmatrix} && \eta = \begin{pmatrix} \eta_0 \\ \eta_1 \\ \eta_2 \end{pmatrix}.
\end{align} 
The higgs generate the following vevs
\begin{align}
\phi = \begin{pmatrix}  0 \\ v_1 \\ 0 \end{pmatrix} && \eta = \begin{pmatrix} v_2  \\ 0 \\ 0 \end{pmatrix}.
\label{eq:vevs}
\end{align} 
which break the $SU(2)_D$ to the center.  A general  scalar potential
does not naively generate the required breaking.  To get the correct
vacuum alignment, we require a scalar potential in the ``dark" sector
to minimizes $\phi\cdot \eta$.   $SU(2)_D$ scalar potential is  
\begin{eqnarray}
V &=& \lambda_1 \biggl( \phi^2 + \lambda_7\,\phi \cdot \eta  - v_1^2\biggr)^2   
+ \lambda_2 \biggl(\eta^2 +  \lambda_8\,\phi \cdot \eta  - v_2^2 \biggr)^2 +  \lambda_3 \biggl( \phi^2 + \eta^2 - v_1^2 - v_2^2 \biggr)^2 \label{eq:potential}  \\  
&+& \lambda_4  (\phi \cdot \eta)^2 + \lambda_5 \,\phi^3 + \lambda_6\, \eta^3. \nonumber
\end{eqnarray}
which generates three new heavy gauge bosons as well as three
additional Higgses.  In addition, we add anomaly free scalar and
fermions with the quantum numbers listed in Table 1. 
\begin{table}[h!]
\begin{center}
  \begin{tabular}{| c | c | c | c | c |}
    \hline
    Particle & $SU(3)_c$ & $SU(2)_L$ & $SU(2)_D$ & $U(1)_Y$   \\ \hline \hline
    Q & 3 & 1 & 2 & 1/3 \\ 
    sQ & 3 & 1 & 2 & 1/3 \\ 
    L & 1 & 2 & 2 &  -1/2\\ 
    sL & 1 & 2 & 2 & -1/2 \\ 
    $\chi$ & 1 & 1 & 2 & 0 \\ 
    s$\chi$ & 1 & 1 & 2 & 0 \\ \hline
    $V_\mu$ & 1 & 1 & 3 & 0 \\   
    $\phi$ & 1 & 1 & 3 & 0 \\    
    $\eta$ & 1 & 1 & 3 & 0 \\    \hline
  \end{tabular}
  \label{tab:partspec}
  \caption{An effective, anomaly free particle spectrum that fills out
    a $(5,2) + (\bar{5},2)$. The ``\textit{s}" prefactor denotes a scalar
    particle.  Here $V_\mu$ are the $SU(2)_D$ gauge bosons.  We assume
    the mass of the $Q$ and s$\chi$ is heavy and integrated out.  The
    rest of the spectrum mediates the decay chain in
    equation~\ref{eq:production}. } 
  \end{center}
  \end{table}
Constructing a supersymmetry UV completion to this effective
lagrangian is straightforward.  Although the details is beyond the
scope of this paper, note a simple way to do so would be to augment
minimum supersymmetric  standard model with chiral superfields with
the charges in Table~\ref{tab:partspec}.  SUSY breaking terms would
then need to be constructed to lift the appropriate particles which
will be integrated out to generate the effective theory.

\subsubsection{Production Rates at the LHC}
\label{subsec:prorate}

\begin{figure}[t]
\centering
\includegraphics[height=2cm,clip]{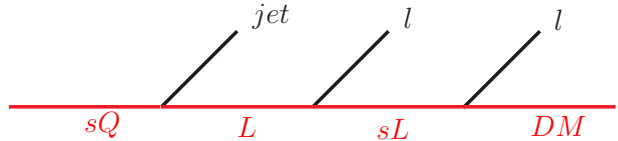}
\caption{The topology of the primary decay chain.  Here  $DM$ is the $\chi$
  particle.  } 
\label{origtopo}
\end{figure}

As an example of the unique decay topologies generated by these models, we consider pair production of new exotic heavy quarks, $p\,p \to \overline{Q}\,Q$.   The leading production mechanism is via QCD
\begin{equation}
p\,p \to sQ^*\,sQ + X \to \overline{q} \,\chi \,\,sQ + X \to \overline{q} \,\chi\,\, q\,\bar{l}\,l\,\bar{\chi} + X
\label{eq:production}
\end{equation}
where $X$ represents the beam remnant and other possible hadronic
activity.  The first $sQ$ decays via $sQ^* \to \overline{q} \,\chi$
and the second decay to $sQ \to q\,\bar{l}\,l\,\bar{\chi} $ which is a
primary decay chain of study.  The charged leptons are $l = e, \mu$.
The signal is for two isolated leptons, two light quark jets and large
amounts of $\etmiss$.  We take a mass spectrum of  
\begin{align}
m_Q = 700\,\,\mathrm{GeV} && m_L = 650\,\,\mathrm{GeV}  && m_{sL} =
300\,\,\mathrm{GeV}  && m_\chi = 100\,\,\mathrm{GeV}   && m_V =
100\,\,\mathrm{GeV}   
\label{eq:masses}
\end{align}
The topology of the primary decay chain is shown in
Figure~\ref{origtopo}.  The partial decay widths for the $sQ$ is  
\begin{align}
\Gamma_1 = \frac{\lambda_1^2\,M_Q}{ 16 \pi}\sqrt{1 - \frac{M_{DM}^2}{ M_Q^2}} &&
\Gamma_2 = \frac{\lambda_2^2\,M_Q}{ 16 \pi}\sqrt{1 - \frac{M_{L}^2}{ M_Q^2}}
\end{align} 
In the analysis, for simplicity, we set all of the Yukawa couplings to
$\lambda_1 = \lambda_2 = \lambda$.  We assume a 100\% branching
fraction of $L \to l \, \mathrm{sL}$ and $\mathrm{sL} \to l\, \chi$.
In this model the gauge boson decays at one loop.  $sL$ is the
lightest partner; thus, fastest the decay rate goes as  
\begin{equation}
\Gamma \sim \frac{g^2 \lambda^4}{16 \pi^2}\, \frac{m^{13} }{ M^8 M_\chi^4}
\end{equation}
where $m$, $M_\chi$ and $M$ are the masses of the gauge boson, dark
matter and sL, respectively.  Here $\lambda$ is the coupling between
the sL, $\chi$ and the SM lepton.  We take $\lambda$ to have a
technically natural  value of $\lambda \sim 0.001$ so the gauge boson
is long-lived.  With the masses given in equation~\ref{eq:masses}, we
have a lifetime of about $10^{-8}$ seconds.   It should be noted that
long-lived particles take about  $\sim \mathcal{O}(1)\times10^{-9}$
seconds to transverse the larger ATLAS detector.  Thus, these gauge
bosons will register as missing energy.  Even though the coupling is
so small, the decay chain proceeds because of the branching fractions.
Finally, the signal is generated with the new gauge bosons, $V$, being
emitted from the decay chain in Figure~\ref{origtopo}.  We list the
topologies generated to order $\alpha$ in
Figures~\ref{Z3topo1}-\ref{Z3topo5} in Appendix B.
In sections~\ref{sec:offshell} and \ref{sec:onshell}, we have
discussed the  
invariant mass distributions for dark matter stabilized with a $Z_2$ or 
$Z_3$ stabilization symmetry with the virtual particles, respectively, off- 
or on-shell.  The present model presents a ``hybrid" between the two 
pictures.  This is because emitting the long-lived gauge boson forces part 
of the decay chain off-shell.  Emitting the new gauge boson also causes 
these diagram to be suppressed because the virtual particles in the decay 
chain must go off-shell.  Because there are three new gauge bosons, the 
overall off-shell suppression is enhanced by a multiplicity factor for each 
boson emitted.

\subsubsection{Extracting the Signal}
\label{subsec:extract}

To get an estimate on the signal we first impose basic acceptance cuts
which are consistent with ATLAS and CMS studies of on- as well as
off-shell SUSY decay chains.~\cite{ATDR,CTDR}  We require 
\begin{eqnarray}
&& |\eta_l| < 2.5, \quad \quad\qquad|\eta_j| < 2.5, \\
&& \Delta R_{ll} > 0.3, \qquad \Delta R_{lj},\,\Delta R_{jj} > 0.4.
\end{eqnarray}
where $l$ and $j$ are lepton and jets.  $\eta_a$  is the
pseudorapidity  of particle $a$.  $\Delta R_{ab}$ is defined as
$\Delta R_{ab} = \sqrt{d\eta^2_{ab} + d\phi_{ab}^2}$ where
$d\eta_{ab}$ and $d\phi_{ab}$ is the difference in the pseudorapidity
and transverse angle of the detector between particles ``$a$" and
``$b$."  As described in the previous section, our example signal has
two leptons and jets as the SM final states.  The primary SM
background for this signal is $\overline{t}t$ decays into two leptons.
Additional backgrounds include QCD, $W +$ jets and $Z +$ jets events.
ATLAS and CMS places additional cuts to reduce this as well as other
SM backgrounds.  The model allows same-sign or opposite-sign dileptons
in the final state.  
Since the purpose of this section is to see the effect of the detector cuts 
on our signal, we choose the more conservative opposite-sign dilepton cuts. 
We adopt cuts consistent with both collaborations by requiring  
\begin{enumerate}
\item Two leptons with $p_T > 20$ GeV
\item  At least one leading jet with $p_T > 100$ GeV and subleading jets with $p_T > 50$ GeV
\item $\etmiss > 100$ GeV and  $\etmiss >  0.2\, M_\mathrm{eff}$
\item Transverse sphericity $S_T > 0.2$.
\end{enumerate}
Here the missing energy ($\etmiss$) is defined as 
\begin{equation}
\vec{\etmiss} = \vec{p_T\!\!\!\!\!\!\slash} \,\,\,\, = - \sum_i \vec{p}_{i\,T}
\end{equation}
and $i$ runs over the transverse momentum, $p_T$, of the visible final
state particles in the event.  The effective transverse mass,
$M_\mathrm{eff}$,  is defined as  
\begin{equation}
M_\mathrm{eff} = \sum_i E_{i\,T} + \etmiss
\end{equation}
where the sum runs over the measured transverse energy, $E_T$, from
the visible particles in the event.  Finally the transverse sphericity
($S_T$) is defined as  
\begin{equation}
S_T = \frac{2 \lambda_2 }{ \lambda_1 + \lambda_2}
\end{equation}
where $\lambda_{1,2}$ are the eigenvalues of the $2 \times 2$ sphericity tensor 
\begin{equation}
S_{ij} = \sum_\kappa p_{\kappa i} p^{\kappa j}
\end{equation}
where $\kappa$ runs over the number of final state jets and leptons.
The other indices, $i$ and $j$, run over the $p_T$ components of each
particle.  $S_{ij}$ is evaluated for the final states with $\eta <
2.5$  and $p_T > 20$ GeV.  $S_T \sim 1$ for approximately spherical
events; QCD events are usually back-to-back with $S_T \sim 0$.
Generally, because our signal has three dark matter candidates per
event, these cuts could be optimized with larger $\etmiss$ cuts.  For
direct comparison with ATLAS and CMS, we simulated our signal with the
cuts above.  The ATLAS collaboration~\cite{ATDR} finds the following
backgrounds for 1 fb$^{-1}$ of integrated luminosity (see Table 2).
\begin{table}[tbp]
\begin{center}
  \begin{tabular}{| c | c | }
    \hline
    Background & Events (1 fb$^{-1}$)\\ \hline \hline
    $t\overline{t}$ & 81.5  \\ 
    $W +$ jets & 1.97  \\ 
    $Z +$ jets & 1.20 \\ 
    QCD &  0 \\ \hline
    Total SM &  84.67 \\ \hline
  \end{tabular}
 \caption{SM backgrounds as computed by~\cite{ATDR}.}
\end{center}
\label{tab:tab2}
\end{table}
In addition to these backgrounds, we have an additional irreducible
background when the $Z_2$-like signal process,
equation~\ref{eq:production}, emits Z boson which decay invisibly.
The invisible branching for Z bosons into neutrinos is
20\%.~\cite{Amsler:2008zzb}  Finally, in our analysis we simulate
calorimetry responses for the energy measurements by adopting Gaussian
smearing~\cite{ATDR} with the following parameters. 
\begin{eqnarray}
{\Delta E_e\over E_e} =  {10\%\over \sqrt {E_e (\mathrm{GeV})}} \oplus 0.7\%,\qquad
{\Delta E_j\over E_j} =  {50\%\over \sqrt {E_j (\mathrm{GeV})}} \oplus 3\%.
\end{eqnarray}

\subsubsection{Results}

\begin{figure}[h]
  \centering
  \includegraphics[width=7.5truecm,height=7.0truecm,clip=true]{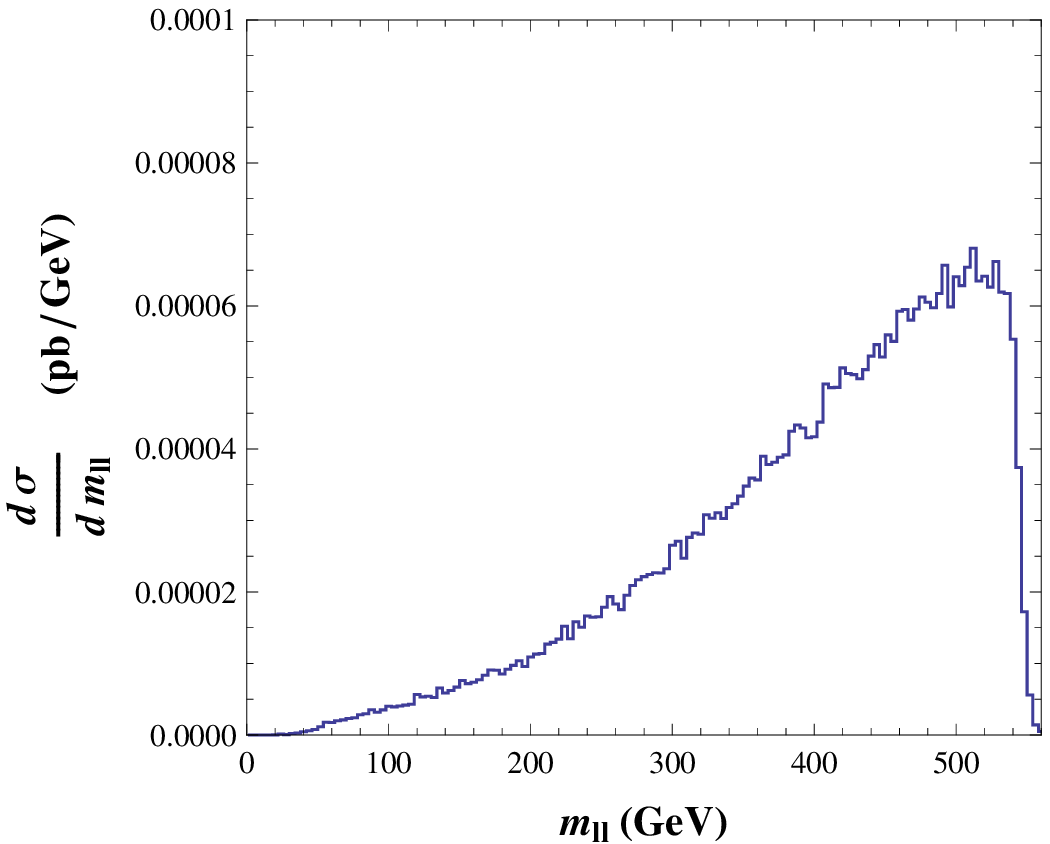} \hspace{0.2cm}
   \includegraphics[width=7.5truecm,height=7.0truecm,clip=true]{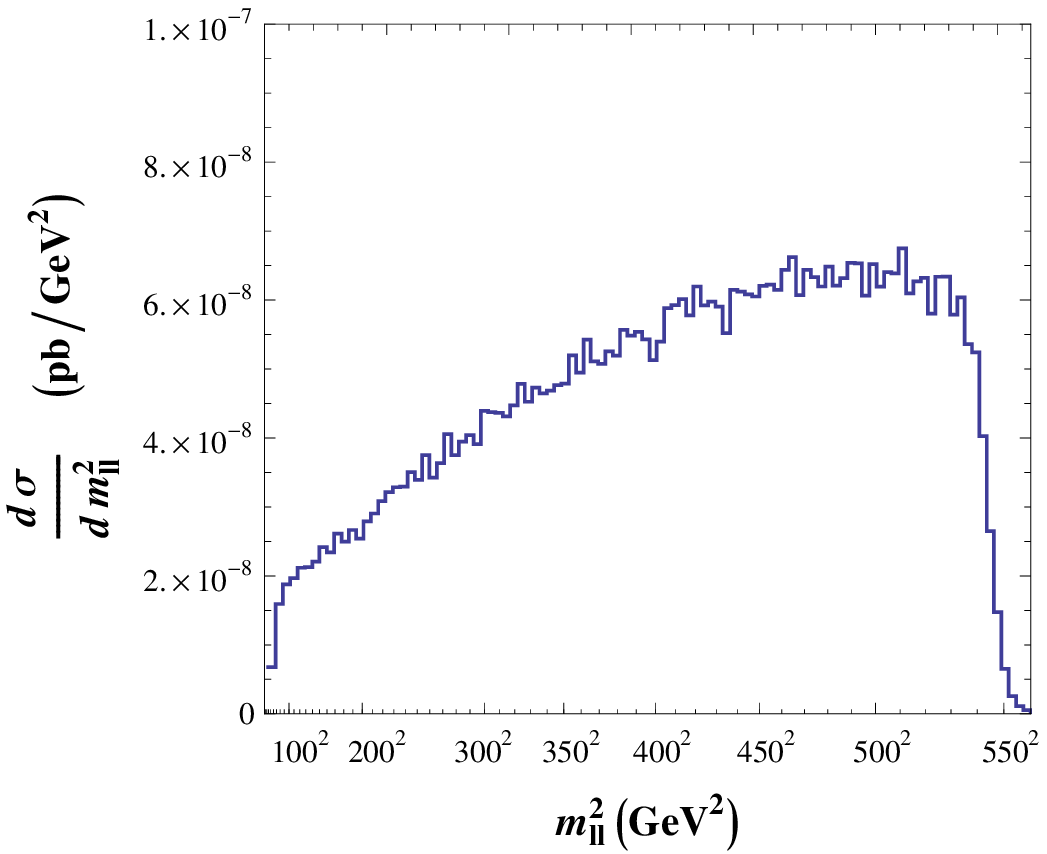}
\caption{Dilepton invariant mass (left panel) and invariant mass
  squared (right panel) distributions for the topology in  Figure
  \ref{origtopo}.  The cuts described in section~\ref{subsec:extract}
  are applied.  
  }    
\label{topo0result}
\end{figure}
\begin{figure}[h!]
  \centering
  \includegraphics[width=7.2truecm,height=6.7truecm,clip=true]{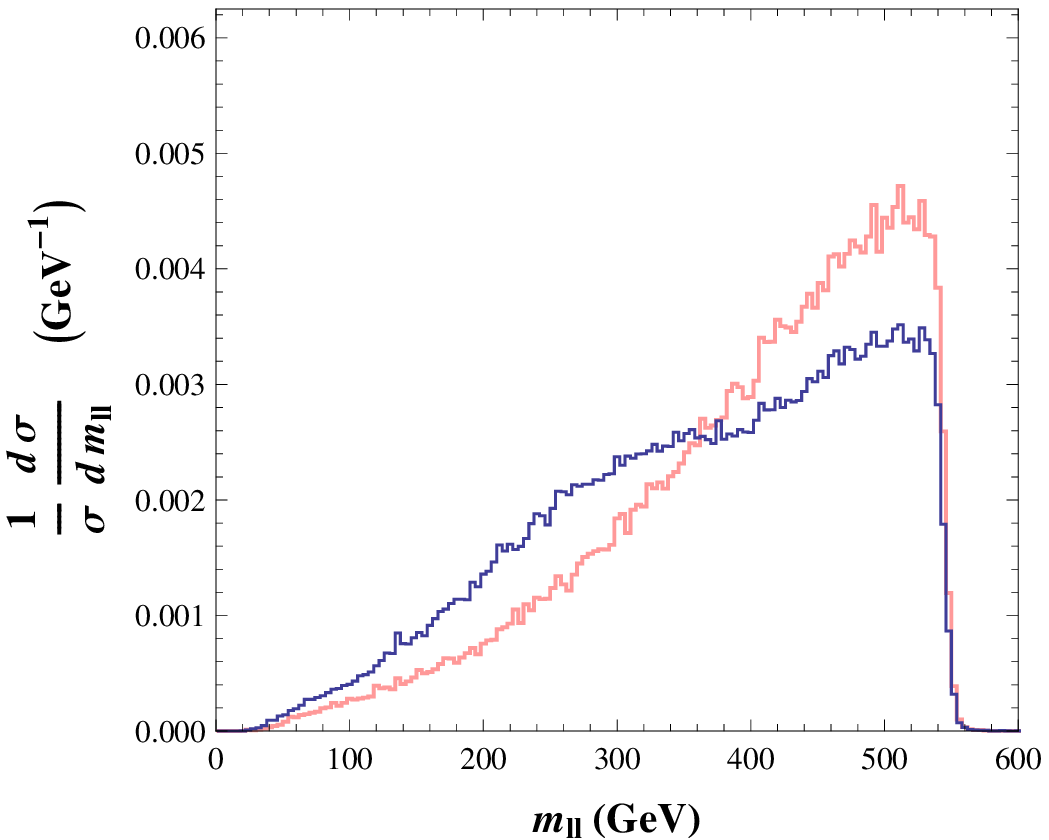} \hspace{0.2cm}
   \includegraphics[width=7.2truecm,height=6.7truecm,clip=true]{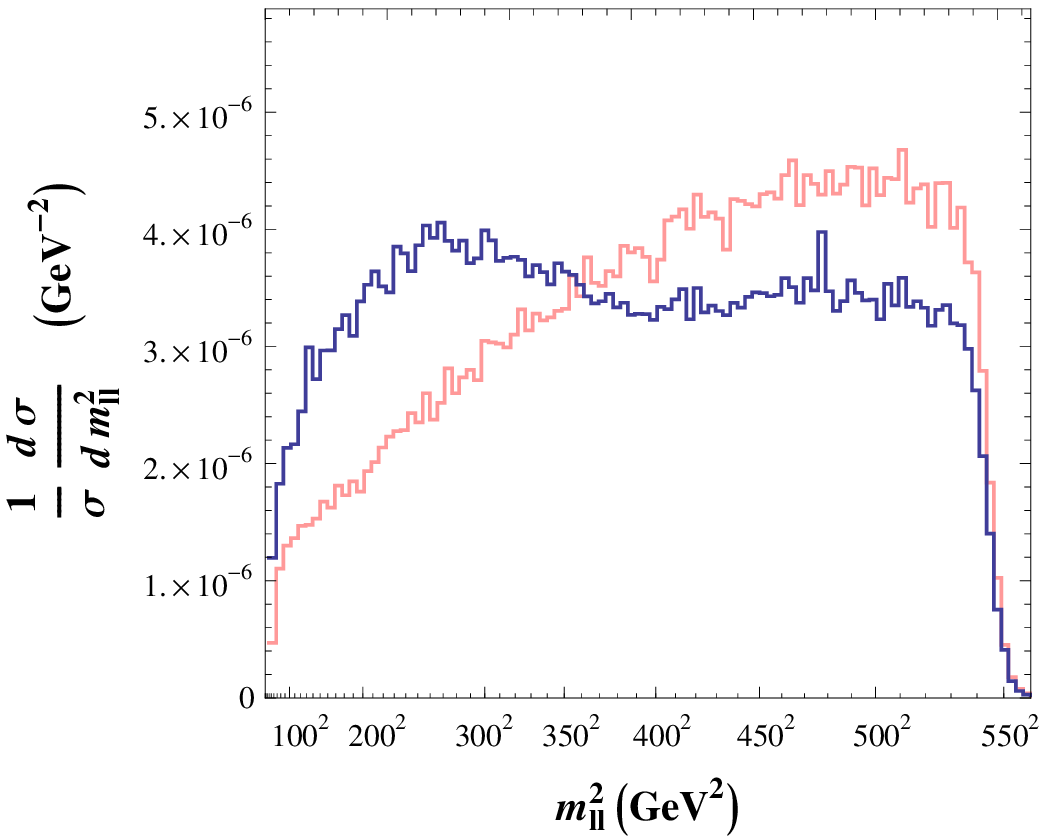}
\caption{Dilepton invariant mass (left panel) and invariant mass
  squared (right panel) distributions but with the first order
  corrections from the long-lived gauge bosons.  The left panel
  features the double kinematic edge.  The edges are roughly separated
  by the 100 GeV gauge boson mass.  The $Z_2$-like signal is also
  plotted for comparison as well as the backgrounds described in
  the figure above.  
  All of the topologies
  generated by the long-lived gauge bosons are listed and individually
  plotted in Appendix B.}    
\label{comboresult}
\end{figure}

We ran our Monte Carlo for the LHC at 14 TeV center-of-mass energy for
the signal process in Figure~\ref{origtopo}.  We used CTEQ 4M parton
distribution functions~\cite{Lai:1996mg}; and, all results are
presented at parton level.  $\alpha_s$ is computed at two-loop level.  At zero order in the $SU(2)_D$ gauge
coupling, the model admits  ``$Z_2$-like" topologies.  We show the
kinematic edge resulting from this topology in
Figure~\ref{topo0result}.  We also include SM the dominant $\bar{t}t$
as well as the irreducible $Z \to \bar{\nu}\nu$ backgrounds.  To order
$\alpha$ in the $SU(2)_D$ coupling, we have six additional diagrams
which generate corrections.  For
completeness, in appendix B, we list all of the
different topologies and plot each correction before interfering the
diagram to get Figure~\ref{comboresult}.  Each diagram listed in the
appendix is instructive for the shape and position for each kinematic
edge.  The kinematic cuts listed in section~\ref{subsec:extract} are
taken; the shapes of the distribution are generally preserved under
the cuts.  The total irreducible background events from $Z \to \bar{\nu} \nu$ and the dileption $\bar{t}t$ channel for 100
fb$^{-1}$ is 
\begin{align}
B_{Z \to \bar{\nu} \nu} =  98.5 && B_{\bar{t}t} =  56630
\end{align}
Additionally total signal events (Figure~\ref{comboresult}) for 100 fb$^{-1}$
\begin{equation}
S =  4440
\end{equation}
which generates the following signal-to-background ratio and statistical significance for signal observability
\begin{align}
S/B = 0.08 && S/\sqrt{B} = 18.6
\end{align}


\subsection{Warped GUT}

We present another very well-motivated $Z_3$ 
model: for more details, see the original 
references \cite{Agashe:2004ci, Agashe:2004bm}.
This model is based on the framework of a warped extra dimension with the SM fields propagating in it
which can address both the Planck-weak and flavor hierarchy problems of the SM \cite{Davoudiasl:2009cd}.
In a grand unified theory (GUT) model within this framework, it was shown that 
\begin{itemize}

\item
a viable DM particle can emerge a {\em spin-off} of suppressing proton decay.

\end{itemize}
Moreover, 
\begin{itemize}

\item
$Z_3$ (rather than $Z_2$) as the symmetry stabilizing DM
arises naturally (due to combination of SM color and baryon number quantum numbers).

\end{itemize}
It turns out that the SM particles resulting from the decay of mother particles in this model
are mostly top quarks and $W$'s. We defer an analysis of the reconstruction of 
these decay chains
to future work. Here we simply give a summary of this model and
the relevant LHC signals.

\subsubsection{Basic framework}

In this framework, the 
SM particles are identified as zero-modes of $5D$ fields, whereas
the heavier modes (i.e., non-trivial excitations of the SM particles 
in the extra dimension) are
denoted by Kaluza-Klein or KK particles and constitute the new physics.
A few TeV KK mass scale can be consistent with electroweak and flavor
precision tests,
at the same time avoiding at least a severe fine-tuning of the weak
scale \cite{Davoudiasl:2009cd}.

An 
extension of the SM gauge group in the bulk to GUT gauge group is motivated
by precision unification of gauge couplings and explanation of
quantization of hypercharge.
In more detail, the
extra/non-SM $5D$ gauge bosons (denoted
generically by $X$) do not have zero-modes by, for example, an appropriate
choice
of boundary conditions. However, these gauge bosons still have
KK modes with a {\em few TeV} mass (i.e., same as SM gauge KK modes),
instead of usual mass of $\sim O \left( 10^{ 15 } \right)$ GeV in $4D$-like
GUTs.
The fermions follow a different story as follows.
The $5D$ fermion fields must of course form complete GUT multiplets.
Usually, an entire SM generation fits in such a complete multiplet(s), for example,
{\bf 16} for $SO(10)$ GUT group, i.e., 
quark-lepton unification is incorporated.
However, if we attempt to identify SM fermions of one generation
as {\em zero-modes} of a complete $5D$ GUT multiplet, then it turns out that
we will get
too fast proton decay via exchange of $X$ between SM quarks and leptons
-- again, with a few TeV
mass.\footnote{It turns out that the couplings of $X$ to SM quarks and leptons
are suppressed -- roughly by powers of SM Yukawa couplings -- due to
the nature of the profiles in the extra dimension of the various particles, but this effect is not sufficient to allow
a few TeV mass for $X$ to be consistent with proton decay.} 

\subsubsection{Split fermion multiplets}

Fortunately, the breaking of GUT gauge group down to SM gauge group by 
boundary condition allows ``split" fermion multiplets (just like for gauge bosons) as follows.
We can choose boundary conditions such that one $5D$ multiplet (labeled ``quark" multiplet) has 
zero-mode only in its quark component, with the lepton-like component having only KK mode
and vice versa
for another multiplet (labeled ``lepton'' multiplet').
Thus SM
quarks and leptons originate from {\em different} $5D$ multiplets, 
avoiding exchange of $X$ gauge bosons between
SM quarks and leptons since such exchange can only couple fermions
(whether zero or KK-modes) within the {\em same} $5D$ fermion multiplet.
In spite of this ``loss'' of quark-lepton unification, the explanation of 
hypercharge quantization is still maintained
since SM quark must still be {\em part} of a complete GUT multiplet.
In fact, such splitting of fermion multiplets results in 
precision unification of couplings in the model where the GUT group is broken down to the
SM on the Planck brane \cite{Agashe:2005vg}. The reason for the modification relative to the running
in the SM (and hence the improvement in the unification) is the
different profiles for quarks and leptons, especially
within the third generation.

\subsubsection{Additional symmetry for proton stability}

It turns out that to 
maintain this suppression of proton decay
at higher orders, we have to impose an extra symmetry,
for example, a gauged $U(1)_B$ [commuting with the GUT group] in the bulk as follows. 
The {\em entire} quark multiplet 
is assigned $B = 1/3$ (i.e., that of the zero-mode contained in this multiplet). Thus
lepton-like states from this multiplet are ``exotic'' in the sense that they have
$B = 1/3$.
Similarly, the entire lepton multiplet is assigned $B=0$, giving
exotic quark-like states (i.e., with $B=0$).
$X$'s are also exotic since they are colored, but have $B=0$.
The 
exoticness is especially striking since these states cannot decay into purely SM: 
explicitly, they are charged under a symmetry
\begin{eqnarray}
\Phi & \rightarrow & e^{ 2 \pi i \left( \frac{ \alpha - \bar{\alpha} }{3}
- B \right) } \Phi
\end{eqnarray}
(where $\alpha$, $\bar{\alpha}$ are the number of color, anti-color
indices on a field $\Phi$),
under which SM is neutral.\footnote{In more detail, $U(1)_B$ has to be broken to avoid zero-mode gauge boson.
We break it 
on the Planck brane so that 
$4$-fermion operators giving proton decay [i.e., violating $U(1)_B$] can only arise on the Planck brane
where they are adequately suppressed.
However, $U(1)_B$ still cannot be broken arbitrarily, i.e., 
a subgroup of $U(1)_B$  must still be preserved in order to forbid
(mass) mixing of lepton and quark multiplets on the Planck brane which will lead to rapid 
proton decay -- for example,
we require that the scalar vev which breaks $U(1)_B$ has
$B=$ integer in which case the above $Z_3$ symmetry is still preserved,
even if $U(1)_B$ is broken.}
Thus the lightest $Z_3$-charged particle (dubbed ``LZP'') is stable (others $Z_3$-charged
particles produced in colliders
or in the early, hot universe decay into it)
and a potential
DM candidate, depending on its couplings.

\subsubsection{Who's the LZP/DM?}

In the model with GUT group broken to the SM on the Planck brane which was the focus
in references \cite{Agashe:2004ci, Agashe:2004bm},
it turns out that,
due to profile of $t_R$ (in turn, based on heaviness of top quark and
constraint from shift in $Z b \bar{b}$ coupling), its exotic GUT partners
are lighter that typical KK scale (say, mass of gauge KK modes).
Thus, it is very likely that 
LZP resides in this multiplet.
In particular, if the GUT group is $SO(10)$, then 
there is a GUT partner of $t_R$ with quantum numbers of right-handed (RH)
neutrino, but with $B=1/3$ (denoted by $\nu^{ \prime }_R$
and its LH Dirac partner, denoted by $\hat{ \nu^{ \prime }_R }$). It has 
been shown that if this $\nu^{ \prime }$ is the LZP\footnote{At leading order, all the GUT partners of 
$t_R$ are degenerate, but higher-order effects can split them.}, then it is a good DM candidate, in the sense that, in some regions of parameter space, it
has the correct relic density upon thermal freeze-out in the early universe: 
see Fig. 5 of reference \cite{Agashe:2004bm} and Figs. 3 and 4 of 
reference \cite{Agashe:2009ja}
(the latter reference studies a modified version of the model outlined here)
and related discussion.
Similarly, the constraints from direct detection of DM can be satisfied: see
Fig. 7 of reference \cite{Agashe:2004bm} and Fig. 3 of reference \cite{Agashe:2009ja}
and related discussion.
Other GUT partners of $t_R$ are then heavier than $\nu^{ \prime }$, but they can still be lighter
than SM gauge KK modes.
And, 
GUT partners of {\em other} SM particles, including $X$-type gauge bosons, 
are as heavy as SM gauge KK modes.

\subsubsection{DM partner at the LHC}

As usual, in order to test this idea at colliders, we consider 
producing the (other than LZP) $Z_3$-charged particles at colliders
(of course, these must be produce in pairs)
and observe their
decays into SM particles and LZP.
Since 
colored and lightest such particles will have largest cross-section at the LHC,
a good candidate  for such a study is
the GUT partner of $t_R$ with
$(t,b)_L$ quantum numbers, denoted by $\left( t^{ \prime }_L, b^{ \prime }_L 
\right)$ [and it's conjugate by $\left( \hat{ t^{ \prime }_L}, \hat{ b^{ \prime }_L }
\right)].$\footnote{Recall that the $(t,b)_L$ and {\em conjugate} of $t_R$
are contained in the same representation, namely, {\bf 16} of $SO(10)$
so that the $\left( t^{ \prime }_L, b^{ \prime }_L 
\right)$ states with $B =-1/3$ are indeed exotic.}
The two states
$t^{ \prime }$ and $b^{ \prime }$ are degenerate before EWSB, but will be split afterward.

\begin{figure}[t]
  \centering
\includegraphics[height=7.5cm,clip]{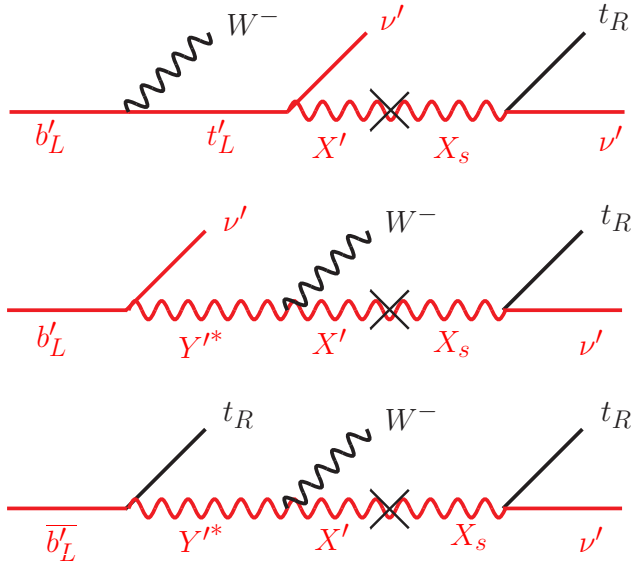}
\caption{Different possible decay chains for $b'$ in the scenario of
  \cite{Agashe:2004bm}. Note the appearance of two DM states in
  two of the possible decay chains, whereas only one DM particle comes
out of the third possibility.}
\label{z3kmod}
\end{figure}

We focus here on $b^{ \prime }$ due to the interesting features
in its decay channels as shown in figure \ref{z3kmod}.
%
$X_s$, the $SU(2)_L$ doublet $X$, $Y$
and another $SU(2)_L$ doublet $X^{ \prime }$, $Y^{ \prime }$
are beyond SM gauge bosons of $SO(10)$, with electric charges
$2/3$, $4/3$ and $1/3$, and $2/3$ and $1/3$, respectively.\footnote{Note that
only the 1st decay channel is shown in Fig. 10 of of reference \cite{Agashe:2004bm}.}
First of all, the 1st decay chain of $b^{ \prime }$ into two DM ($\nu^{ \prime }_R$) and $tW$
does have the topology needed to give a cusp in the invariant mass of SM/visible particles
(of course assuming on-shell intermediate particles; see comment below).
Second,  the 
2nd process above involves the same final state, but with a different topology
and thus is relevant for obtaining multiple edges (again, with on-shell intermediate particles),
even with two DM in final state.
However, it is clear that the intermediate particles ($X$-type gauge bosons)
in the above processes 
are actually {\em off}-shell so that these features are not so useful for us here.
Finally, while there does not seem to be a decay of 
$b^{ \prime }$  into $tW$ and one DM which would be
relevant for obtaining double edge with off-shell intermediate particles (when combined with the
1st and 2nd processes above with two DM),
there is a decay to $tWW$ and one DM (i.e., with an ``extra'' $W$)
as in the 3rd process above which might play this role.
Thus, decay of $b^{ \prime }$ will exhibit a 
double edge due to presence of one DM and two DM
in final state [along the lines of the discussion of Eq. (\ref{4and4body})]:
$M_{ b^{ \prime } } - 2 \; M_{ \nu^{ \prime } } - m_t - m_W$ (for 1st and 2nd processes) and 
$M_{ b^{ \prime } } - M_{ \nu^{ \prime } } - 2 \;  m_t - m_W$ (for 3rd process).

Note that, in a non-minimal model, for example if $SO(10)$ is broken on the
TeV brane instead of the Planck brane, the $X$'s bosons might be lighter so that
the intermediate particles in decay chains similar to those above might be on-shell.
However, 
then we might as well study production of the lighter $X$'s (instead of the
exotic fermions) which are also colored 
and
which will decay into LZP via {\em off}-shell GUT partners of SM fermions.


\section{Conclusions and Outlook}

Many extensions of the SM contain stable WIMPs which can be viable DM candidates. 
Most of these models stabilize the DM with a parity ($Z_2$) symmetry. 
However, in the spirit that 
all fundamental particles in nature are defined by how they transform under different symmetries, these
models actually correspond to one type of candidate!
On the 
other hand,  any continuous or discrete global symmetry can be adopted for DM stabilization; and, 
DM candidates stabilized by a parity symmetry and, e.g., a $Z_3$ symmetry are different.

This possibility of other than $Z_2$ symmetries stabilizing
DM is more than just of academic interest; the nature of the stabilization symmetry has important implications on collider searches for DM.  At colliders other particles (heavier than DM) which are charged under the same symmetry which stabilizes the DM candidate(s) can be produced, decaying into
DM and SM particles. Such events will generate decay topologies and modes that are determined, in part, by the stabilization symmetry. Thus the analysis  of such ``missing energy" signals will present a hint not only for the 
existence of the DM but also the nature of its stabilization symmetry.  

For example, the decay of a {\em single} such mother particle can contain one or two
DM in the case of a $Z_3$ symmetry, but only one in the case of $Z_2$ symmetry.
We showed that in many cases simple kinematic observables, such as invariant mass distributions of the visible particles of such
decay chain, could then characterize the stabilizing symmetry.  
Specifically, 
when a mother particle decay via off-shell intermediate states into the same visible particles
along with one and two DM (for the case of $Z_3$ symmetry), it may be possible to observe a {\it double-edge} in the distribution of these visible particles
(vs. single edge for $Z_2$ symmetry).  
In fact, the difference between the location of the edges will be a
direct measure of the mass of the dark matter particle for $Z_3$ models.   On the other
hand, when the intermediate particles are on-shell, we also pointed
out the possibility of a very distinctive feature appearing in the
invariant mass distribution of {\em two} visible particles in the case of
$Z_3$ symmetry: a cusp dividing the distribution into two regions.
This happens when two DM particles emerge from the same chain, with
one of these DM particles  being situated in-between the two SM particles.


\textbf{Signal Fakes:}  We point out that models with parity stabilized dark matter can naively
fake {\it double kinematic edges} and {\it cusps}. 
An example of a faked cusp comes from a decay chain with on-shell intermediate particles and 
involving a SM neutrino(s) in the final state. 
If a neutrino is emitted between the two other, visible SM particles,
then we obtain the new topology considered above so that
the invariant mass distribution of the two visible SM particles 
can give a cusp.   
As another example, a double kinematic edge also comes from two
different mother particles charged under $Z_2$ which could be produced
in separate events~\cite{Baer:2008ih}, but with same visible decay 
products\footnote{Note that, assuming pair production,
multiple mothers in $Z_2$ cannot fake
different edges in two distributions that we discussed in section 
\ref{different}.} 
Also, in this paper, we considered decays of mother particle into DM and SM particles
occurring inside the detector.  However, a long-lived, parity odd particle in a 
$Z_2$ model that decays outside of the detector can fake our $Z_3$ signal.  For example,
a single mother can decay into such a particle and SM particles.  
The same mother can also decay to the same SM final state and the 
DM.  A combination of these two decay chains can generate a double edge.
A closely related scenario is that there are actually two (absolutely) stable 
DM particles in a $Z_2$ model \cite{multiple}, again giving double edges 
from two decay
chains (along the lines just mentioned).
Finally, if there are {\em more than} two SM particles involved in the decay of a single mother, the analysis 
of the cusp in the invariant mass distribution will be more
complicated, but nevertheless we expect that in general the same type
of arguments made here should be able to be implemented in this
case. As an example, a possible \textit{pseudo}-faking situation will
arise when three visible particles are emitted in a decay cascade in
the $Z_2$ case.  
However, we mentioned how this type of faking can be resolved by considering
all possible pair invariant masses.  
  
\textbf{Future Considerations:}  Of course, in any given event, there will be two such mother particles present\footnote{However, studying the decay of only one mother (as we have mostly done
in this paper) is still relevant, for example, if this decay
involves a ``clean'' (SM) final state (vs. a dirtier one from the other mother). Of course, we also need
to determine {\em experimentally} which SM particles  
in the event came from this decay chain, which is possible, for example, if this mother
is boosted in the laboratory frame so that the decay products are roughly
collimated.} (three mothers
is a possibility in $Z_3$ case, but it is phase-space
suppressed). 
The assumption of a $Z_2$ symmetry thus points to an eventual
emergence of two invisible particles for every new
physics event in the collider. 
On the other hand, models where dark matter is stabilized with,
e.g. $Z_3$ symmetry, can have two, three, or four dark matter
candidates in an event.  
In an ongoing work, we construct a variant of the  $m_{T2}$ and $m_{TX}$
variables~\cite{Lester:1999tx} to use the
information on both decay chains in the full event. The goals 
of such techniques are to establish that the DM is stabilized by a $Z_3$ symmetry
in other situations which were not discussed here, e.g., when a cusp-like topology is absent,  and to 
better eliminate the fakes of $Z_3$ signal by $Z_2$ models.  We also are studying other techniques to eliminate the fakes described above.

In all, we emphasize that parity symmetries are not necessarily the only way to stabilize the DM and 
we showed (via a few example cases) that 
models with other stabilization symmetries generally have testable consequences at the LHC, i.e.,
can potentially be distinguished from the parity case.  The reader
should regard this work as a first step into a more complete study of
{\it beyond} $Z_2$ stabilized dark matter scenarios.

\section*{Acknowledgments}

We would like to thank I.~Hinchliffe, A.~Katz, G.~Servant, M.D.~Shapiro, R.~Sundrum and B.~Tweedie for valuable
discussions.  K.A. was supported in part by NSF grant No.~PHY-0652363.  D.W. is supported in part by a University of California Presidential Fellowship. 


\section*{Appendix}

\subsection*{A. The distribution for the new topology}
\label{app:distnewtopo}

Most of the intermediate steps in the derivation of the cusp in Eq. (\ref{cusp})
are similar to
the analysis in reference \cite{Miller:2005zp} of the reaction in
Eq. (\ref{thrZ2}), except for the fact that a DM (i.e., massive) particle
is situated in-between two SM particles in the new topology (See Eq.~(\ref{newtopo})). Based on the algebra and the
notations found in reference \cite{Miller:2005zp}, we will derive a few useful
relations. 

Basically, the invariant mass formed by the two SM particles in this topology is given by 
\begin{eqnarray}
m_{ca}^2=(p_c+p_a)^2=2E_c E_a(1-\cos \theta_{ca})
\end{eqnarray}
where $\theta_{ca}$ is the opening angle between two visible
particles. Note that this relation is always valid in any frame so
that we can rewrite the above relation as  

\begin{eqnarray}
m_{ca}^2=2E_c^{(B)} E_a^{(B)}(1-\cos \theta_{ca}^{(B)}).\label{invmass}
\end{eqnarray}
Here and henceforth the (particle) superscripts on $\theta$'s (in this case B) imply that those angles
are measured in the rest
frame of the corresponding particle. Using energy-momentum conservation, we can
easily obtain the energies for $a$, DM, and $c$, which are measured in the rest frame
of particle $B$.  

\begin{eqnarray}
E_a^{(B)}&=&\frac{m_B^2-m_A^2}{2m_B} \\
E_{ \mathrm{DM} }^{(B)}&=&\frac{m_C^2-m_B^2-m_{ \mathrm{DM} }^2}{2m_B} \\
E_c^{(B)}&=&\frac{(m_D^2-m_C^2)m_B}{m_B^2+m_C^2-m_{ \mathrm{DM} }^2-\lambda^{1/2}(m_C^2,m_B^2,m_{ \mathrm{DM} }^2)\cos \theta_{c \; { \mathrm{DM} } }^{(B)}}
\end{eqnarray}
Inserting these relations into
Eq. (\ref{invmass}), we obtain 
\begin{eqnarray}
m_{ca}^2=2\cdot
\frac{(m_D^2-m_C^2)m_B}{m_B^2+m_C^2-m_{ \mathrm{DM} }^2-\lambda^{1/2}(m_C^2,m_B^2,m_{ \mathrm{DM} }^2)\cos
  \theta_{c \; { \mathrm{DM} } }^{(B)}}\cdot \frac{m_B^2-m_A^2}{2m_B}(1-\cos
\theta_{ca}^{(B)}). 
\end{eqnarray}  
We easily see that the maximum of $m_{ca}^2$ occurs when $\cos
\theta_{c \; { \mathrm{DM} } }^{(B)}=1$ and $\cos \theta_{ca}^{(B)}=-1$. We want to
express the invariant mass $m_{ca}$ in terms of variables which have flat
distributions: 
this is the case for $\cos
\theta_{ca}^{(B)}$, but not for $\cos
\theta_{c \; { \mathrm{DM} }}^{(B)}$. So, we need
to express $\cos \theta_{c \; { \mathrm{DM} }}^{(B)}$ in terms of $\cos
\theta_{c \; { \mathrm{DM} }}^{(C)}$ (i.e., the same angle in the rest frame of particle
$C$) for which the distribution is also flat. This relation can be found by calculating $m_{c \; { \mathrm{DM} } }^2$ in the rest
frames of particle $C$ and $B$:
\begin{eqnarray}
m_{c \; { \mathrm{DM} }}^2&=&m_{ \mathrm{DM} }^2+2E_c^{(C)}E_{ \mathrm{DM} }^{(C)}-2E_c^{(C)}\sqrt{(E_{ \mathrm{DM} }^{(C)})^2-m_{ \mathrm{DM} }^2}\cos \theta_{c \; { \mathrm{DM} }}^{(C)}\label{mcbinc} \\
&=&m_{ \mathrm{DM} }^2+2E_c^{(B)}E_{ \mathrm{DM} }^{(B)}-2E_c^{(B)}\sqrt{(E_{ \mathrm{DM} }^{(B)})^2-m_{ \mathrm{DM} }^2}\cos \theta_{c \; { \mathrm{DM} } }^{(B)}\label{mcbinb}
\end{eqnarray}
Again, the energy-momentum conservation in the rest frame of $C$ gives the following relations:
\begin{eqnarray}
E_{ \mathrm{DM} }^{(C)}&=&\frac{m_C^2-m_B^2+m_{ \mathrm{DM} }^2}{2m_C} \\
E_c^{(C)}&=&\frac{m_D^2-m_C^2}{2m_C}
\end{eqnarray}
Substitution of $E_{ \mathrm{DM} }$ and $E_c$ in the rest frame of $C$ and $B$ into
Eq. (\ref{mcbinc}) and Eq. (\ref{mcbinb}) gives the relation between
$\cos \theta_{c { \mathrm{DM} }}^{(B)}$ and $\cos \theta_{c { \mathrm{DM} } }^{(C)}$:
\begin{eqnarray}
\hspace{-1cm}\frac{2m_B^2}{m_C^2+m_B^2-m_{ \mathrm{DM} }^2-\lambda^{1/2}(m_C^2,m_B^2,m_{ \mathrm{DM} }^2)\cos
  \theta_{c \; { \mathrm{DM} }
  }^{(B)}}=\left(1-\frac{m_C^2-m_B^2+m_{ \mathrm{DM}
  }^2-\lambda^{1/2}(m_C^2,m_B^2,m_{ \mathrm{DM} }^2)\cos 
  \theta_{c \; { \mathrm{DM} } }^{(C)}}{2m_C^2}\right)\label{reltwovariable} 
\end{eqnarray}
Next, we introduce the variables $u$ and $v$:  
\begin{eqnarray}
u\equiv \frac{1-\cos \theta_{c \; { \mathrm{DM} } }^{(C)}}{2},\;\;\; v\equiv \frac{1-\cos \theta_{ca}^{(B)}}{2}
\end{eqnarray}
and using
Eq. (\ref{reltwovariable}), we express $m_{ca}^2$ in terms of $u$ and $v$:
\begin{eqnarray}
m_{ca}^2=(m_{ca}^{\textnormal{max}})^2(1-\alpha u)v\label{invinflat}
\end{eqnarray}
where
\begin{eqnarray}
(m_{ca}^{\textnormal{max}})^2&=&\frac{2(m_D^2-m_C^2)(m_B^2-m_A^2)}{m_B^2+m_C^2-m_{ \mathrm{DM} }^2-\lambda^{1/2}(m_C^2,m_B^2,m_{ \mathrm{DM} }^2)}.
\end{eqnarray}
Note that the differential distributions for $u$ and $v$ ($0\leq u,v\leq 1$) are also flat:
\begin{eqnarray}
\frac{1}{\Gamma}\frac{\partial^2\Gamma}{\partial u\partial v}=\theta(u)\theta(1-u)\theta(v)\theta(1-v)
\end{eqnarray}
where $\theta(x)$ is the usual step function. Replacing $u$ and $v$ by
$u$ and $m_{ca}^2$ by using Eq. (\ref{invinflat}) gives the
differential distribution 
\begin{eqnarray}
\frac{1}{\Gamma}\frac{\partial^2\Gamma}{\partial u\partial m_{ca}^2}=\hat{\theta}\left(\frac{m_{ca}^2}{(m_{ca}^{\textnormal{max}})^2(1-\alpha u)}\right)\frac{\hat{\theta}(u)}{(m_{ca}^{\textnormal{max}})^2(1-\alpha u)}
\end{eqnarray}
where a ``top-hat'' function $\hat{\theta}(x)\equiv
\theta(x)\theta(1-x)$. The next step is to integrate over $u$ to find
the distribution in $m_{ca}^2$: 
\begin{eqnarray}
\frac{1}{\Gamma}\frac{\partial^2\Gamma}{\partial m_{ca}^2}
&=&\int_{-\infty}^{\infty}\frac{1}{\Gamma}\frac{\partial^2\Gamma}{\partial u\partial m_{ca}^2}du \nonumber \\
&=&\int_0^1 \hat{\theta}\left(\frac{m_{ca}^2}{(m_{ca}^{\textnormal{max}})^2(1-\alpha u)}\right)\frac{1}{(m_{ca}^{\textnormal{max}})^2(1-\alpha u)}du \nonumber \\
&=&\int_0^{u_{\textnormal{max}}}\frac{1}{(m_{ca}^{\textnormal{max}})^2(1-\alpha u)}du
\end{eqnarray}
for $0\leq m_{ca} \leq m_{ca}^{\textnormal{max}}$, where
\begin{eqnarray}
u_{\textnormal{max}}=\textnormal{Max}\left(1,\;\frac{1}{\alpha}\left[1-\frac{m_{ca}^2}{(m_{ca}^{\textnormal{max}})^2}\right]\right).
\end{eqnarray}
Now the above integral is easy to evaluate, and we finally obtain the distribution which was
given earlier in
Eq. (\ref{cusp}): 
\begin{eqnarray}
\frac{1}{\Gamma}\frac{\partial^2\Gamma}{\partial m_{ca}^2}&=&
\left\{
\begin{array}{l}
\frac{1}{(m_{ca}^{\textnormal{max}})^2 \alpha} \ln \frac{m_C^2}{m_B^2} \hspace{1cm} \textnormal{for\hspace{.5cm} \ }  0<m_{ca}<\sqrt{1-\alpha} m_{ca}^{\textnormal{max}} \cr
\cr
\frac{1}{(m_{ca}^{\textnormal{max}})^2 \alpha} \ln \frac{(m_{ca}^{\textnormal{max}})^2}{m_{ca}^2} \;\;\; \textnormal{for}\hspace{.5cm} \ \sqrt{1-\alpha} m_{ca}^{\textnormal{max}}<m_{ca}<m_{ca}^{\textnormal{max}}.
\end{array}\right.
\end{eqnarray}


\subsection*{B. Signal Topologies from Section~\ref{subsec:prorate}}
\label{app:signaltopo}

In addition to the decay chain in figure~\ref{origtopo} for the model
presented in section~\ref{subsec:stabilmodel}, there are additional
corrections by the long-lived $SU(2)_D$ gauge bosons.  The masses are
listed in Eq.~\ref{eq:masses}.  As described above, the virtual
particles in the decay chain go slightly off-shell when emitting the
new gauge boson; however, because there are three of them, the
additional multiplicity factor helps to ameliorate this suppression.
In this appendix for each topology, we plot the invariant mass
distributions with the cuts in section~\ref{subsec:extract} (see Figure~\ref{topo1result}-\ref{topo6result}). 

\begin{figure}[t]
\centering
\includegraphics[height=1.8cm,clip]{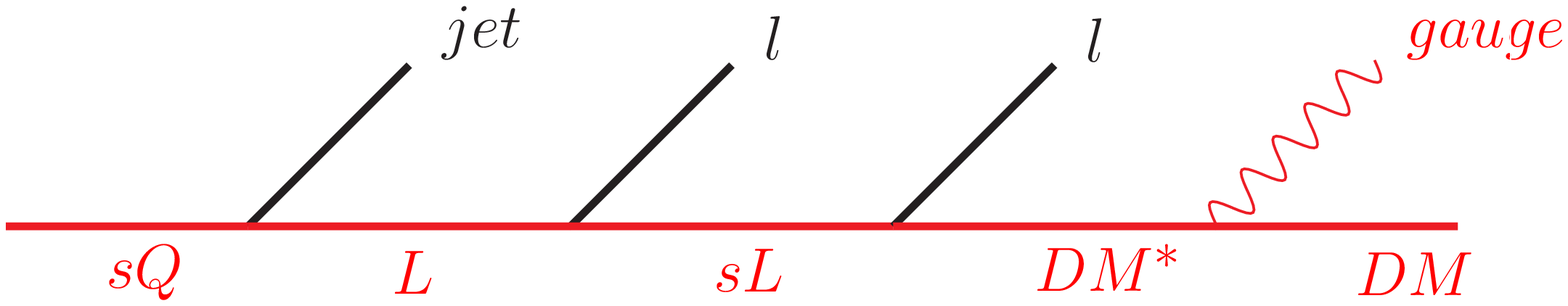}
\vspace{-1.8cm}
\bea
\label{Z3topo1}
\eea

\vspace{0.8cm}
\includegraphics[height=1.95cm,clip]{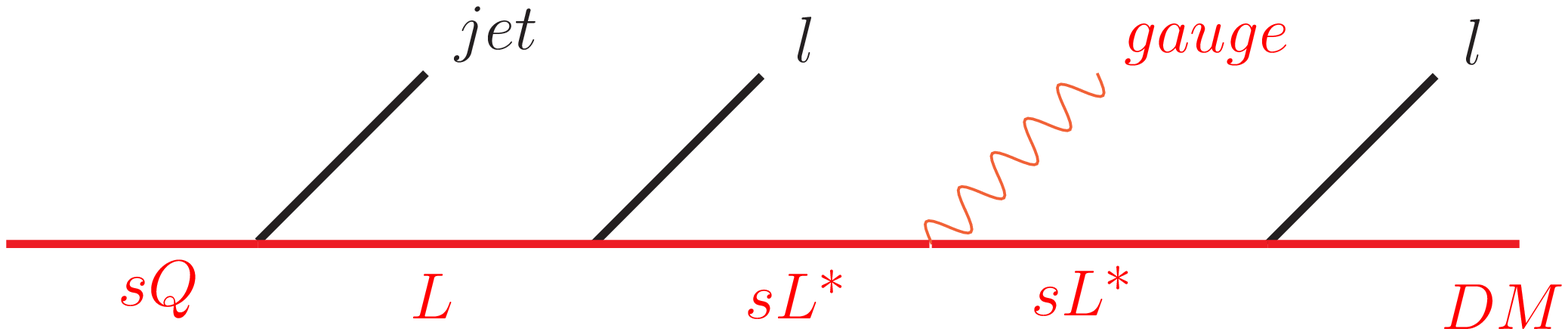}
\vspace{-1.8cm}
\bea
\label{Z3topo2}
\eea

\vspace{0.8cm}
\includegraphics[height=1.97cm,clip]{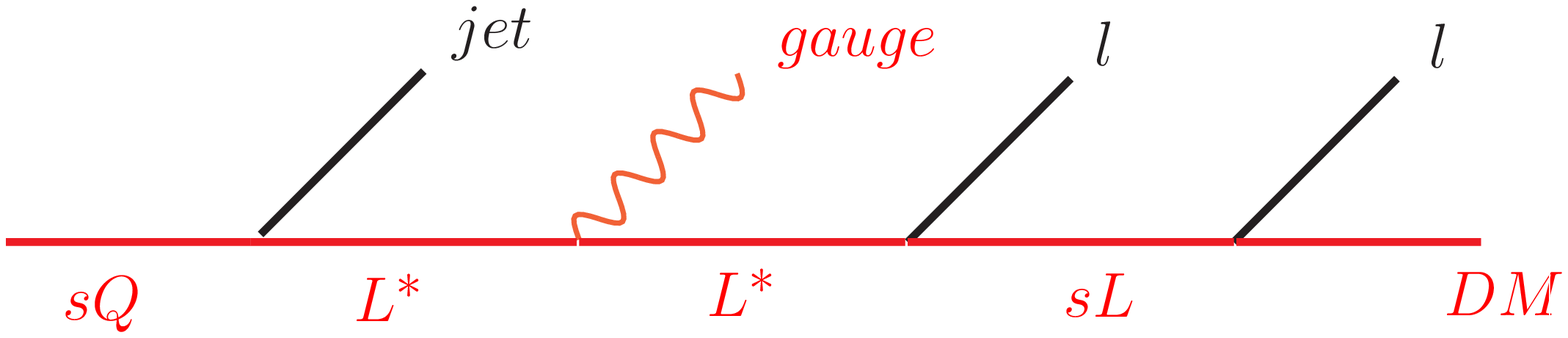}
\vspace{-1.8cm}
\bea
\label{Z3topo3}
\eea

\vspace{0.8cm}
\includegraphics[height=2.0cm,clip]{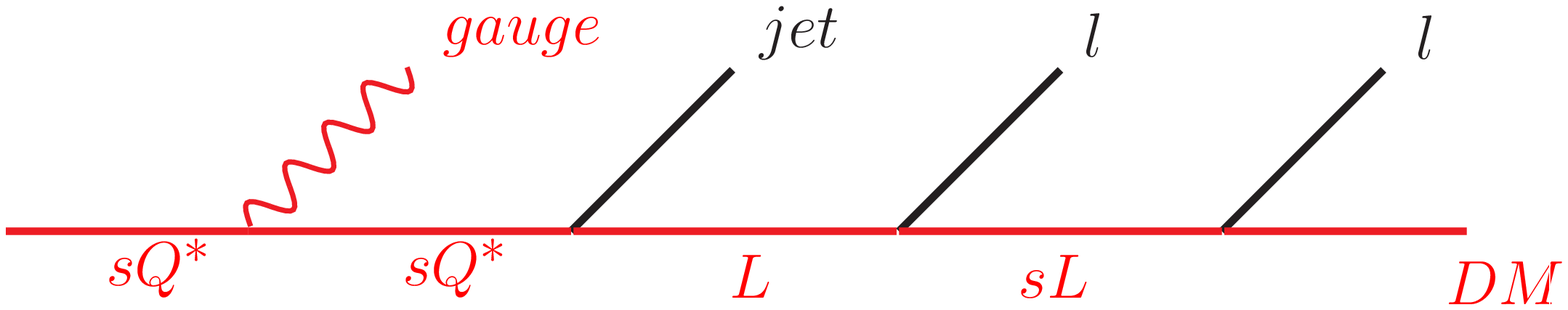}
\vspace{-1.8cm}
\bea
\label{Z3topo4}
\eea
\vspace{0.4cm}
\end{figure}
\begin{figure}
Figures~\ref{Z3topo1}-\ref{Z3topo4}:   Order $\alpha$ corrections by
the $SU(2)_D$ gauge bosons to the decay chain in
Figure~\ref{origtopo}.  See the model in
section~\ref{subsec:stabilmodel}. \vspace{.2cm} 
\end{figure}

\vspace{.2cm}

\begin{figure}[h!]
\centering
\includegraphics[height=2.1cm,clip]{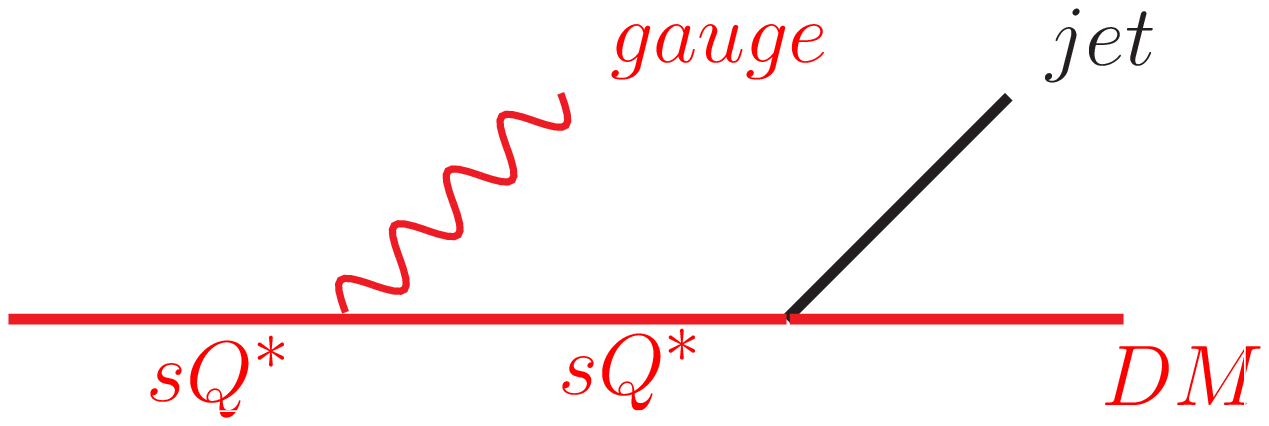}
\hspace{0.5cm}
\includegraphics[height=1.95cm,clip]{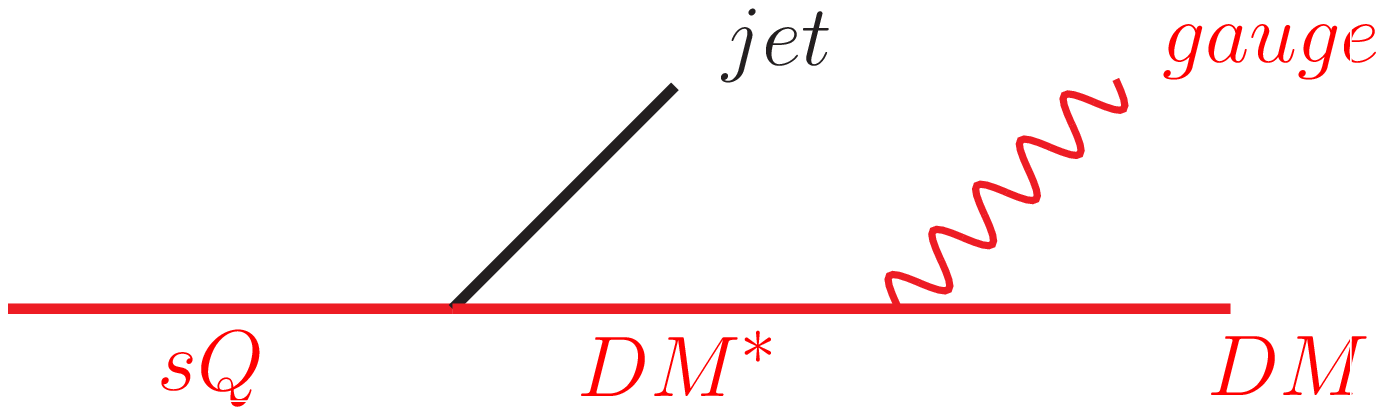}
\vspace{-1.8cm}
\bea
\label{Z3topo5}
\eea
\centering
\vspace{0.6cm}
\end{figure}
\begin{figure}
Figure~\ref{Z3topo5}:   Order $\alpha$ corrections by the $SU(2)_D$
gauge bosons to the decay chain in Figure~\ref{origtopo}.  See the
model in section~\ref{subsec:stabilmodel}. 
\end{figure}

\vspace{10cm}
\begin{figure}[h!]
  \centering
  \includegraphics[width=7.2truecm,height=6.7truecm,clip=true]{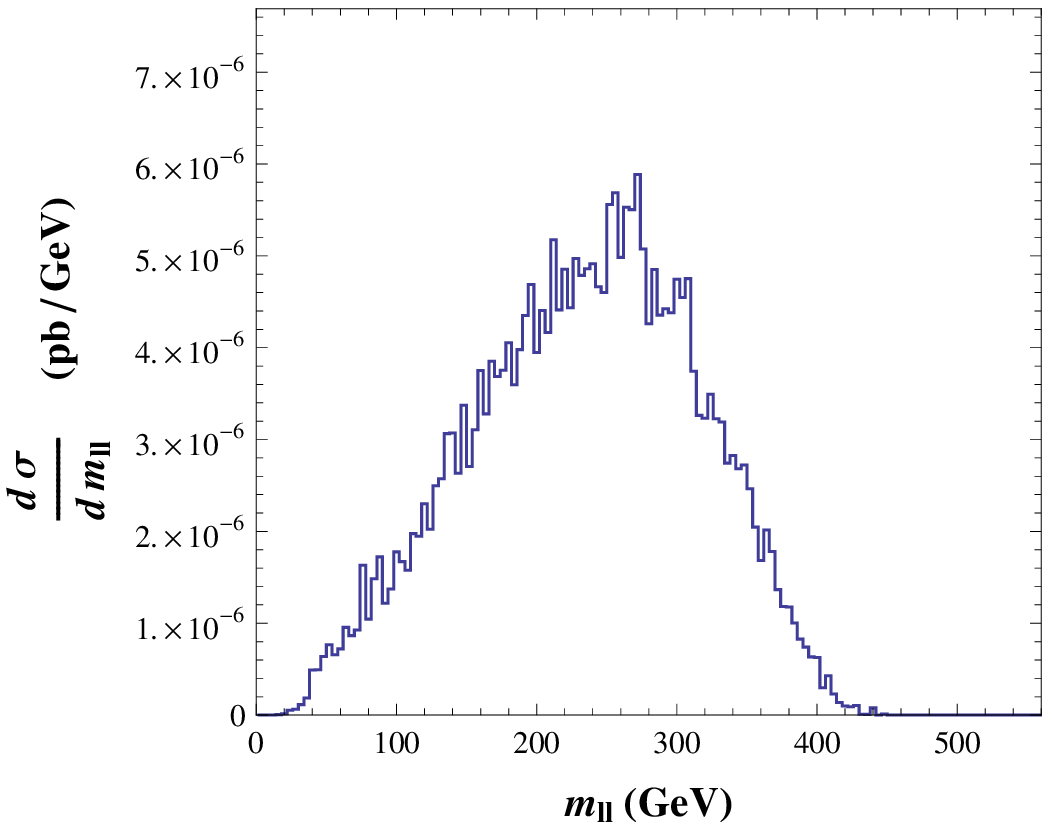} \hspace{0.2cm}
   \includegraphics[width=7.2truecm,height=6.7truecm,clip=true]{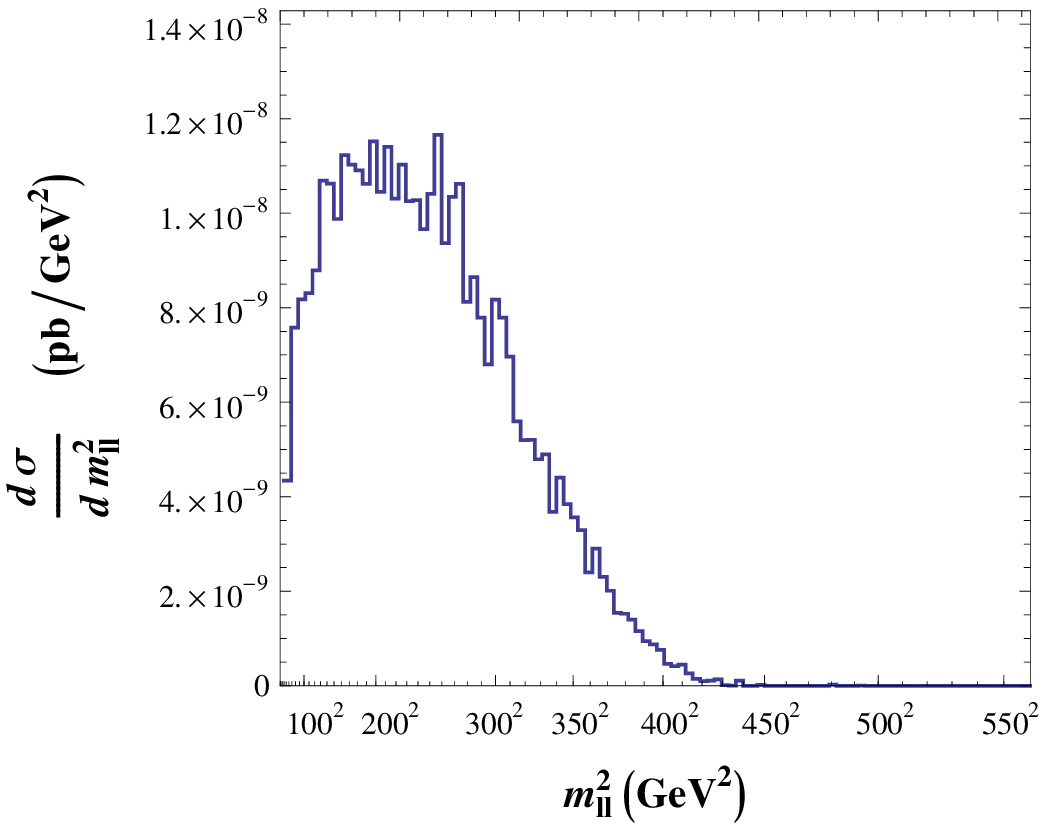}
\caption{The plots corresponding to the topology of Figure~\ref{Z3topo1}.}   
\label{topo1result}
\end{figure}
\begin{figure}[h!]
  \centering
  \includegraphics[width=7.2truecm,height=6.7truecm,clip=true]{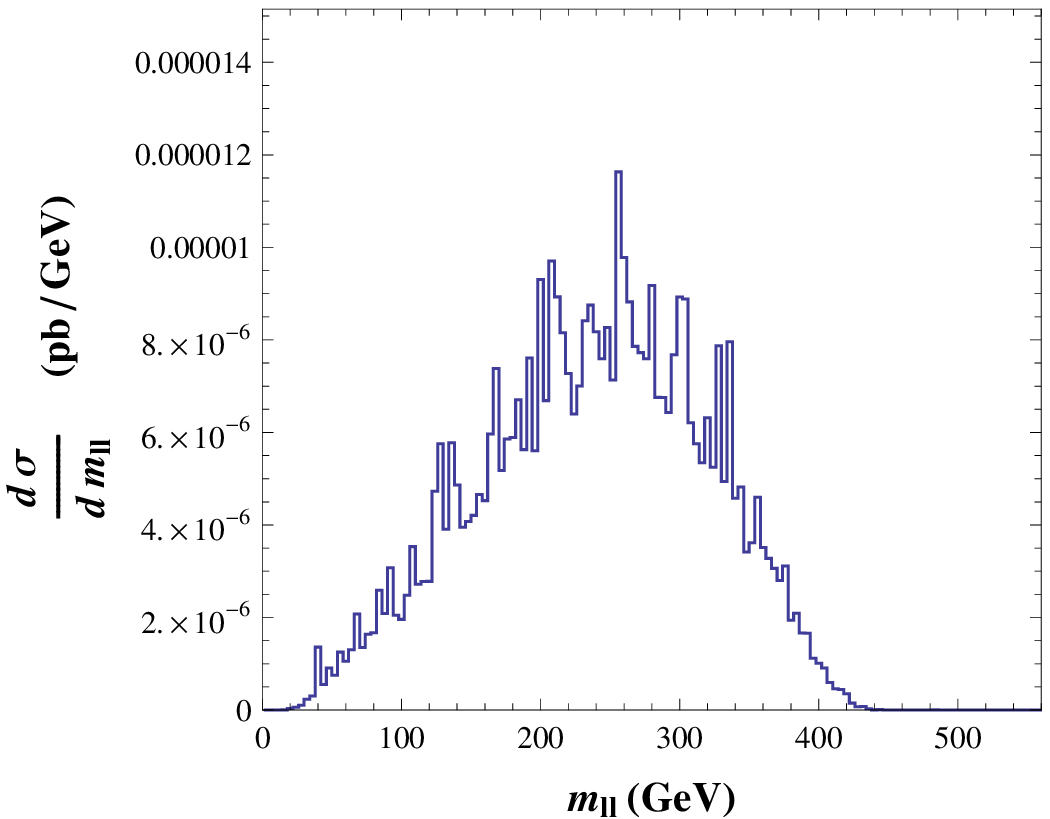} \hspace{0.2cm}
   \includegraphics[width=7.2truecm,height=6.7truecm,clip=true]{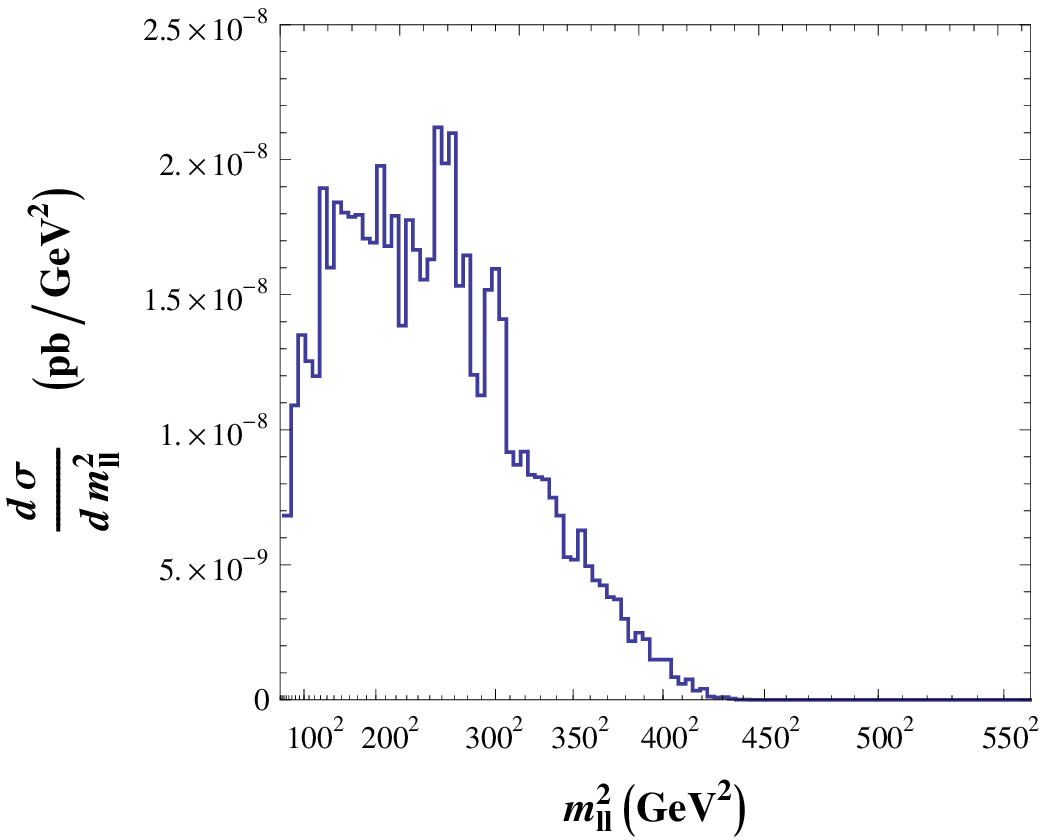}
\caption{The plots corresponding to the topology of Figure~\ref{Z3topo2}.}   
\label{topo2result}
\end{figure}
\begin{figure}[h!]
  \centering
  \includegraphics[width=7.2truecm,height=6.7truecm,clip=true]{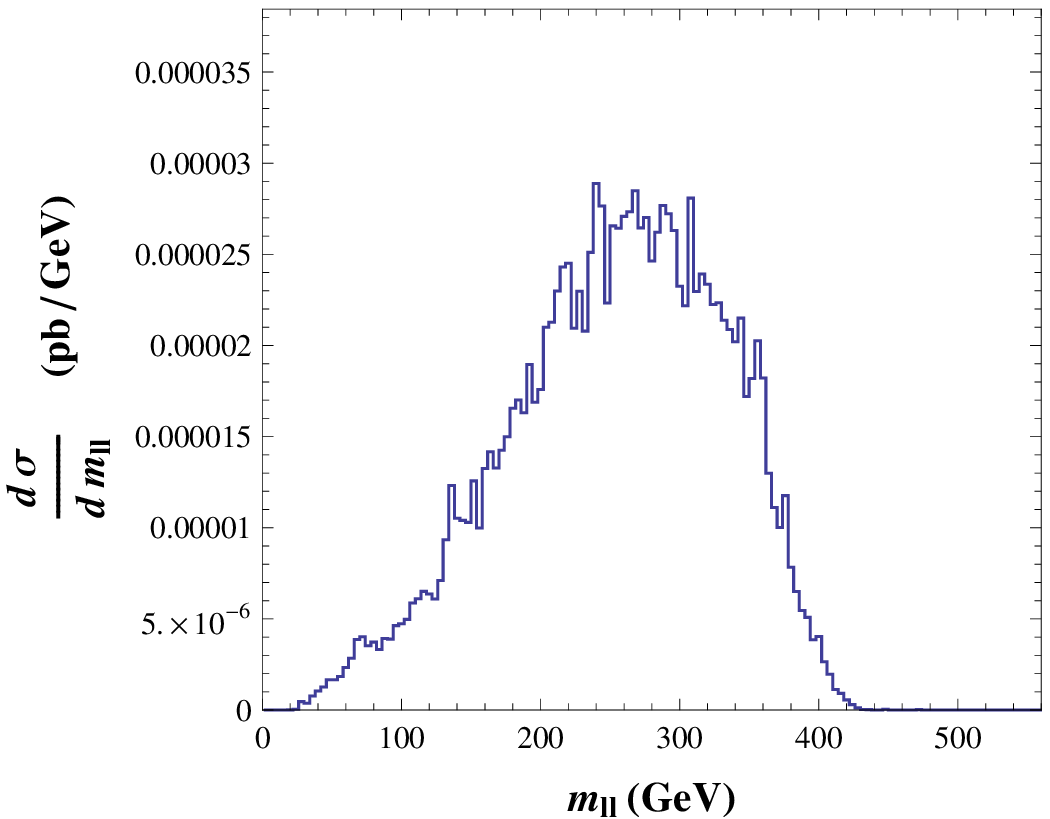} \hspace{0.2cm}
   \includegraphics[width=7.2truecm,height=6.7truecm,clip=true]{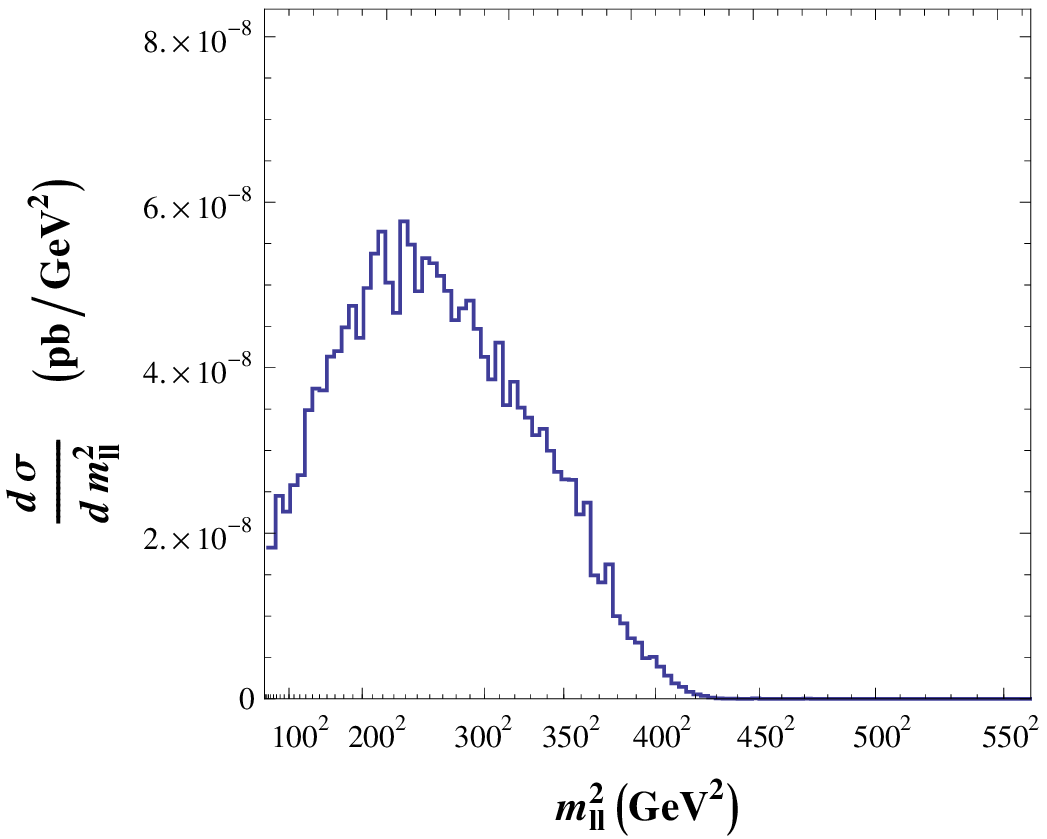}
\caption{The plots corresponding to the topology of Figure~\ref{Z3topo3}.}   
\label{topo3result}
\end{figure}
\begin{figure}[h!]
  \centering
  \includegraphics[width=7.2truecm,height=6.7truecm,clip=true]{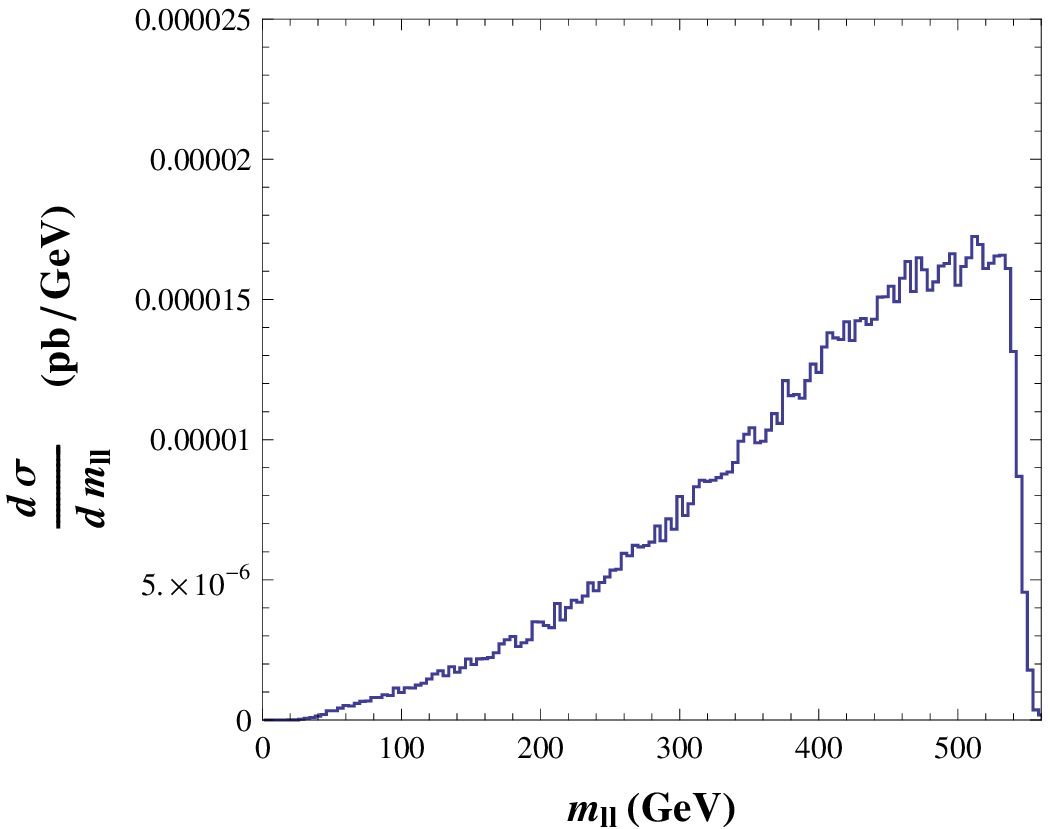} \hspace{0.2cm}
   \includegraphics[width=7.2truecm,height=6.7truecm,clip=true]{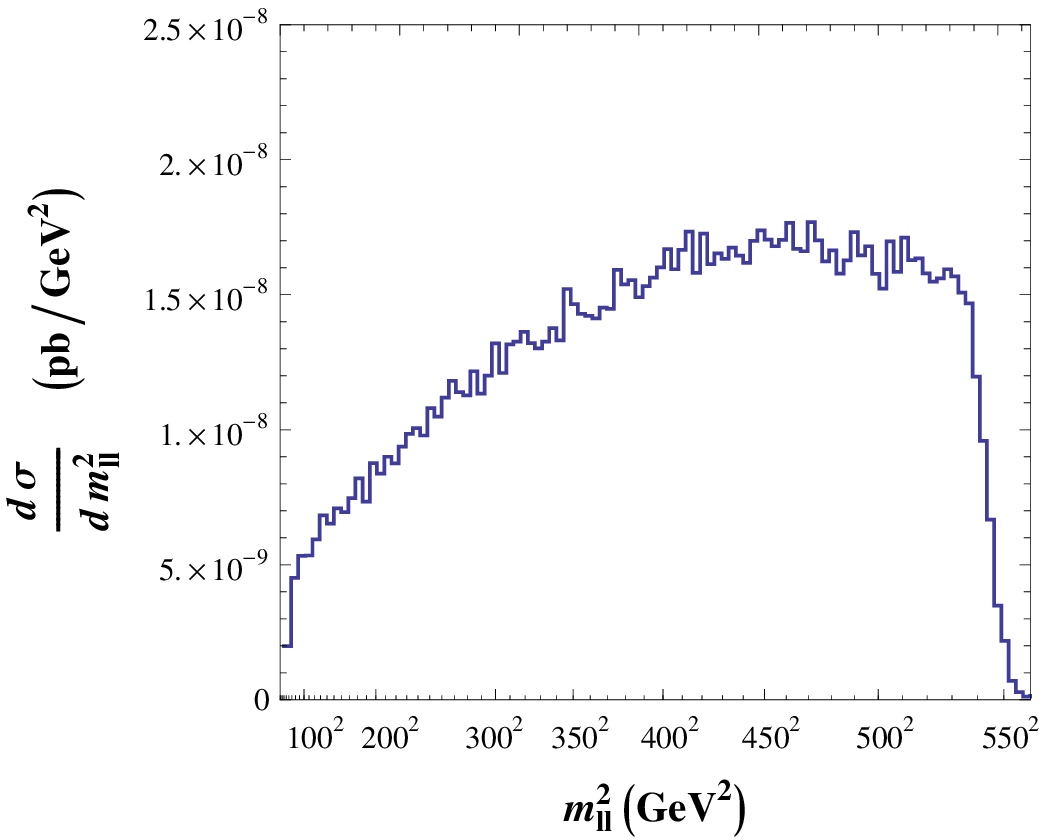}
\caption{The plots corresponding to the topology of Figure~\ref{Z3topo4}.}   
\label{topo4result}
\end{figure}

\begin{figure}[h!]
  \centering
  \includegraphics[width=7.2truecm,height=6.7truecm,clip=true]{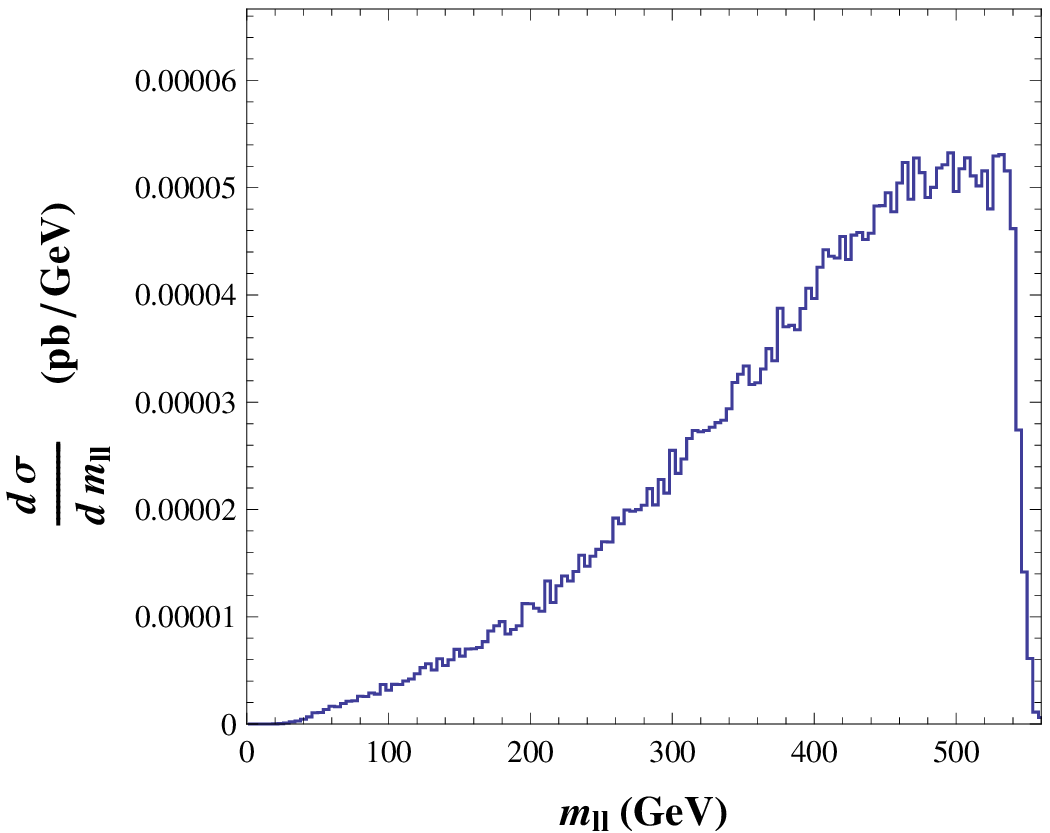} \hspace{0.2cm}
   \includegraphics[width=7.2truecm,height=6.7truecm,clip=true]{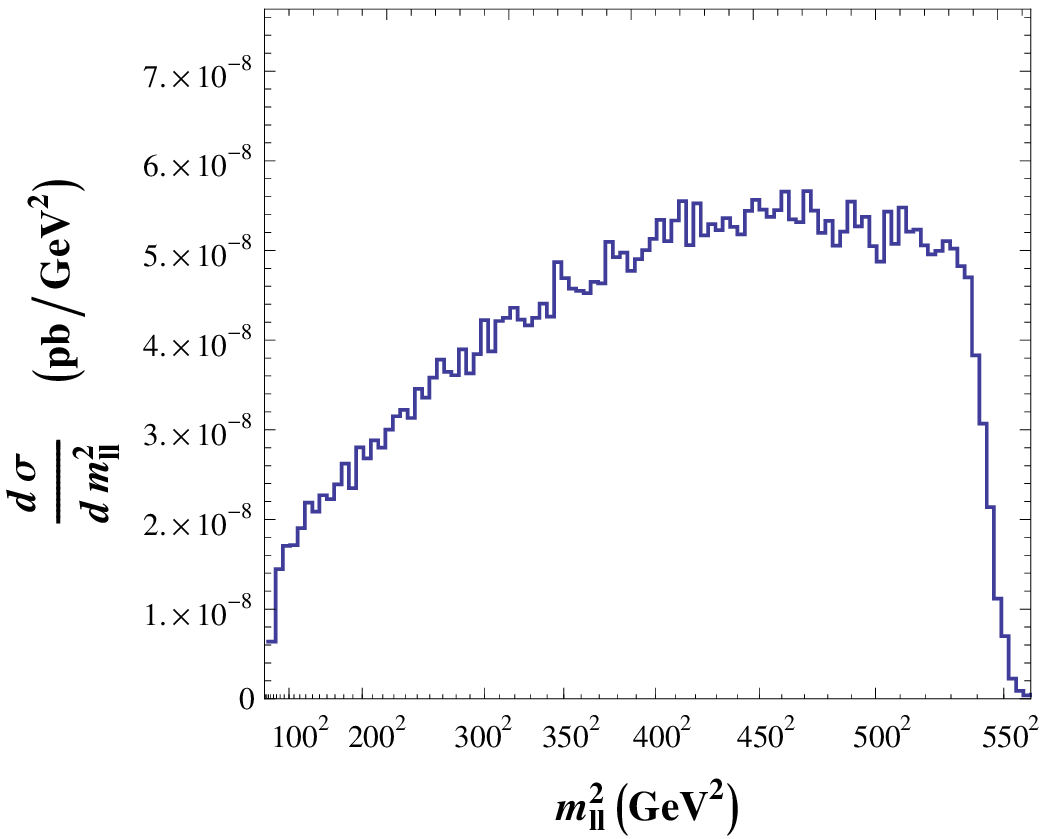}
\caption{The plots corresponding to the topology of Figure~\ref{Z3topo5}.}   
\label{topo5result}
\end{figure}

\begin{figure}[h!]
  \centering
  \includegraphics[width=7.2truecm,height=6.7truecm,clip=true]{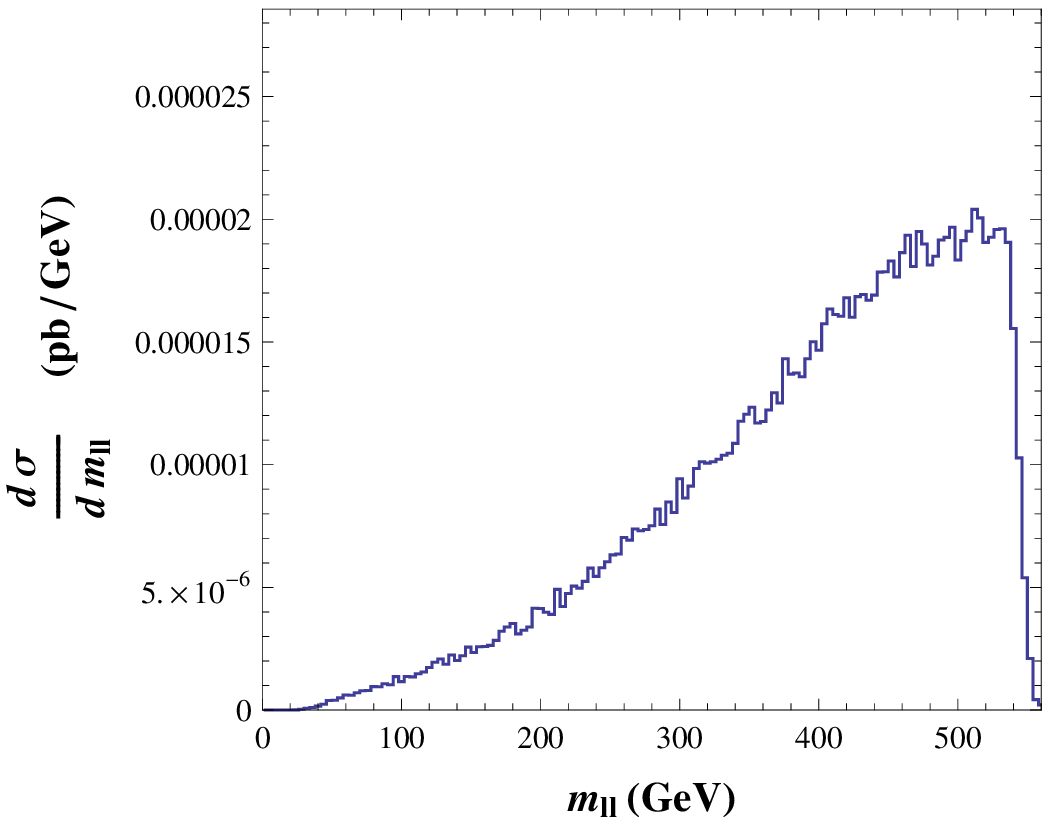} \hspace{0.2cm}
   \includegraphics[width=7.2truecm,height=6.7truecm,clip=true]{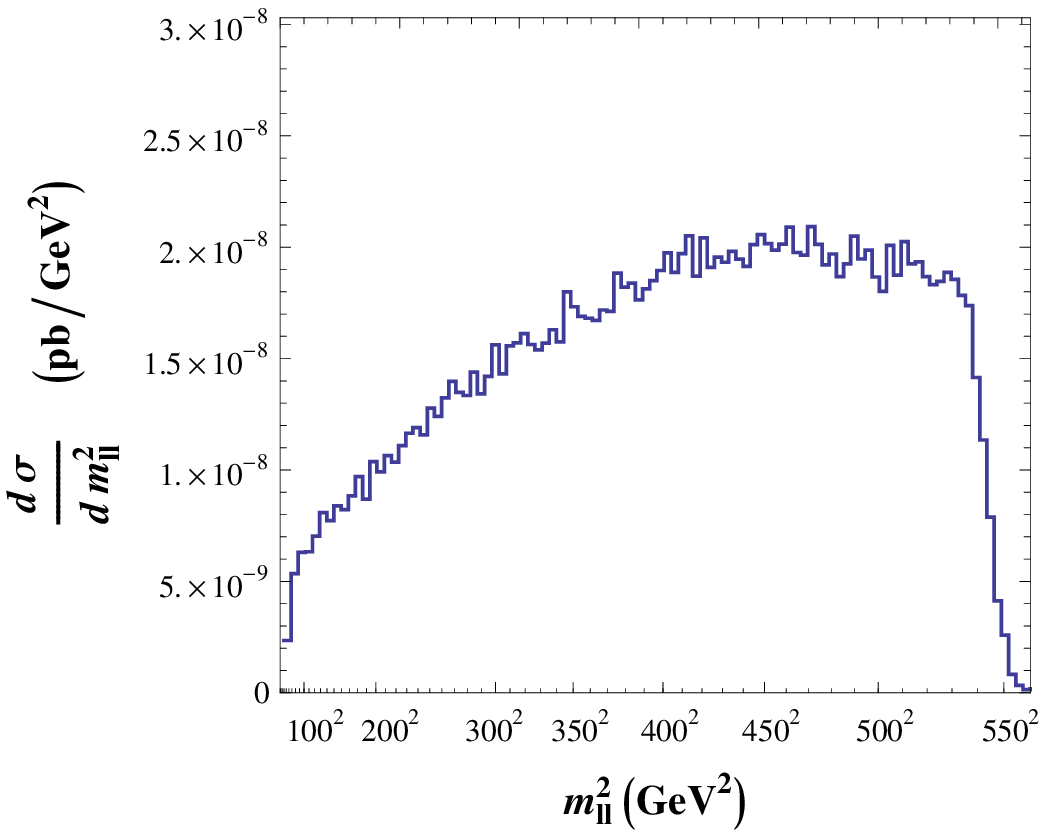}
\caption{The plots corresponding to the topology of Figure~\ref{Z3topo5}.}   
\label{topo6result}
\end{figure}


\begin{thebibliography}{99}

\bibitem{Bertone:2004pz}
  See, for example, G.~Bertone, D.~Hooper and J.~Silk,
  Phys.\ Rept.\  {\bf 405}, 279 (2005)
[arXiv:hep-ph/0404175].


\bibitem{Hinshaw:2008kr}
  G.~Hinshaw {\it et al.}  [WMAP Collaboration],
  arXiv:0803.0732 [astro-ph].

\bibitem{Kolb:1990vq}
  E.~W.~Kolb and M.~S.~Turner,
  Front.\ Phys.\  {\bf 69}, 1 (1990).



\bibitem{Lester:1999tx}
  C.~G.~Lester and D.~J.~Summers,
  Phys.\ Lett.\  B {\bf 463}, 99 (1999)
  [arXiv:hep-ph/9906349];
%
  A.~J.~Barr, C.~G.~Lester, M.~A.~Parker, B.~C.~Allanach and P.~Richardson,
  JHEP {\bf 0303}, 045 (2003)
  [arXiv:hep-ph/0208214];
 %
  A.~Barr, C.~Lester and P.~Stephens,
  J.\ Phys.\ G {\bf 29}, 2343 (2003)
  [arXiv:hep-ph/0304226];
%
  W.~S.~Cho, K.~Choi, Y.~G.~Kim and C.~B.~Park,
  Phys.\ Rev.\ Lett.\  {\bf 100}, 171801 (2008)
  [arXiv:0709.0288 [hep-ph]];
%
  B.~Gripaios,
  JHEP {\bf 0802}, 053 (2008)
  [arXiv:0709.2740 [hep-ph]];
%
  A.~J.~Barr, B.~Gripaios and C.~G.~Lester,
  JHEP {\bf 0802}, 014 (2008)
  [arXiv:0711.4008 [hep-ph]];
%
  W.~S.~Cho, K.~Choi, Y.~G.~Kim and C.~B.~Park,
  JHEP {\bf 0802}, 035 (2008)
  [arXiv:0711.4526 [hep-ph]].

\bibitem{Cheng:2008mg}
  H.~C.~Cheng, D.~Engelhardt, J.~F.~Gunion, Z.~Han and B.~McElrath,
  Phys.\ Rev.\ Lett.\  {\bf 100}, 252001 (2008)
  [arXiv:0802.4290 [hep-ph]];
%
  H.~C.~Cheng, J.~F.~Gunion, Z.~Han and B.~McElrath,
 Phys.\ Rev.\  D {\bf 80}, 035020 (2009)
 [arXiv:0905.1344 [hep-ph]].

\bibitem{Matchev:2009iw}
  K.~T.~Matchev, F.~Moortgat, L.~Pape and M.~Park,
  JHEP {\bf 0908}, 104 (2009)
  [arXiv:0906.2417 [hep-ph]];
%
  P.~Konar, K.~Kong, K.~T.~Matchev and M.~Park,
  arXiv:0911.4126 [hep-ph].

\bibitem{Han:2009ss}
  T.~Han, I.~W.~Kim and J.~Song,
  arXiv:0906.5009 [hep-ph].


  




\bibitem{Jungman:1995df}
  G.~Jungman, M.~Kamionkowski and K.~Griest,
  Phys.\ Rept.\  {\bf 267}, 195 (1996)
  [arXiv:hep-ph/9506380].


\bibitem{Lee:2008pc}
  H.~S.~Lee,
  Phys.\ Lett.\  B {\bf 663}, 255 (2008)
  [arXiv:0802.0506 [hep-ph]].

\bibitem{Cheng:2003ju}
  H.~C.~Cheng and I.~Low,
  JHEP {\bf 0309}, 051 (2003) and
%
  JHEP {\bf 0408}, 061 (2004)
  [arXiv:hep-ph/0405243].

\bibitem{Servant:2002aq}
  G.~Servant and T.~M.~P.~Tait,
  Nucl.\ Phys.\  B {\bf 650}, 391 (2003);
%
H.~C.~Cheng, J.~L.~Feng and K.~T.~Matchev,
  Phys.\ Rev.\ Lett.\  {\bf 89}, 211301 (2002)
  [arXiv:hep-ph/0207125].
%

\bibitem{Agashe:2007jb}
  K.~Agashe, A.~Falkowski, I.~Low and G.~Servant,
  JHEP {\bf 0804}, 027 (2008).
  





\bibitem{ATDR}
The ATLAS Collaboration, CERN-LHCC-99-015.

\bibitem{CTDR}
The CMS Collaboration, J. Phys. G: Nucl. Part. Phys. 34 995-1579 (2007).

  
\bibitem{Walker:2009en}
  D.~G.~E.~Walker,
  arXiv:0907.3146 [hep-ph].

\bibitem{Walker:2009ei}
  D.~G.~E.~Walker,
  arXiv:0907.3142 [hep-ph].
  

\bibitem{Agashe:2004ci}
  K.~Agashe and G.~Servant,
  Phys.\ Rev.\ Lett.\  {\bf 93}, 231805 (2004)
  [arXiv:hep-ph/0403143].

\bibitem{Agashe:2004bm}
  K.~Agashe and G.~Servant,
  JCAP {\bf 0502}, 002 (2005)
  [arXiv:hep-ph/0411254].


\bibitem{Ma:2007gq}
  E.~Ma,
  Phys.\ Lett.\  B {\bf 662}, 49 (2008)
  [arXiv:0708.3371 [hep-ph]].



\bibitem{Byckling:1973bk}
  E.~Byckling and K.~Kajantie,
  Particle Kinematics (John Wiley \& Sons, 1973).

\bibitem{Miller:2005zp}
  D.~J.~Miller, P.~Osland and A.~R.~Raklev,
  JHEP {\bf 0603}, 034 (2006)
  [arXiv:hep-ph/0510356].
  
\bibitem{Kraml:2005kb}
  S.~Kraml and A.~R.~Raklev,
  Phys.\ Rev.\  D {\bf 73}, 075002 (2006)
  [arXiv:hep-ph/0512284].

\bibitem{Wang:2006hk}
  L.~T.~Wang and I.~Yavin,
  JHEP {\bf 0704}, 032 (2007)
  [arXiv:hep-ph/0605296].

 
\bibitem{Alwall:2007st}
  J.~Alwall {\it et al.},
  JHEP {\bf 0709}, 028 (2007)
  [arXiv:0706.2334 [hep-ph]].
  
  \bibitem{Amsler:2008zzb}
  C.~Amsler {\it et al.}  [Particle Data Group],
  Phys.\ Lett.\  B {\bf 667}, 1 (2008).
  
  \bibitem{Lai:1996mg}
  H.~L.~Lai {\it et al.},
  Phys.\ Rev.\  D {\bf 55}, 1280 (1997)
  [arXiv:hep-ph/9606399].

  

  
  
\bibitem{Davoudiasl:2009cd}For a review and the
original references, see 
  H.~Davoudiasl, S.~Gopalakrishna, E.~Ponton and J.~Santiago,
  arXiv:0908.1968 [hep-ph].


\bibitem{Agashe:2005vg}
  K.~Agashe, R.~Contino and R.~Sundrum,
  Phys.\ Rev.\ Lett.\  {\bf 95}, 171804 (2005)
  [arXiv:hep-ph/0502222].


\bibitem{Agashe:2009ja}
  K.~Agashe, K.~Blum, S.~J.~Lee and G.~Perez,
  Phys.\ Rev.\  D {\bf 81}, 075012 (2010)
  [arXiv:0912.3070 [hep-ph]].


\bibitem{Baer:2008ih}
For a supersymmetric example see,
  H.~Baer, A.~Mustafayev, E.~K.~Park and X.~Tata,
  JHEP {\bf 0805}, 058 (2008)
  [arXiv:0802.3384 [hep-ph]].
  


 
 \bibitem{multiple} For some recent studies, see, for example, 
  K.~M.~Zurek,
  Phys.\ Rev.\  D {\bf 79}, 115002 (2009)
  [arXiv:0811.4429 [hep-ph]];
 %
  F.~Chen, J.~M.~Cline and A.~R.~Frey,
  Phys.\ Rev.\  D {\bf 80}, 083516 (2009)
  [arXiv:0907.4746 [hep-ph]];
%
  D.~Feldman, Z.~Liu, P.~Nath and G.~Peim,
  Phys.\ Rev.\  D {\bf 81}, 095017 (2010)
  [arXiv:1004.0649 [hep-ph]];
%
  M.~Cirelli and J.~M.~Cline,
  arXiv:1005.1779 [hep-ph].


  

\end{thebibliography}
\end{document}